\documentclass{egpubl}
\usepackage{hpg23}
 
\SpecialIssuePaper         

\CGFccby

\usepackage[T1]{fontenc}
\usepackage{dfadobe}  

\usepackage{cite}  

\BibtexOrBiblatex
\electronicVersion
\PrintedOrElectronic

\ifpdf \usepackage[pdftex]{graphicx} \pdfcompresslevel=9
\else \usepackage[dvips]{graphicx} \fi

\usepackage{booktabs} 
\usepackage{multirow}
\usepackage{xcolor}
\usepackage{subcaption}
\usepackage{ulem}
\usepackage{xcolor}
\usepackage{makecell}
\usepackage{pgfplots}
\pgfplotsset{compat=1.14}
\usepackage{amsfonts}
\usepackage{amsmath}

\usepackage{egweblnk} 

\title[Neural Intersection Function]%
      {Neural Intersection Function}

\author[S. Fujieda \& C.\,C. Kao \& T. Harada]
{\parbox{\textwidth}{\centering S. Fujieda\orcid{0000-0002-2472-7365}
          C.\,C. Kao\orcid{0000-0002-7631-2284}
          T. Harada \orcid{0000-0001-5158-8455}
         }
         \\
 {\parbox{\textwidth}{\centering Advanced Micro Devices, Inc.
        }
 }
 }

\begin{document}

\teaser{
  \centering
    \begin{subfigure}[b]{0.325\textwidth}
        \centering
        \includegraphics[width=\textwidth]{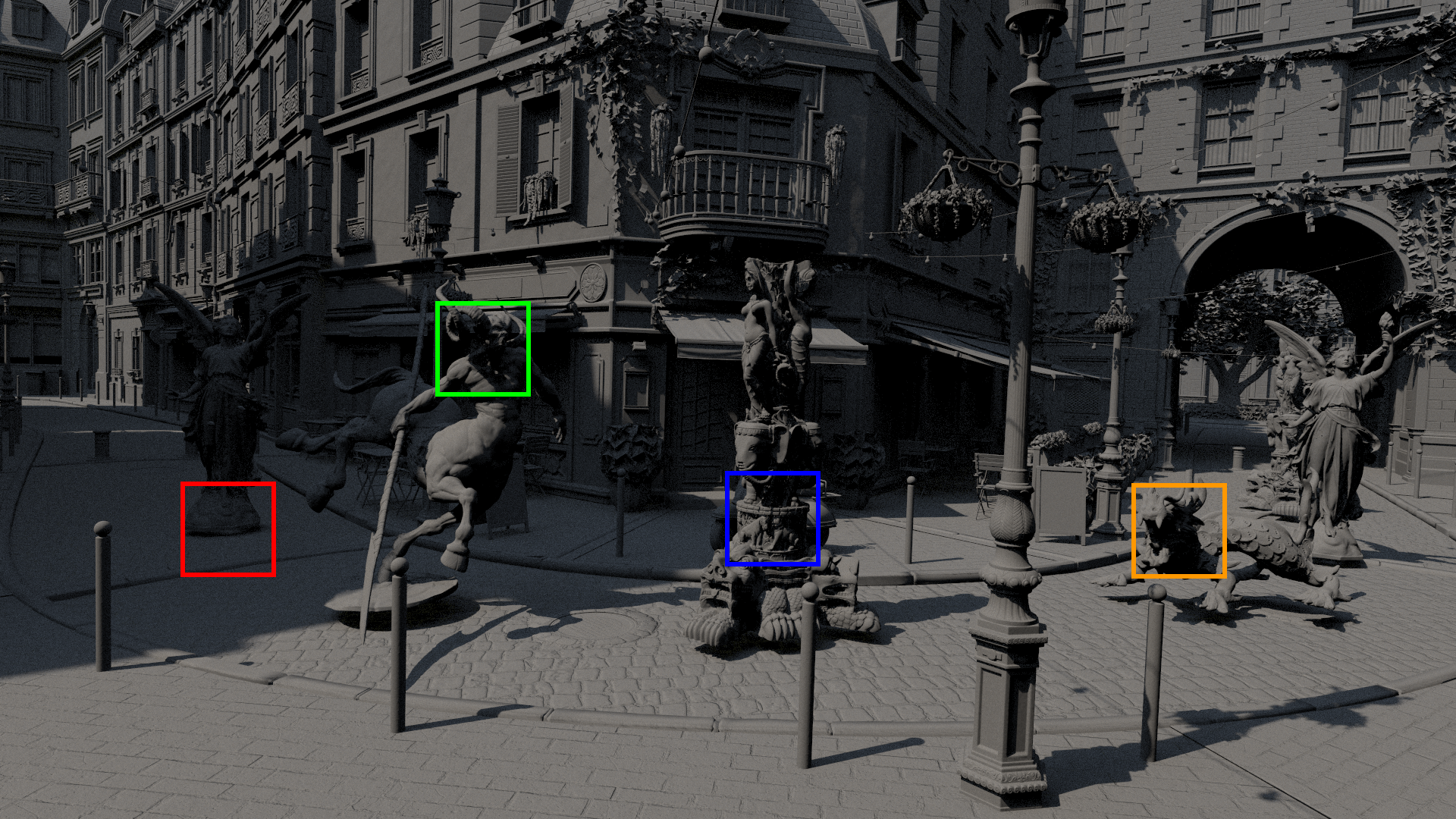}
    \end{subfigure}
    \hfill
    \begin{subfigure}[b]{0.325\textwidth}
        \centering
        \includegraphics[width=\textwidth]{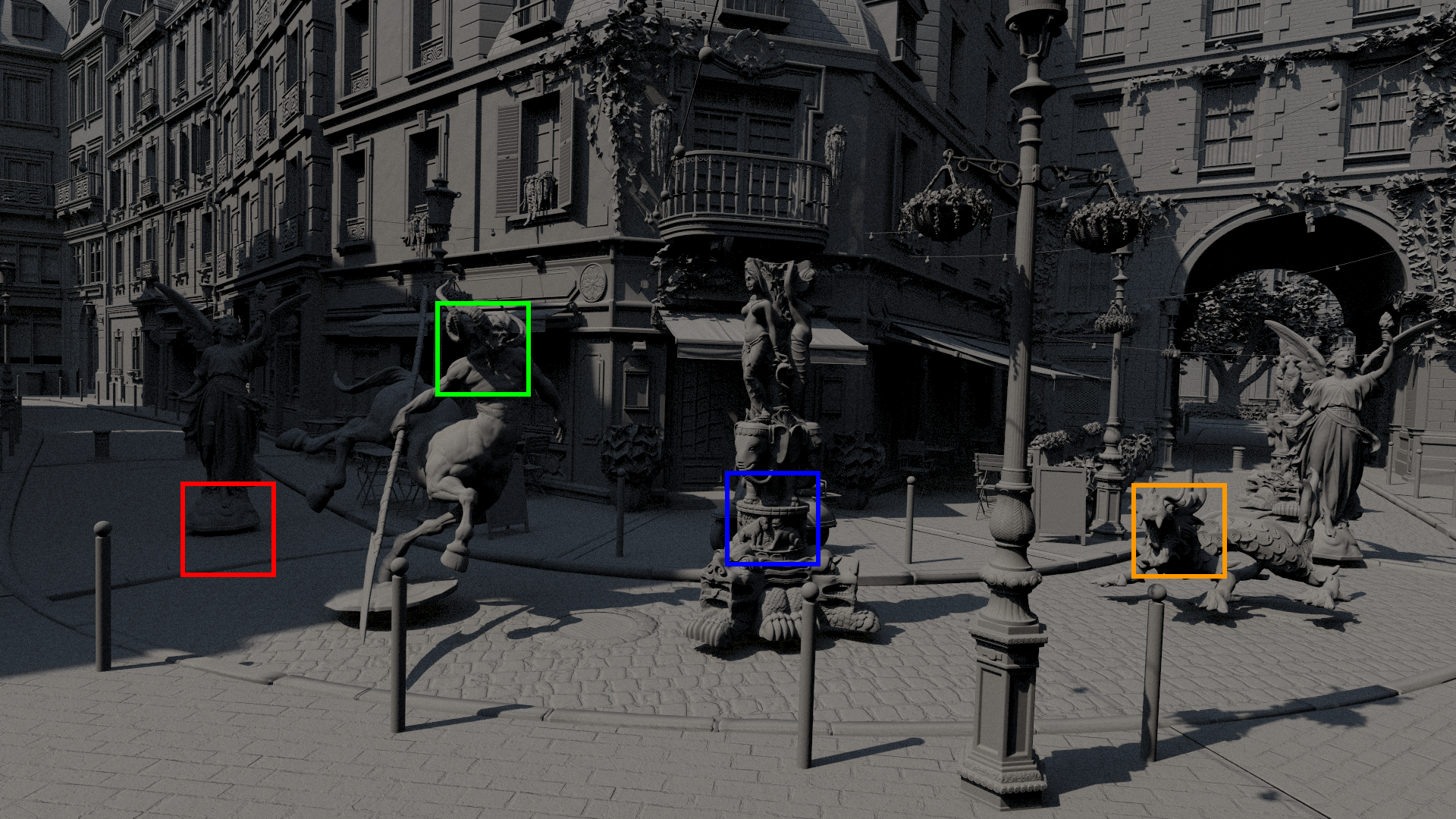}
    \end{subfigure}
    \hfill
    \begin{subfigure}[b]{0.325\textwidth}
        \centering
        \includegraphics[width=\textwidth]{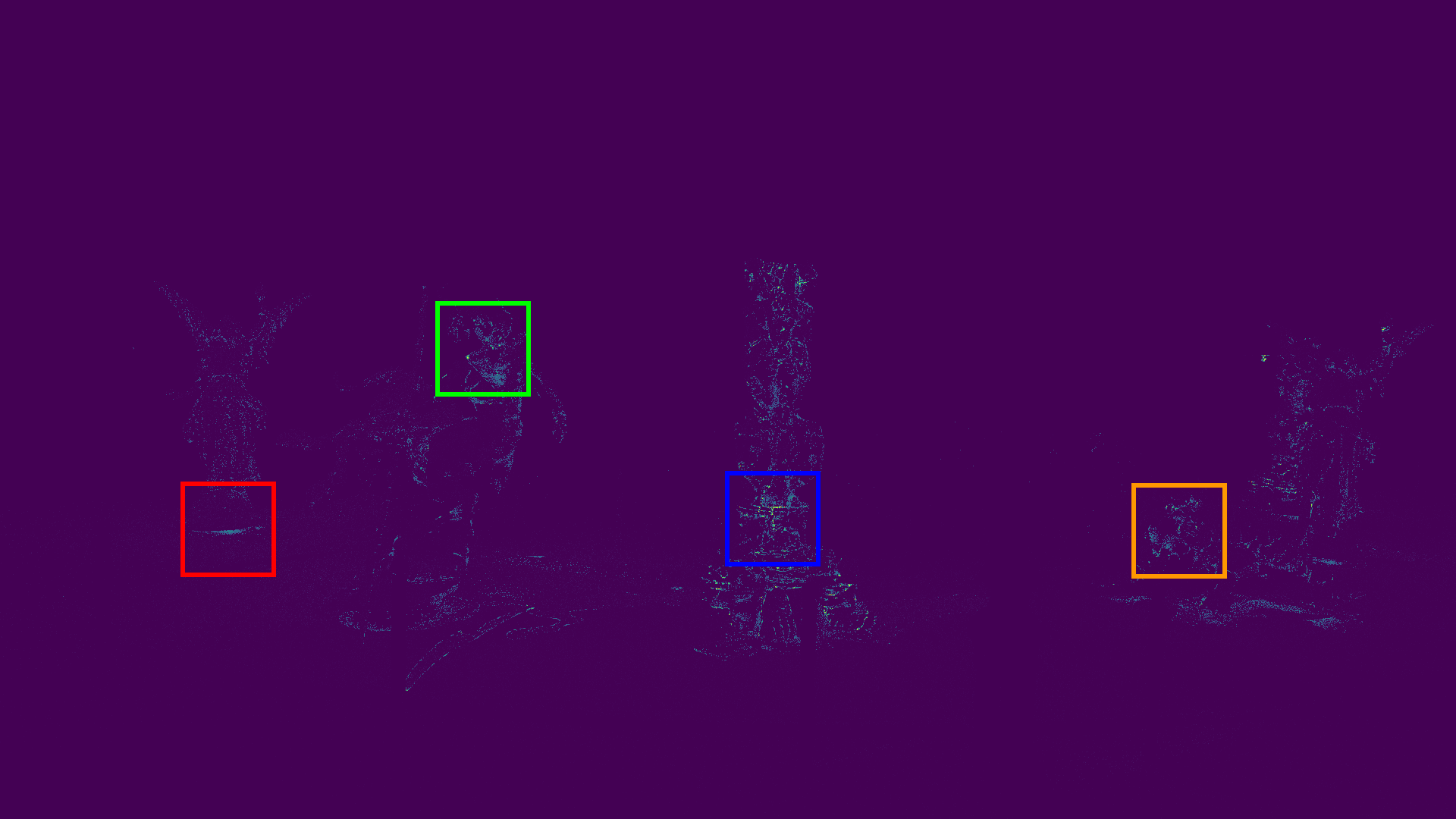}
    \end{subfigure}
    \begin{subfigure}[b]{0.325\textwidth}
        \centering
        \includegraphics[width=\textwidth]{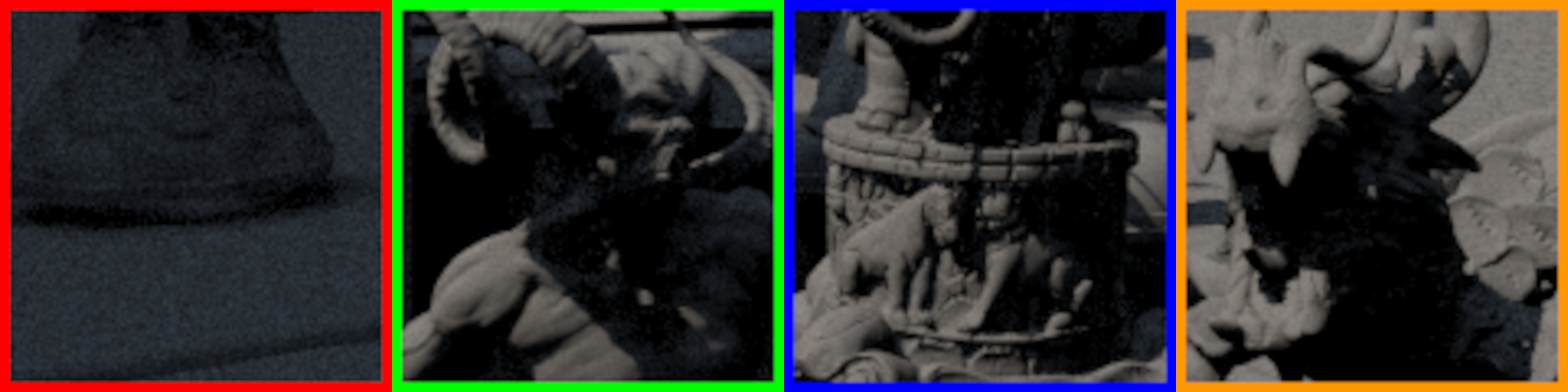}
        \caption{}
    \end{subfigure}
    \hfill
    \begin{subfigure}[b]{0.325\textwidth}
        \centering
        \includegraphics[width=\textwidth]{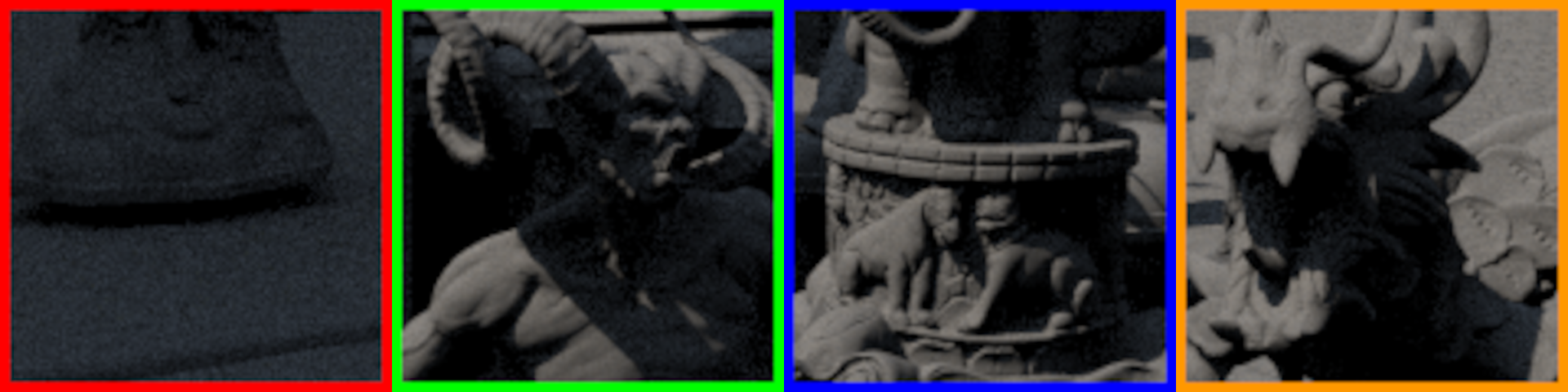}
        \caption{}
    \end{subfigure}
    \hfill
    \begin{subfigure}[b]{0.325\textwidth}
        \centering
        \includegraphics[width=\textwidth]{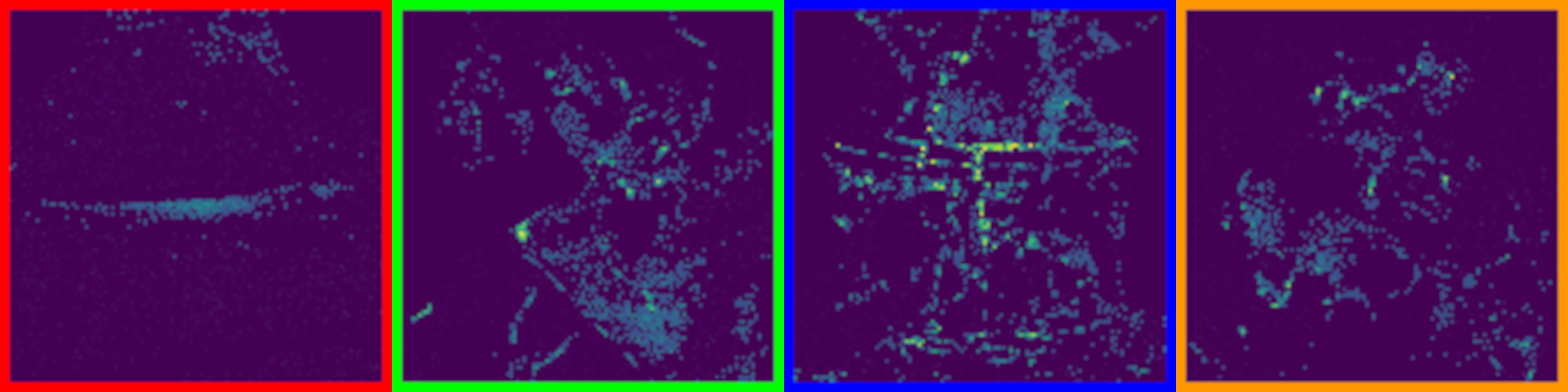}
        \caption{}
    \end{subfigure}
  \caption{(a) A rendered image using Neural Intersection Function after $64$ training samples per pixel. (b) A rendered image using ray tracing with BVH. (c) Difference $\times 3$ between (a) and (b). PSNR is $39.11$ dB. The scene has 30M triangles rendered at $1920 \times 1080$ on AMD Radeon\textsuperscript{\texttrademark} RX 7900 XT. Secondary ray casting times are $4.54$ ms and $5.27$ ms in (a) and (b), respectively. }
  \label{fig:teaser}
}

\maketitle
\begin{abstract}
The ray casting operation in the Monte Carlo ray tracing algorithm usually adopts a bounding volume hierarchy (BVH) to accelerate the process of finding intersections to evaluate visibility. However, its characteristics are irregular, with divergence in memory access and branch execution, so it cannot achieve maximum efficiency on GPUs. This paper proposes a novel Neural Intersection Function based on a multilayer perceptron whose core operation contains only dense matrix multiplication with predictable memory access. Our method is the first solution integrating the neural network-based approach and BVH-based ray tracing pipeline into one unified rendering framework. We can evaluate the visibility and occlusion of secondary rays without traversing the most irregular and time-consuming part of the BVH and thus accelerate ray casting.
The experiments show the proposed method can reduce the secondary ray casting time for direct illumination by up to $35 \%$
compared to a BVH-based implementation and still preserve the image quality. 


\end{abstract}  
\section{Introduction}
Monte Carlo ray tracing has been studied for decades and remains an active research topic. Ray casting is the core operation of Monte Carlo ray tracing to perform visibility tests from a given position in a scene.
A computationally expensive operation, ray casting is often accelerated with a bounding volume hierarchy (BVH) to find intersections~\cite{https://doi.org/10.1111/cgf.142662}.
There have been many attempts and explorations to improve the performance of BVH-based ray casting using technologies emerging from both software and hardware~\cite{Meister2022BVH, navi3, nvidiartx}.

However, these implementations are often non-ideal to be carried out on Single-Instruction Multiple-Threads (SIMT) architectures, such as GPUs, where the instructions in all threads are executed in lock-step and, therefore, cannot achieve maximum efficiency. 
The BVH traversal is an irregular algorithm, which implies divergence in memory access and branch execution. While irregular algorithms contain operations that cannot be handled efficiently on GPUs and cause the performance of GPUs to decrease~\cite{DBLP:journals/vc/KaoH18}, neural network (NN) execution, especially in the case of fully connected networks, is regarded as a regular algorithm because its core operation comprises mainly dense matrix multiplications with a predictable memory access pattern, which is GPU-friendly.
Thus, we could expect the ray casting performance on GPUs to improve if the BVH traversal can be replaced with a NN.

For this purpose, we propose a novel method called Neural Intersection Function (NIF). Unlike the conventional methods, it employs a NN to evaluate visibility from a point to a specific direction instead of traversing the entire BVH tree down to its bottom level. Thus, it avoids executing the most irregular part of the algorithm on GPUs.
In this paper, we demonstrate its feasibility by analyzing performance and image quality when using NIF only for secondary ray casting for direct illumination.
Extensions to other types of rays remain our interesting future work.

The paper makes the following contributions:

\begin{itemize}
    \item We introduce Neural Intersection Function (NIF), a novel method based on a multilayer perceptron (MLP) to accelerate ray casting as an alternative to BVH-based methods.

    \item We demonstrate that NIF can handle rays cast in a scene for rendering using two distinct types of NNs and grids storing latent vectors with a carefully selected input parameterization.

    \item We experimentally prove that the adoption of NIF into the ray tracing pipeline to compute direct lighting can improve performance and validate that our approach can preserve image quality.

\end{itemize}

The advantages of NIF compared to a BVH-based approach are: 

\begin{itemize}
    \item Improved computational efficiency on GPUs thanks to less divergent execution and memory access.
    \item Constant memory footprint, independent of objects' geometric complexity, makes our method more advantageous for a complex model with a larger number of triangles. 
    \item Constant execution time if the number of rays intersecting against the AABB of an object is the same. 
    
\end{itemize}

\section{Related Work}


There are many attempts to represent 3D shapes with implicit neural representations (INRs) based on the multilayer perceptron (MLP) architectures. 
These neural representations encode the geometric information by learning the mapping from a given position in 3D space to other properties at that location which usually denotes the distance to the surface of the shape~\cite{Park_2019_CVPR, 9710631} or the density and emitted radiance~\cite{mildenhall2020nerf}.
On the other hand, a similar concept utilizes a network trained to model the occupancy function which acts as a binary classifier~\cite{Occupancy_Networks, 8953765}.
A 3D surface determined from the methodologies above is regarded as an isosurface. An extra step is required to extract the 3D mesh of desired quality and resolution from the represented isosurface~\cite{chibane20ifnet}.


Unlike the main application of INRs, whose purpose is to reconstruct the 3D surface, NIF aims to approximate the function of the visibility test for given spatial positions. Thus, it is not determined by any isosurface of objects. Instead, our method encodes the visibility by learning the hit information associated with the corresponding hit object utilizing multiple feature grids.
Despite the remarkable results from INRs, it remains a challenge to train a NN that can correctly capture the attribute of 3D shapes and, at the same time, perform the computation efficiently when adopted in applications. Previous studies have proposed numerous methodologies to enhance the quality of the representation or the performance. From many ideas and techniques, the core concept can be categorized into two approaches. One is to utilize a global network to represent the 3D shape and slice the surface into small patches to reduce the difficulty of fitting a complex surface into a single global network~\cite{Tretschk2020PatchNets, genova2019learning}. Another approach is to partition the 3D space spatially into local regions and then train multiple INRs for each region~\cite{9607613, martel2021acorn}.

Inspired by both approaches, we propose a combined solution in NIF: we categorize the rays based on whether they originated from outside or inside of any Axis Aligned Bounding Box (AABB) of objects and train two networks, outer and inner, respectively. In other words, we do not construct networks for each object but train two networks to capture the characteristics of all objects. However, for each network, we utilize feature grids that take the AABB of the hit object into consideration to aid in fitting and training the network. Specifically, we condition the network locally by transforming the latent code so that it becomes coordinate-dependent~\cite{10.1111:cgf.14505, jiang2020local}.


Input parametrization plays an important role and is a decisive factor in the quality of the network. However, to train a network that can accurately map rays to the hit information, it is not ideal to concatenate the positions and directions of rays as the input for the network. The reason is due to ray aliasing: assume that we have a ray $r$ which is represented by its position and direction as $(\mathbf{p}, \mathbf{d})$, if we move the ray's origin from $\mathbf{p}$ along its direction $\mathbf{d}$ to another point $\mathbf{p'}$, this becomes an aliased ray and the hit result caused by them should be identical. However, in reality, there is no guarantee that the network would still be able to produce the same output since the input of the two, $(\mathbf{p}, \mathbf{d})$ and $(\mathbf{p'}, \mathbf{d})$, would be different. 

To overcome the challenge, previous studies have explored different methods to parameterize rays. For example, Sitzmann et al. have demonstrated how to train an MLP to handle rays with arbitrary origins and directions by transforming them to Plücker coordinates~\cite{sitzmann2021lfns}. Furthermore, Feng et al. have extended this concept to incorporate the correspondence of the surface by introducing the \textit{foot} notation, which is also designed to be invariant to changing the ray position along the ray direction~\cite{10.1007/978-3-031-20062-5_9}.
Following a similar concept, our approach also aims to identify aliased rays that would result in identical hit points. Instead of representing the rays in Plücker coordinates, we transform the positions to a representation with respect to the hit object as follows: if a ray originated from outside of the AABB of the hit object, we translate its position to the intersection point of the ray with the AABB. On the other hand, in the self-occlusion case where the ray originated from inside of the AABB, it is mapped onto a unique position of the surface. As a result, aliased rays would be encoded identically.

\section{Design of the Neural Intersection Function} \label{NIF}

\begin{figure*}
     \centering
     \begin{subfigure}[b]{\columnwidth}
         \centering
         \includegraphics[width=\textwidth]{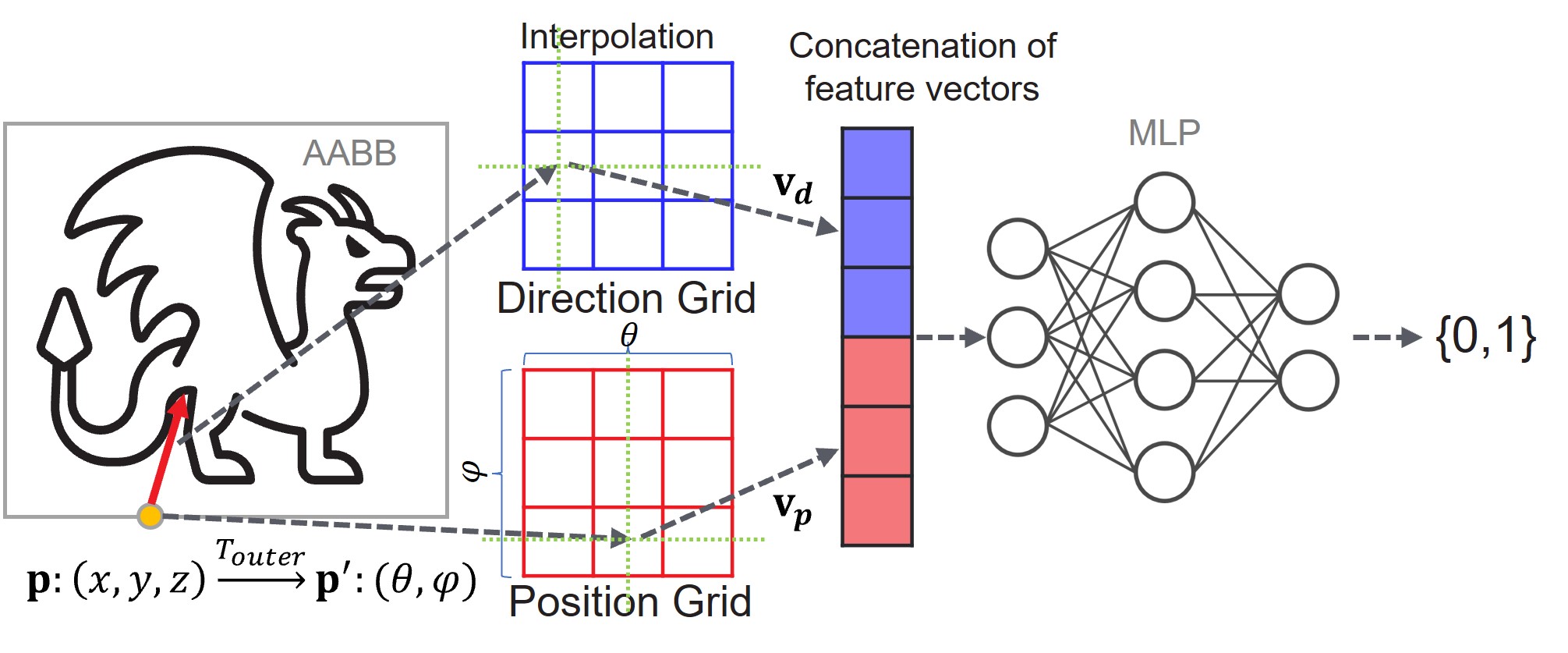}
         \caption{\small{NIF outer network}}
         \label{fig:nifOuter}
     \end{subfigure}
     \hfill
     \begin{subfigure}[b]{0.9\columnwidth}
         \centering
         \includegraphics[width=\textwidth]{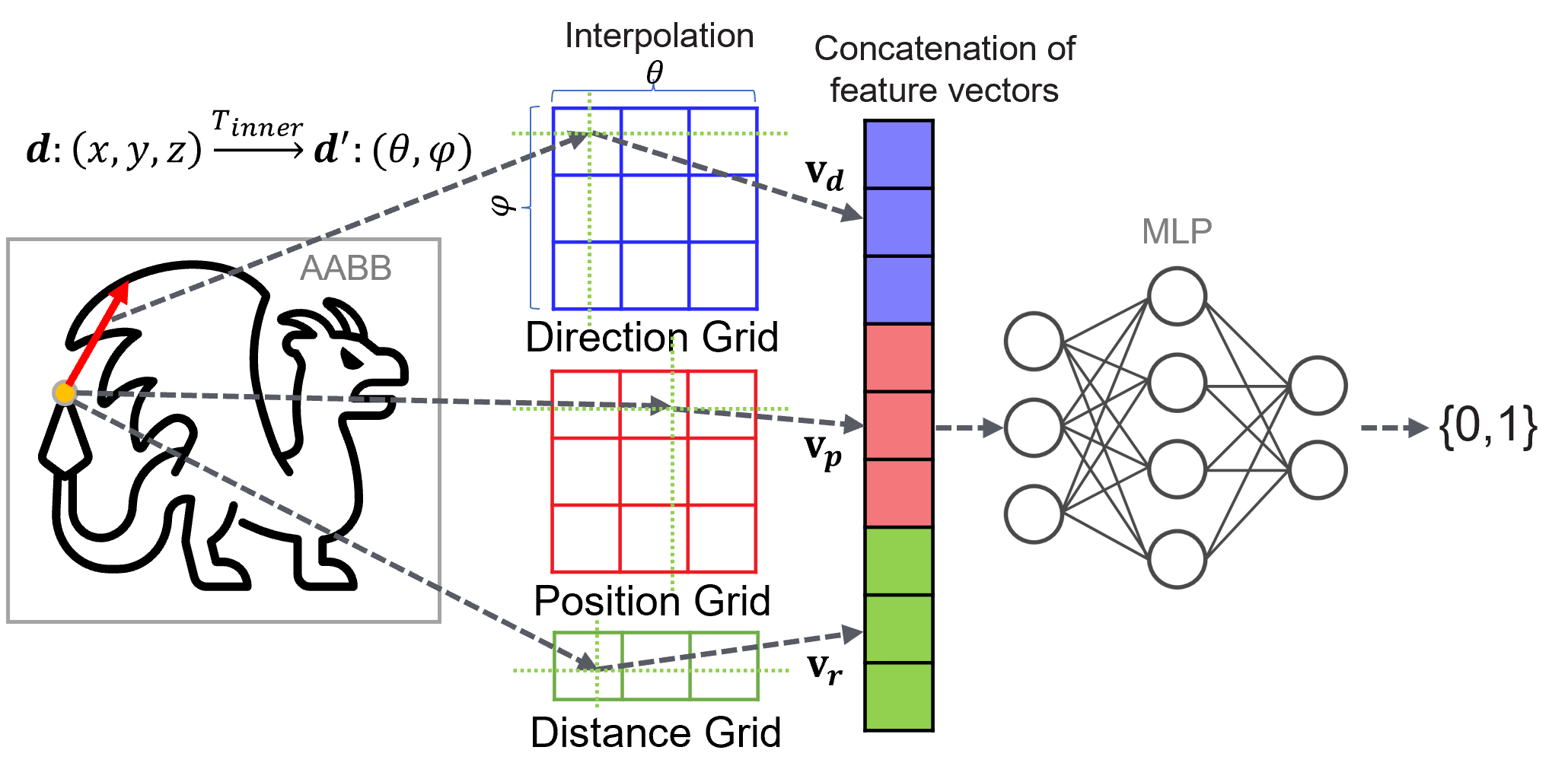}
         \caption{NIF inner network}
         \label{fig:nifInner}
     \end{subfigure}
        \caption{(a) The outer network of NIF. Starting from the left of the figure, the original 3D position $\mathbf{p}$ is converted into a 2D spherical coordinate $\mathbf{p'}$ by the transformation function $T_{outer}$. After that, $\mathbf{p'}$ is used to retrieve the corresponding feature vector $\mathbf v_{p}$ from the grid. The final content of the feature vector is bi-linearly interpolated by considering the neighbor indices. Direction is handled by the same logic to retrieve $\mathbf v_{d}$. Finally, the feature vectors are concatenated to form the input for MLP. During the backpropagation, those trainable feature vectors are also updated. (b) The inner network adopts a similar architecture with an additional feature vector $\mathbf v_{r}$ derived from the distance.}
        \label{fig:nif}
\end{figure*}

The concept of the visibility test is to evaluate the occupancy from a given position in 3D space $\mathbf{p} \in \mathbb{R}^3$ along a direction $\mathbf{d}$. The result is a binary value $\{0, 1\}$, where $0$ denotes it is occluded by an object and $1$ represents it is clear (i.e. visible). Namely, the concept can be formulated as the following function in Equation~\ref{eq:viz}:

\begin{equation}
f: \{(\mathbf p, \mathbf d) | \mathbf{p}, \mathbf{d} \in \mathbb{R}^3 \} \to \{ 0, 1\}
\label{eq:viz}
\end{equation}

Our motivation is to approximate this function by training NNs which can map the input to a visible probability between $0$ and $1$. To improve accuracy, we must eradicate the problem caused by aliased rays where two rays result in the same intersection hit point but are represented differently. This is done by conditioning the inputs with the Axis Aligned Bounding Box (AABB).
The AABB of an object partitions the space geometrically into two regions known as outside or inside of the AABB. During the ray casting phase, we can also classify rays as either outside or inside based on their origin positions. We create two NNs for each of them, denoted as outer and inner.

\subsection{Outer Network for Ray Cast from Outside of AABBs}

There are infinite numbers of rays with distinct origins and directions that can be cast from outside of an AABB toward the object it encapsulates. However, the rays which have different origins but travel along the same direction would eventually hit the same intersection point on the AABB. Therefore, instead of using the origins of rays, we can better represent them by computing the intersection of a ray with the AABB. With this parameterization, rays that lie on the same line are mapped to an identical representation, which reduces aliasing. Furthermore, since AABBs are concave, we can convert the position $\mathbf{p}$ from the 3D Cartesian coordinate to a 2D spherical coordinate $\mathbf{p'}$ with a bijective mapping to further reduce the data dimension. Similarly, the direction is also converted from $\mathbf d$ to $\mathbf{d'}$.
Equation~\ref{eq:t_outer} formulates the transformation function:

\begin{multline}
\label{eq:t_outer}
T_{outer}: \{(\mathbf{p}, \mathbf{d}) | \mathbf{p}, \mathbf{d} \in \mathbb{R}^3 \} \\ \to
\{  (\mathbf{p'}, \mathbf{d'}) |  \mathbf{p'}, \mathbf{d'} \in \mathbb{R}^2 \land \mathbf{p'} \in \mathbf{AABB} \}
\end{multline}

However, we have found that supplying those converted values directly to the network for training is not optimal. This is because complex geometries usually incorporate geometrically high-frequency details, which makes it difficult for the network to learn effectively~\cite{mildenhall2020nerf, kuznetsov2021neumip}. In order to improve accuracy further, we adopt a representation learning technique~\cite{bengio2013representation} and utilize grid encoding to automatically discover a set of features that can describe data compactly and yet expressively.
The way grid encoding functions is to first construct two-dimensional grids whose cells store trainable latent vectors (i.e. feature vectors) for the position and the direction per object. The converted values, $\mathbf{p'}$ and $\mathbf{d'}$, are now used as indices of the grids to retrieve the corresponding feature vectors, $\mathbf v_{p}$ and $\mathbf v_{d}$, respectively. The feature vectors become the inputs for the MLP and are simultaneously optimized together with the weights of the MLP through back-propagation~\cite{Park_2019_CVPR}. Furthermore, the final contents of feature vectors from grid cells are bi-linearly interpolated and concatenated before being supplied to the MLP, which is shown in Fig.~\ref{fig:nifOuter}. By conditioning the network's input on the feature vectors fetched from grids constructed for each object, this formulation allows modeling the visibility result of multiple objects with a single neural network. To summarize, the function of NIF for the outer network is formulated in Equation \ref{eq:nif_outer}:

\begin{multline}
\label{eq:nif_outer}
NIF_{outer}: \{(\mathbf v_{p}, \mathbf v_{d}) | (\mathbf v_{p}, \mathbf v_{d}) = Grid(\mathbf{p'}, \mathbf{d'}) \\ \land (\mathbf{p'}, \mathbf d') \in Range(T_{outer}) \} \to \{ 0, 1\}
\end{multline}

\subsection{Inner Network for Ray Cast from Inside of AABBs}

When a primary ray hits an object enclosed in a valid AABB, the corresponding secondary ray is generated from the inside. Similar to the transformation method in the outer network, we convert the position and the direction from 3D to 2D. The conversion could be viewed as a procedural UV mapping with spherical vertex projection since the hit point position will always be on the surface of an object. However, considering only the position and the direction is not enough because there could have more than one intersection point caused by the secondary ray if the object is concave. Therefore, we need to record the distance from the center of an AABB to the hit point as well. Equation~\ref{eq:t_inner} describes the transformation function for the inner network:

\begin{equation}
\label{eq:t_inner}
T_{inner}: \{(\mathbf p, \mathbf d) | \mathbf{p}, \mathbf{d} \in \mathbb{R}^3 \} \to  \{  (\mathbf{p'}, \mathbf{d'}, \mathbf{r'}) |  \mathbf{p'}, \mathbf{d'} \in \mathbb{R}^2, \mathbf{r'} \in \mathbb{R} \}
\end{equation}

After the conversion, the converted values are also encoded with grids to get the feature vectors $\mathbf v_{p}$ and $\mathbf v_{d}$. In addition, we incorporate the information derived from the distance as a feature vector $\mathbf v_{r}$ which is retrieved from a 1D grid by using the normalized distance $\mathbf{r'}$ as the index. The feature vectors $\mathbf v_{p}$, $\mathbf v_{d}$ and $\mathbf v_{r}$ are also concatenated to form the input for the MLP, which is shown in Fig.~\ref{fig:nifInner}.
As a result, the function of NIF for the inner network can be formulated in Equation~\ref{eq:nif_inner}:

\begin{multline}
\label{eq:nif_inner}
NIF_{inner}: \{(\mathbf v_{p}, \mathbf v_{d}, \mathbf v_{r}) | (\mathbf v_{p}, \mathbf v_{d}, \mathbf v_{r}) = Grid(\mathbf{p'}, \mathbf{d'}, \mathbf{r'}) \\ \land (\mathbf{p'}, \mathbf d', \mathbf r') \in Range(T_{inner}) \} \to \{ 0, 1\}
\end{multline}

Note that we can further extend this network by encoding auxiliary data at the hit point as the output of the network. In this example, we only need to use NIF for the occlusion query. Therefore, a value between 0 and 1 is used to represent the occupancy. It is also possible to add other surface properties, such as shading normal or texture coordinates. We demonstrate the applicability of extending NIF to support primary ray casting later in this paper.

\section{Neural Intersection Function in Ray Tracing Pipeline}

\begin{figure}
     \centering
     \begin{subfigure}[b]{0.475\columnwidth}
         \centering
         \includegraphics[width=\textwidth]{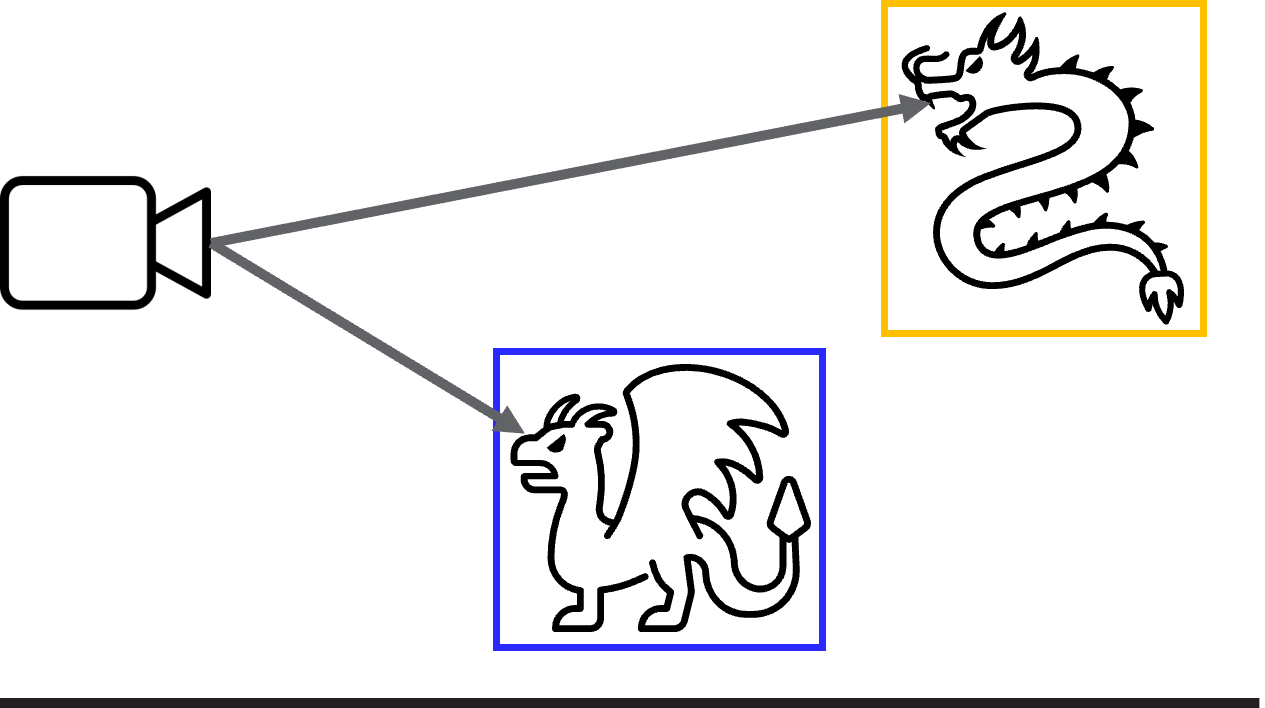}
         \caption{}
         \label{fig:pipeline:0}
     \end{subfigure}
     \hfill
     \begin{subfigure}[b]{0.475\columnwidth}
         \centering
         \includegraphics[width=\textwidth]{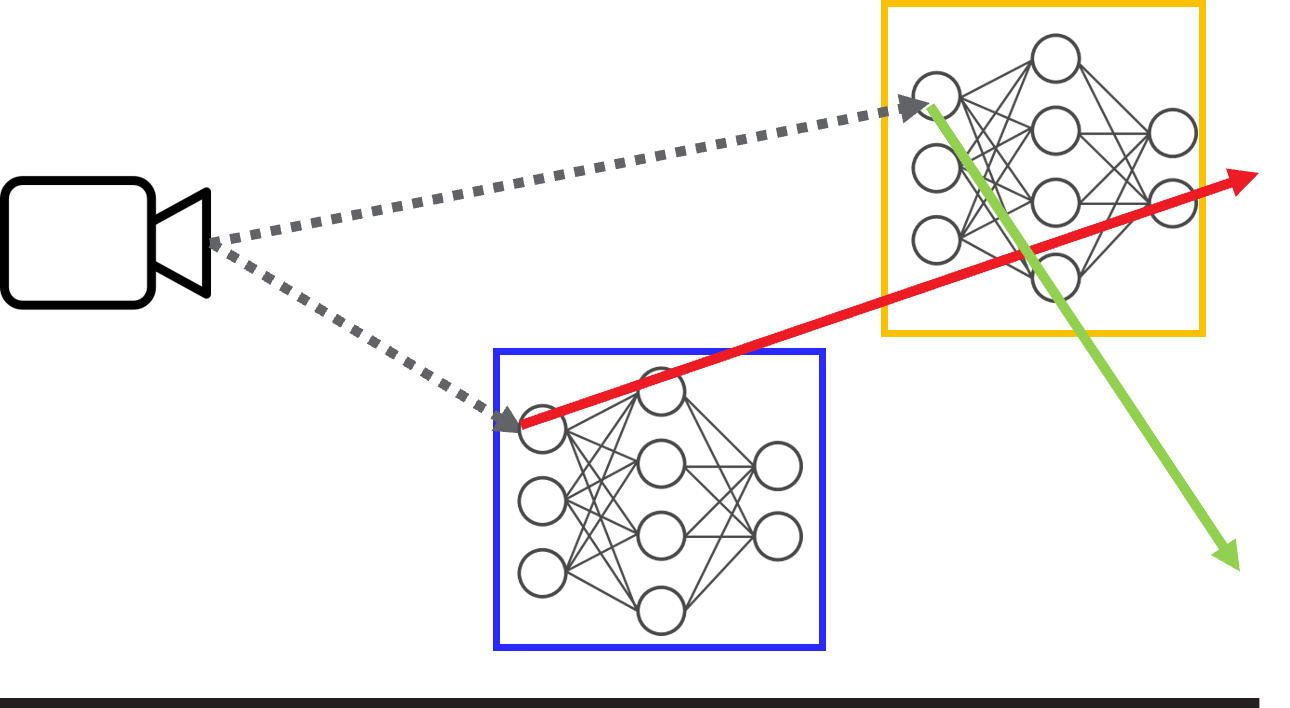}
         \caption{}
         \label{fig:pipeline:1}
     \end{subfigure}     
        \caption{Rendering pipeline using NIF. (a) Primary rays are cast from the camera as usual. (b) Secondary rays are generated and cast against the scene. Objects are replaced with NIF in this step. Thus, we do not traverse BVHs for objects. The red ray starting from the blue AABB first checks the inner intersection of the object where the origin is located. Then it checks the intersection against the object in the orange AABB.}
        \label{fig:pipeline}
\end{figure}

Here we discuss the integration of NIF to a regular ray-tracing pipeline where shading and lighting are executed, and a two-level BVH is used. 
To the best of our knowledge, our proposed method is the first approach to embedding a neural network-based method into such a regular ray-tracing framework.

Although NIF is applicable to all types of rays, errors of NIF used for primary ray casting directly result in visible artifacts and the accuracy is not enough compared to BVH ray tracing as shown in Fig.~\ref{fig:gridResCmp}. Thus, we only limit the use of NIF for secondary ray casting for direct illumination in this paper.
Fig. \ref{fig:pipeline} is an illustration of the processes in Sec.~\ref{NIFPipeline} where objects are replaced with NIF after primary ray casting as shown in Fig. \ref{fig:pipeline:1}.

\subsection{Embedding NIF to a Ray-Tracing Pipeline} \label{NIFPipeline}

To achieve high performance in finding the intersections of rays within the scene, spatial acceleration structures such as a BVH are usually employed. Instead of building a single BVH over all geometric primitives in the scene, we construct the data structure into two levels where the geometric primitives of each object are grouped into separate bounding volumes, each contains their own BVH also known as a bottom-level BVH, and with a top-level BVH built over all these volumes. Generally, the more complex the object is, the more divergent it would be when traversing the bottom-level BVH which leads to performance degradation. To tackle this issue, NIF is designed as a replacement for bottom-level BVHs.

As NIF replaces the bottom-level BVH, the remaining part of the rendering pipeline stays the same. Since the input of NIF is derived from transformation functions as shown in Equations \ref{eq:t_outer} and \ref{eq:t_inner} and thus depends on the AABB, we need to traverse the top-level BVH to search for a ray intersecting an AABB. The ray casting starts with the top-level BVH traversal as usual until it hits a leaf node. When an intersection is found, instead of diving into the bottom-level BVH, we query from NIF to let it infer the intersection result of the ray to the object. If an occlusion is reported, the intersection will be recorded as the current closest hit point. After processing the object, we return to the top-level BVH traversal and execute NIF at the leaf node. This process repeats until we finish traversing the top-level BVH. 

To optimize the execution pattern and to minimize the impact of branch divergence, we divide these two stages, the top-level BVH traversal and the execution of NIF, into different kernel executions. This way, we can achieve maximum GPU occupancy and could also tune the size of the thread group and the shared memory for individual kernels to find the optimal configuration. As a result, the overall GPU utilization increases. The entire execution flow functions as follows:  at first, we only traverse the top-level BVH and store the input data when an intersection is found as described in Sec.~\ref{NIF}. This step is computationally inexpensive since it only traverses a relatively small BVH and checks intersections with AABBs. After all the input data are gathered during the traversal stage, we move to NN execution. At this stage, we invoke NIF for the outer network and then for the inner network sequentially. Specifically, there is a kernel being executed for feature grid look-ups and concatenation of the latent vectors, followed by the NN inference execution which predicts the status of occlusions. Since we only need to train one single NN (two in total for the outer and the inner network) that can handle all objects in the scene, this step is effective. Note that this step would become computationally expensive if we had to prepare a single NN for each object, as we need to gather the rays intersecting against each object and execute them one by one. 

\subsection{Training} \label{training}

NIF is trained with rays that are generated from the current viewpoint (camera) while rendering a scene based on a predefined configuration and with a first few samples per pixel (spp).
During this rendering phase, we collect the ray-AABB intersection results in order to train the parameters in NIF afterward.
Alternatively, it is also possible to train NIF separately and in an offline setting using a larger set of rays that could be generated from multiple viewpoints covering a wider range, provided that the structure of the MLP in NIF is adjusted accordingly. 

\section{Implementation}

The NNs in NIF are implemented from scratch using AMD C++ Heterogeneous-Compute Interface for Portability (HIP)~\cite{hip} in order to train and run inference on GPUs. Each NN is implemented in the form of an MLP. Inspired by the previous work from Müller et al.~\cite{10.1145/3450626.3459812}, we implemented the NNs to fuse operations from all layers to minimize memory transfer overhead. By design, a single thread block only needs to load the weight matrix once in a single inference execution and store the data in the shared memory. Our NN framework supports both 32-bit and 16-bit precision of floating points for the network weights.

As described in Sec.~\ref{NIFPipeline}, to maximize GPU efficiencies, the execution is implemented in a way that is composed of two parts: the first part traverses the top-level BVH and stores the data, and the second part invokes the NNs for inference.

\paragraph*{Architecture}

The outer network in NIF contains two hidden layers with $64$ nodes in each of them, whereas the inner network comprises three hidden layers with $48$ nodes each. Every hidden layer is followed by a leaky ReLU activation function except for the last one. For the last layer, we used a sigmoid activation function to output a visible probability between 0 and 1.
In this paper, we consider the ray is occluded if the output probability is less than $0.5$.
For the grid encoding, we used two-dimensional grids for the position and the direction, and a one-dimensional grid for the distance as described in Sec.~\ref{NIF}.
These grids have a resolution $R$ per dimension. Each cell in the grids stores the latent vector with dimensions of $N$.
Table~\ref{tab:arch} shows the values for these hyperparameters used in our experiments.
These values are chosen experimentally, which we will describe in Sec.~\ref{hyperparam}.

\begin{table}
    \small
    \caption{Hyperparameters of feature grids. $R$ and $N$ are the resolution of the grids, the dimension of the latent vectors, respectively. }
    \label{tab:arch}
    \centering
    \begin{tabular}{ l p{15mm} | p{15mm} p{15mm} } 
        \toprule
         & \multicolumn{1}{c|}{NIF outer network} & \multicolumn{2}{c}{NIF inner network} \\
         & \multicolumn{1}{c|}{  2D Grid} & \multicolumn{1}{r}{ 2D Grid} & \multicolumn{1}{r}{  1D Grid} \\ 
        \midrule
        $R$ & \multicolumn{1}{||r|}{256} & \multicolumn{1}{r}{128} & \multicolumn{1}{r}{128} \\
        $N$ & \multicolumn{1}{||r|}{3} & \multicolumn{1}{r}{5} & \multicolumn{1}{r}{3} \\
        \bottomrule
    \end{tabular}
\end{table}

\paragraph*{Initialization}
The weights of NNs are initialized with Xavier initialization procedure~\cite{Glorot2010UnderstandingTD}.
The latent vectors in grid cells are initialized with the uniform distribution $\mathcal{U}(-10^{-4}, 10^{-4})$.

\paragraph*{Training}
To generate the training samples (i.e. ray-AABB intersection information), we used ray directions generated with light importance sampling according to radiant flux where a tabled CDF is computed and sampled using binary search.
Using these samples, we jointly trained NNs and feature grids by applying the Adam optimizer, where we set ${\beta}_1 = 0.9$, ${\beta}_2 = 0.999$ and $\epsilon = 10^{-15}$, with the $\mathcal{L}^2$ loss function~\cite{Adam}. 
We used a learning rate of $0.005$, a batch size of $2^{11}$ in NIF for the outer network and $2^{12}$ in NIF for the inner network.
These hyperparameters are also chosen experimentally.

\paragraph*{Memory Footprint}
We utilized half-precision floating points as much as possible.
As each cell in a grid stores latent vectors in half, gradient vectors in 32-bit float, and first and second moments for the Adam optimizer in 32-bit float, a two-dimensional grid with resolution $R^2$ requires $14 R^2$ bytes. Thus the grids for outer and inner networks consume $28R^2$ and $28R^2+14R$ bytes, respectively. If we move the training to an offline process and free all the memory required for training, they only require $4R^2$ and $4R^2+2R$ for the runtime. Specifically, they are $256$ kB for the outer and $65$ kB for the inner with $R=256$ and $R=128$ which we used for our implementation.
For brevity, the memory footprint for the grids described here is for a single object though the grids are prepared for each object.
On the other hand, the network is shared, thus the memory footprint for the network is constant even if we put more objects in the scene.
Specifically, they are $18$ kB for the outer network and $21$ kB for the inner network with the settings for our implementation for the runtime. Note that we are only allocating about $321$ kB even for a model with $10$M triangles, which is a huge compression ratio. 

\paragraph*{Primary Visibility Computation}
When NIF is adopted to compute the visibility of secondary rays, we need a solution to find the shading points directly visible from the camera. In this paper, we selected ray tracing to compute these but our method is not restricted. That is to say, it is also possible to use rasterization to compute primary visibility. 

\section{Results}\label{sec:results}
We demonstrate the results of our proposed method, NIF, from various aspects. NIF is implemented in our progressive path tracer written in HIP. 
To evaluate the performance, we render images with $1920 \times 1080$ screen resolution on 
AMD Radeon\textsuperscript{\texttrademark} RX 7900 XTX GPU and measure the GPU kernel execution time with AMD Radeon\textsuperscript{\texttrademark} GPU Profiler.
The image quality is evaluated with the peak signal-to-noise ratio (PSNR) where a higher value indicates a better prediction quality.
We show the results of our method only for the static scenes in this paper. However, it should be possible to extend it to dynamic scenes using the training method described in~\ref{training} so that NN can adapt to the new viewpoint. This is our interesting and important future work.

\subsection{Hyperparameters on Grids} \label{hyperparam}
\begin{figure*}
\centering
\setlength{\tabcolsep}{0.002\linewidth}
\begin{tabular}{cccc}

\includegraphics[width=0.26\linewidth]{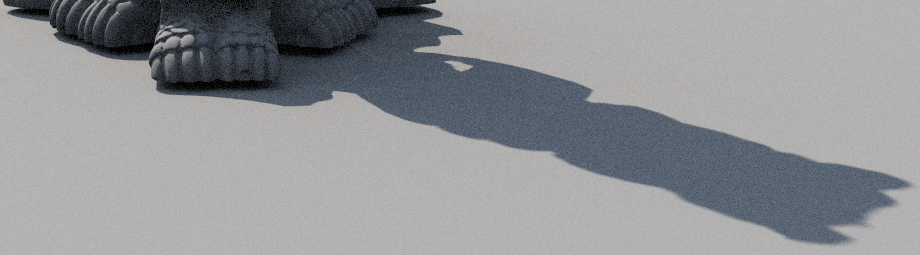}&
\includegraphics[width=0.26\linewidth]{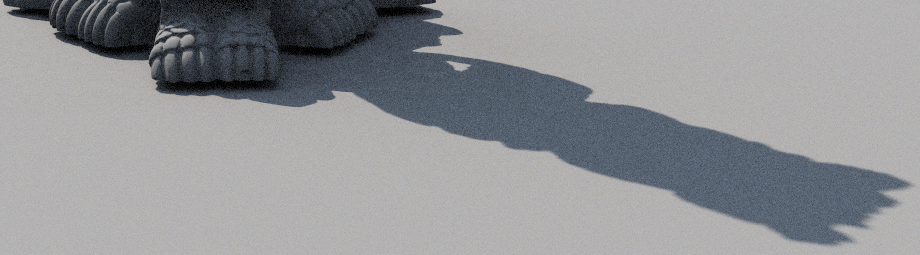}&
\includegraphics[width=0.26\linewidth]{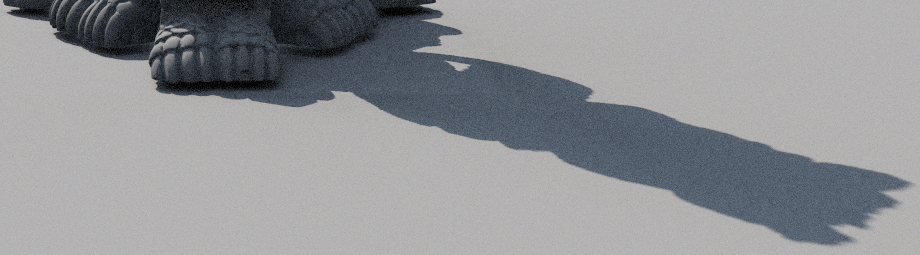}&
\multirow{2}{*}[10.0mm]{\includegraphics[width=0.18\linewidth]{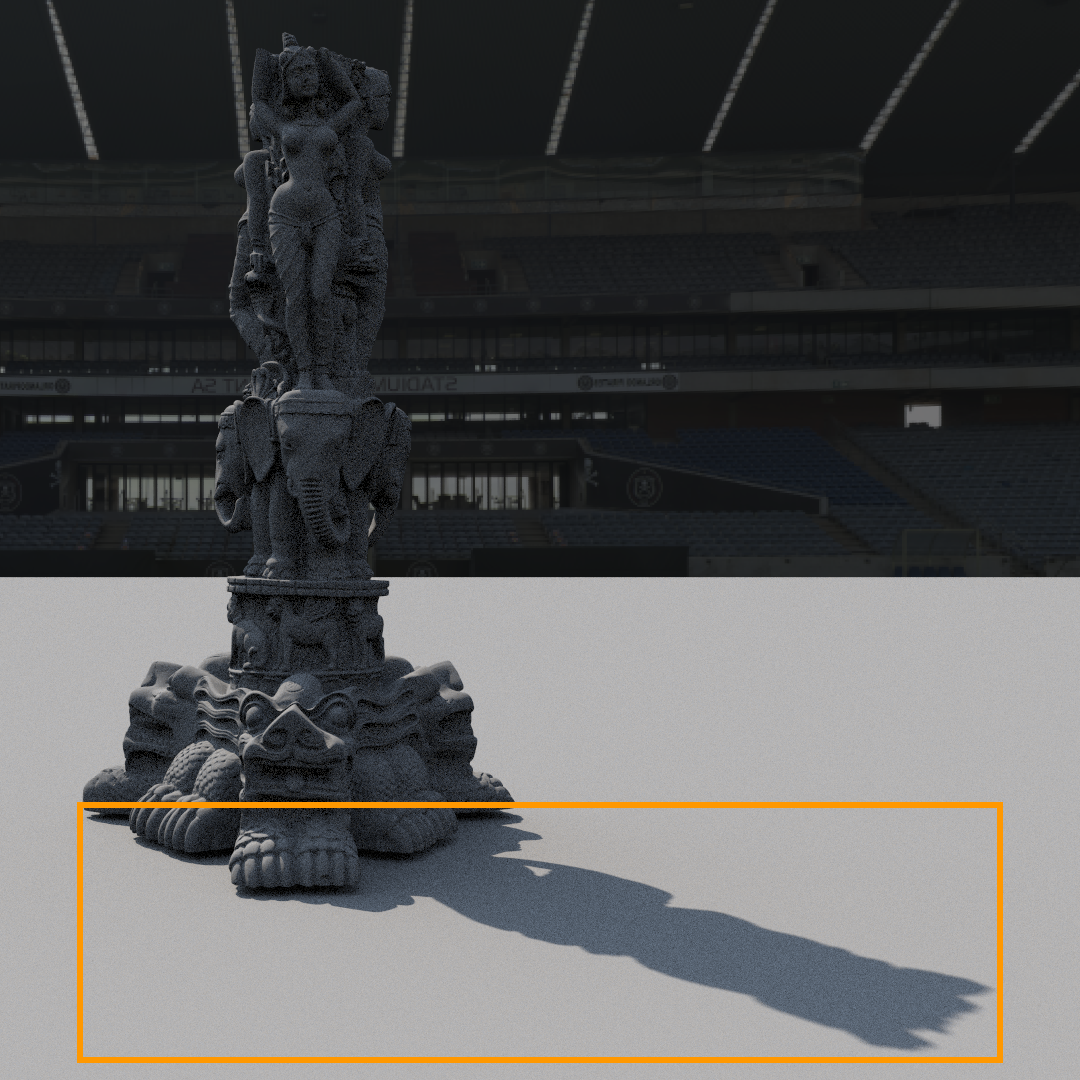}} \\

\includegraphics[width=0.26\linewidth]{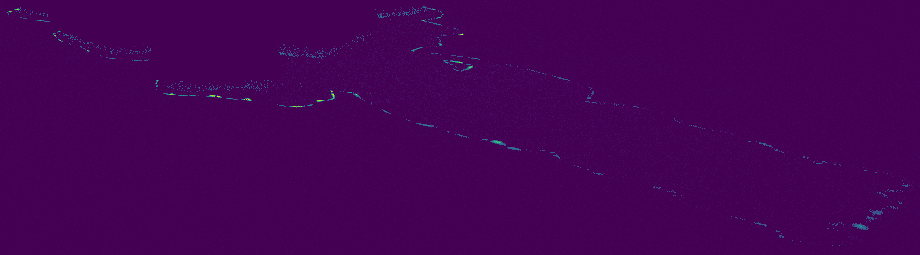}&
\includegraphics[width=0.26\linewidth]{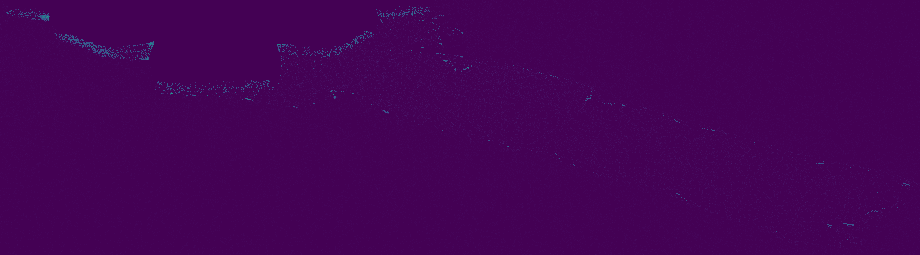}&
\includegraphics[width=0.26\linewidth]{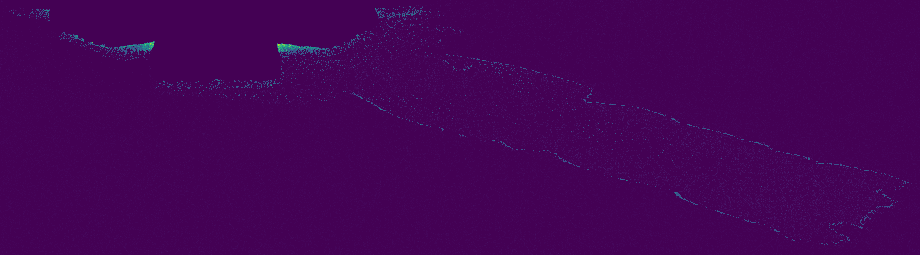}& \\

\small{$131.91$ microseconds} & \small{$135.49$ microseconds} & \small{$153.09$ microseconds} & \\

\small{PSNR: $48.39$ dB} & \small{PSNR: $49.75$ dB} & \small{PSNR: $45.70$ dB} & \\

\small{(a) $R = 64$} & \small{(b) $R = 256$} & \small{(c) $R = 1024$} & \small{Reference} \\
\end{tabular}
\caption{Visualizations of differences with various resolutions in grids for the outer network with the \textsc{Statuette} scene. The second row shows errors $\times 3$ in each rendered image in the first row. All images are rendered after 128 training spp. The inference time and PSNR are shown below each image.}
\label{fig:Outer2DGridRes}
\end{figure*}
\begin{figure}
    \centering
\setlength{\tabcolsep}{0.002\linewidth}
\begin{tabular}{cccccc}
\raisebox{0.08\linewidth}{\rotatebox[origin = c]{-90}{\textsf{\scriptsize{\textsc{Bunny} (Normal)}}}}&
\includegraphics[width=0.18\linewidth]{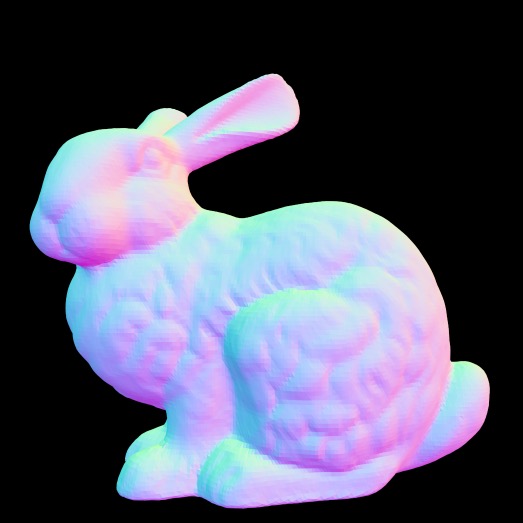} &
\includegraphics[width=0.18\linewidth]{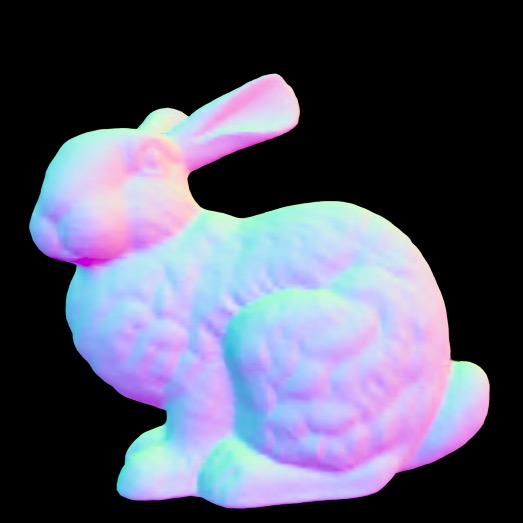} &
\includegraphics[width=0.18\linewidth]{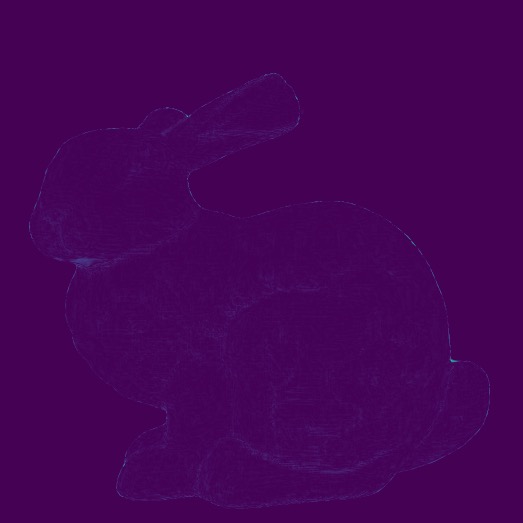} &
\includegraphics[width=0.18\linewidth]{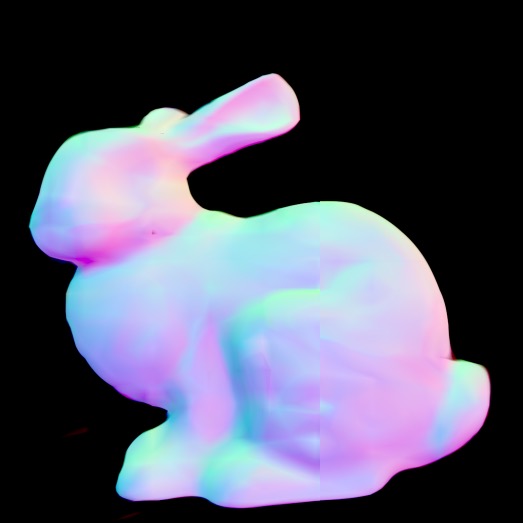} &
\includegraphics[width=0.18\linewidth]{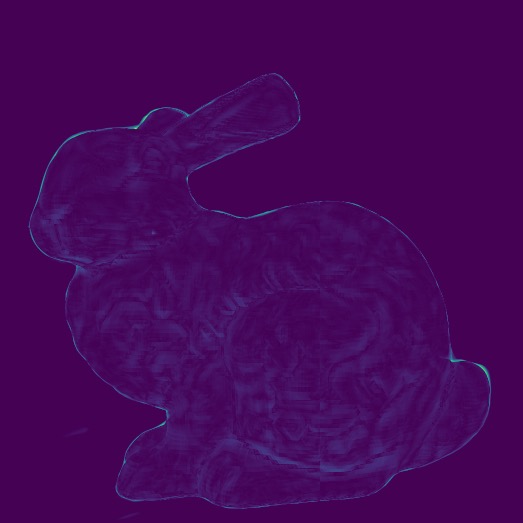} \\

\raisebox{0.08\linewidth}{\rotatebox[origin = c]{-90}{\textsf{\scriptsize{\textsc{Bunny} (Depth)}}}}&
\includegraphics[width=0.18\linewidth]{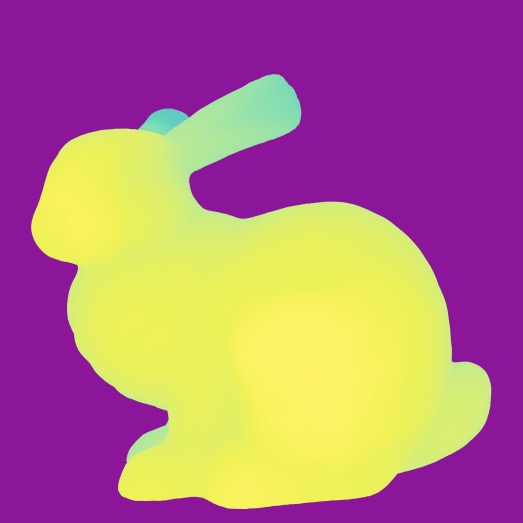} &
\includegraphics[width=0.18\linewidth]{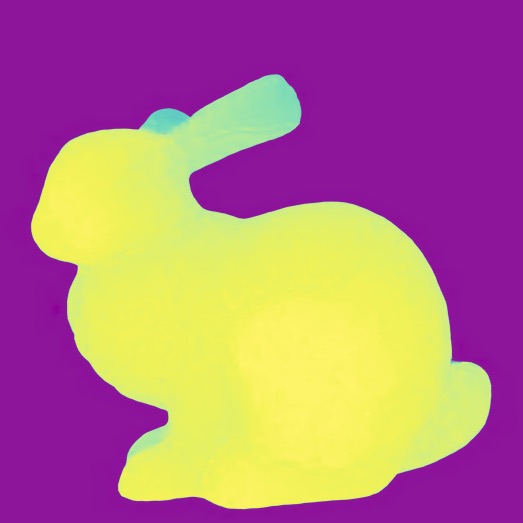} &
\includegraphics[width=0.18\linewidth]{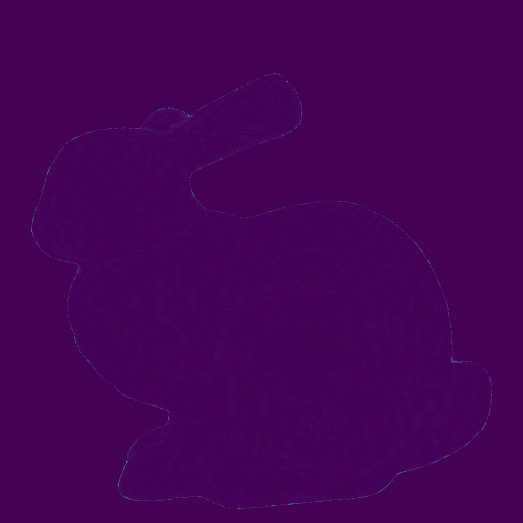} &
\includegraphics[width=0.18\linewidth]{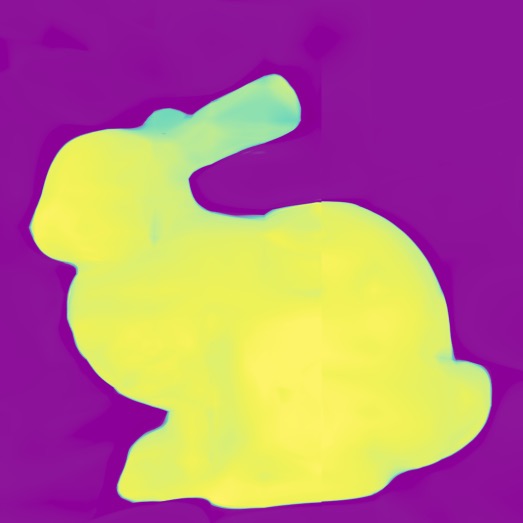} &
\includegraphics[width=0.18\linewidth]{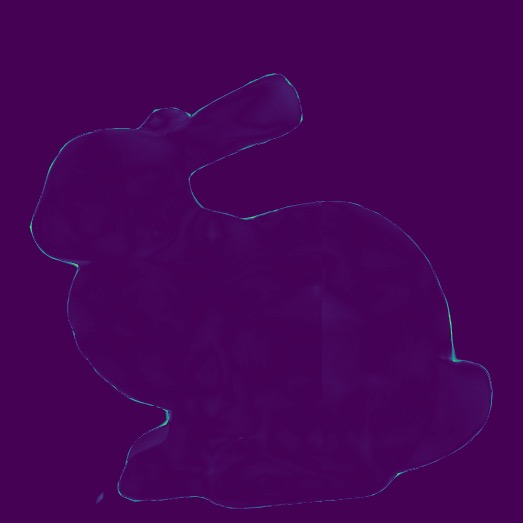} \\

\raisebox{0.08\linewidth}{\rotatebox[origin = c]{-90}{\textsf{\scriptsize{\textsc{Dragon} (Normal)}}}}&
\includegraphics[width=0.18\linewidth]{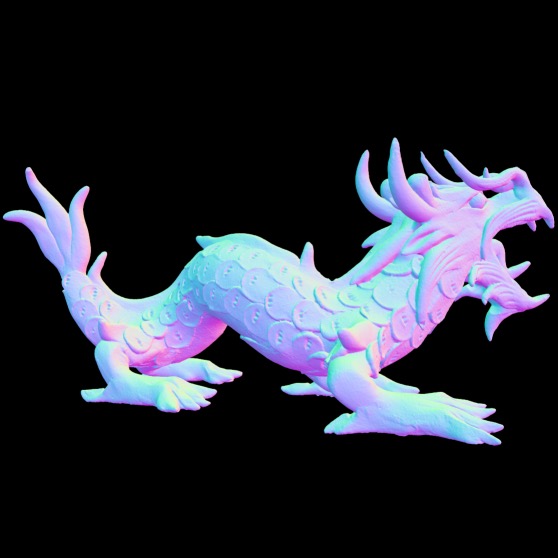} &
\includegraphics[width=0.18\linewidth]{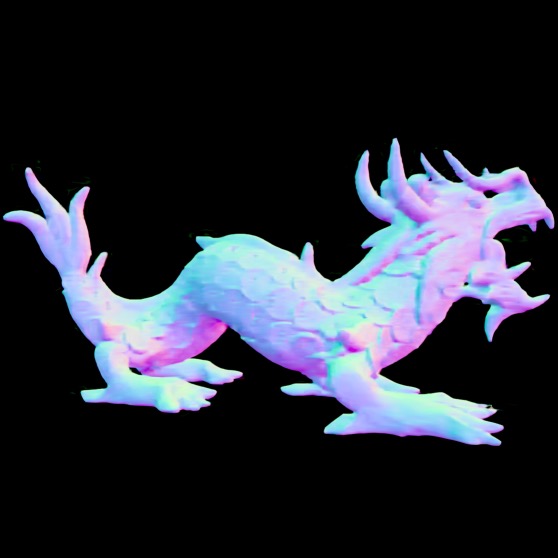} &
\includegraphics[width=0.18\linewidth]{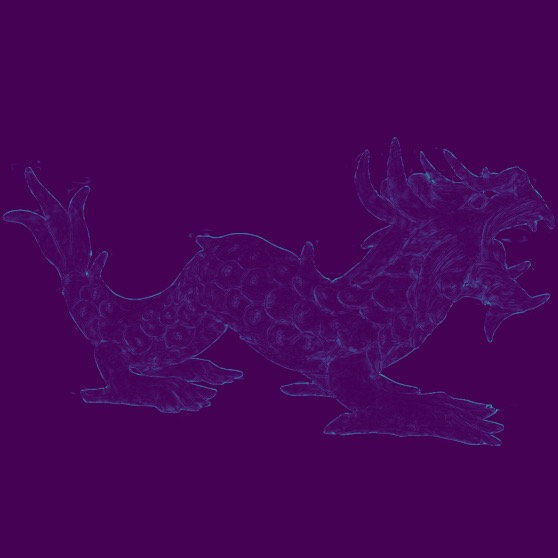} &
\includegraphics[width=0.18\linewidth]{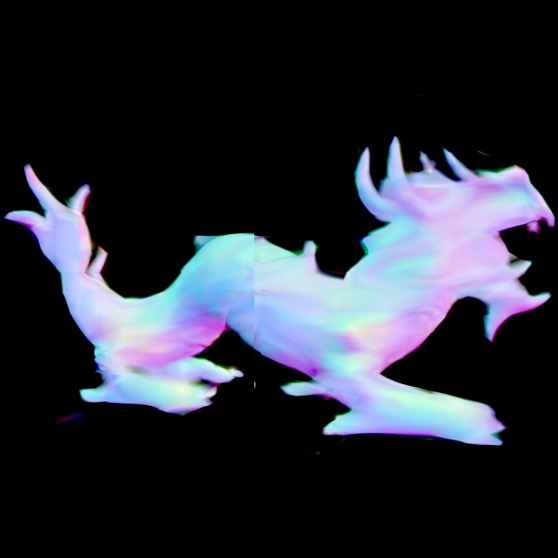} &
\includegraphics[width=0.18\linewidth]{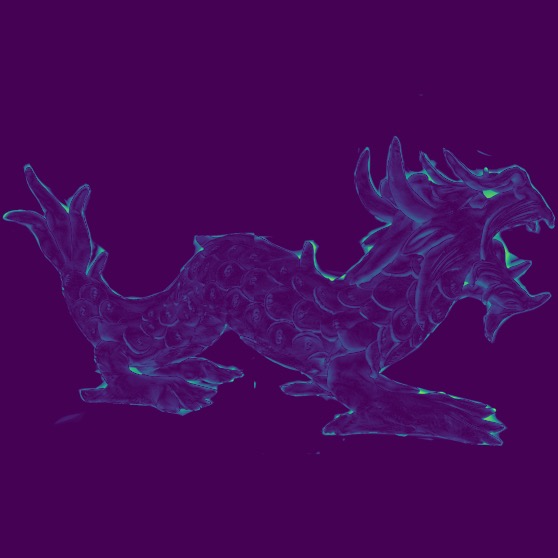} \\

\raisebox{0.08\linewidth}{\rotatebox[origin = c]{-90}{\textsf{\scriptsize{\textsc{Dragon} (Depth)}}}}&
\includegraphics[width=0.18\linewidth]{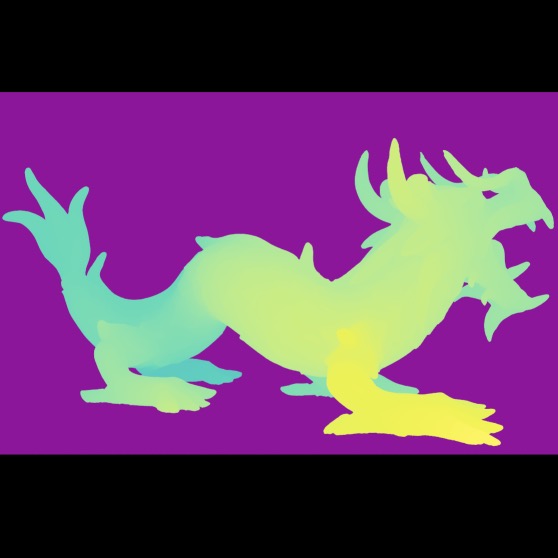} &
\includegraphics[width=0.18\linewidth]{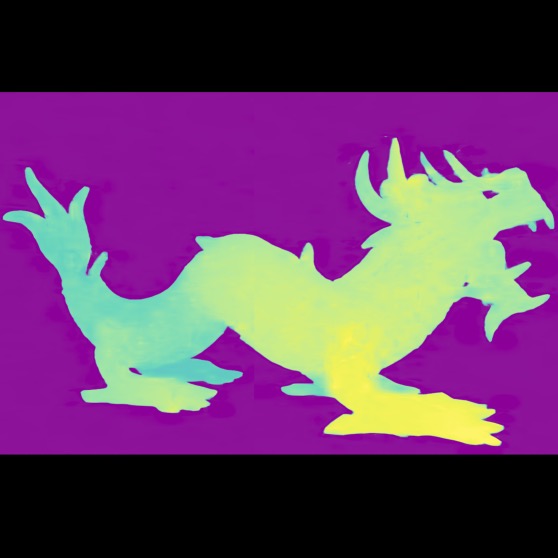} &
\includegraphics[width=0.18\linewidth]{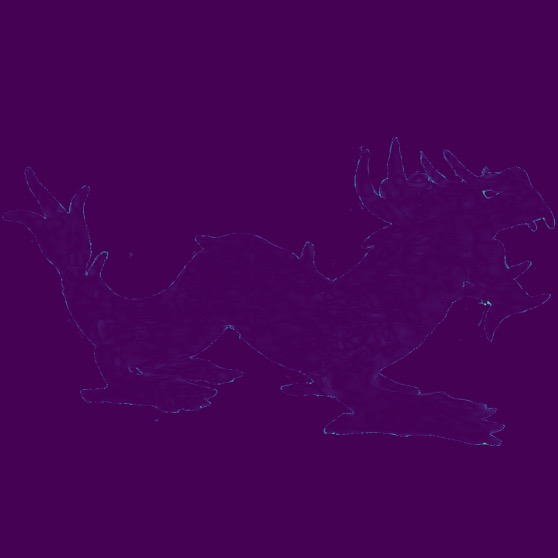} &
\includegraphics[width=0.18\linewidth]{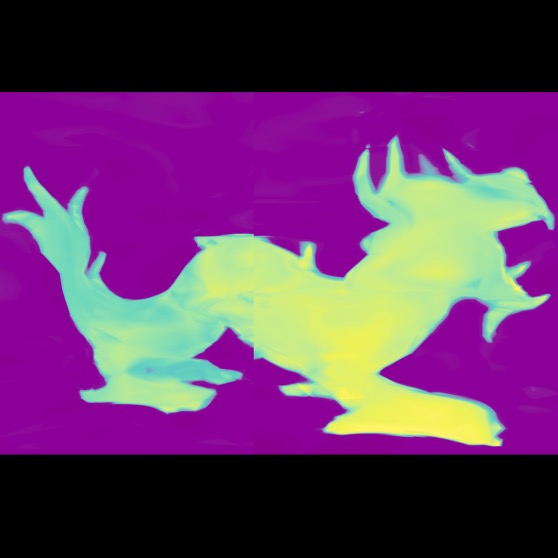} &
\includegraphics[width=0.18\linewidth]{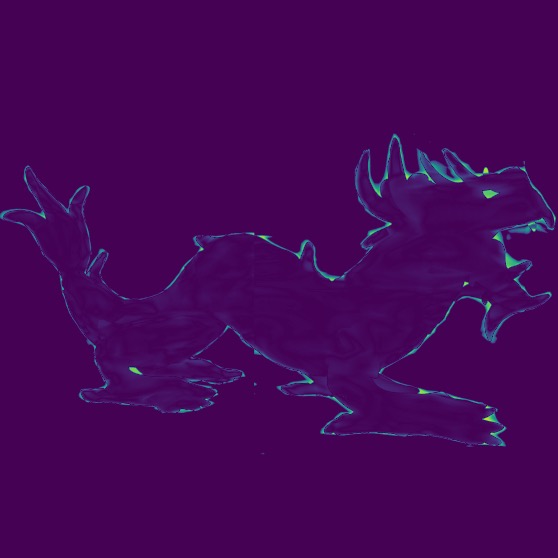} \\

&\small{(a) Reference}&\small{(b) $R = 256$}&\small{(c) Error of (b)}&\small{(d) $R = 32$}&\small{(e) Error of (d)}
\end{tabular}
    \caption{Comparison of grid resolutions when NIF is used for primary ray casting. All images are rendered after 32 training spp. The top and bottom rows of each model show shading normal and depth from the camera.}
    \label{fig:gridResCmp}
\end{figure}

Feature grids have hyperparameters $R$ (grid resolution) and $N$ (dimension of latent vectors), as shown in Table~\ref{tab:arch}, which balance quality and performance.
Here, we analyzed how these hyperparameters affect the results by tuning them.

We investigate various combinations of $R$ and $N$ in NIF for the outer and inner network independently to evaluate the differences coming from one network.
Thus, when analyzing NIF for the outer network, we traverse BVHs for objects instead of executing NIF for the rays originating from the inside of AABBs, and vice versa.
First, we analyze the impact of $R$ in NIF for the outer network in Fig.~\ref{fig:Outer2DGridRes}.
Generally, grids with higher resolution $R$ take a longer execution time for inference. But for the quality, PSNR peaks at $R=256$ and starts to decrease with even higher resolutions.
Visualizations of differences show that with $R = 64$, errors are significant in the shadow of the edge of the object while with $R = 1024$, higher errors can be seen close to the object boundary.
This is because the number of samples hitting near the object boundary is relatively small and with higher resolutions, more training samples are required to train a feature grid since each cell encodes more precise regions.
Requiring more training samples is not suitable for our case of online training described in Sec.~\ref{training}.
Thus, we choose to use $R = 256$ in this paper to evaluate the performance.

Additionally, in Fig.~\ref{fig:gridResCmp}, we also evaluate the impact of $R$ when NIF is used for primary ray casting.
To use NIF for primary ray casting, we set shading normal and depth from the camera for the output of the networks instead of zero or one as discussed in Sec.~\ref{NIF}.
As shown in Fig.~\ref{fig:gridResCmp}, errors can be significantly reduced using the higher resolution both for shading normal and depth.
These results illustrate that using higher resolutions can improve image quality in any case where NIF is used for primary and secondary ray casting though it requires more training samples.
Furthermore, we analyze the effectiveness of the dimensions of latent vectors in the outer network by tuning $N$ from 3 to 7 with the fixated $R = 256$. In this analysis, we find that a higher value of $N$ results in a higher PSNR
despite taking more time for inference.
In this paper, we choose $N = 3$ to prioritize performance, but generally, it is recommended to tweak $N$ to balance performance and quality in different use cases.
Based on the analysis, we select $R = 256$ and $N = 3$ for NIF for the outer network as our default settings.

\begin{figure}
    \centering
\setlength{\tabcolsep}{0.002\linewidth}
\begin{tabular}{ccccc}
\raisebox{0.06\linewidth}{\rotatebox[origin = c]{-90}{\textsf{\scriptsize{Uniform}}}}&
\includegraphics[width=0.23\linewidth]{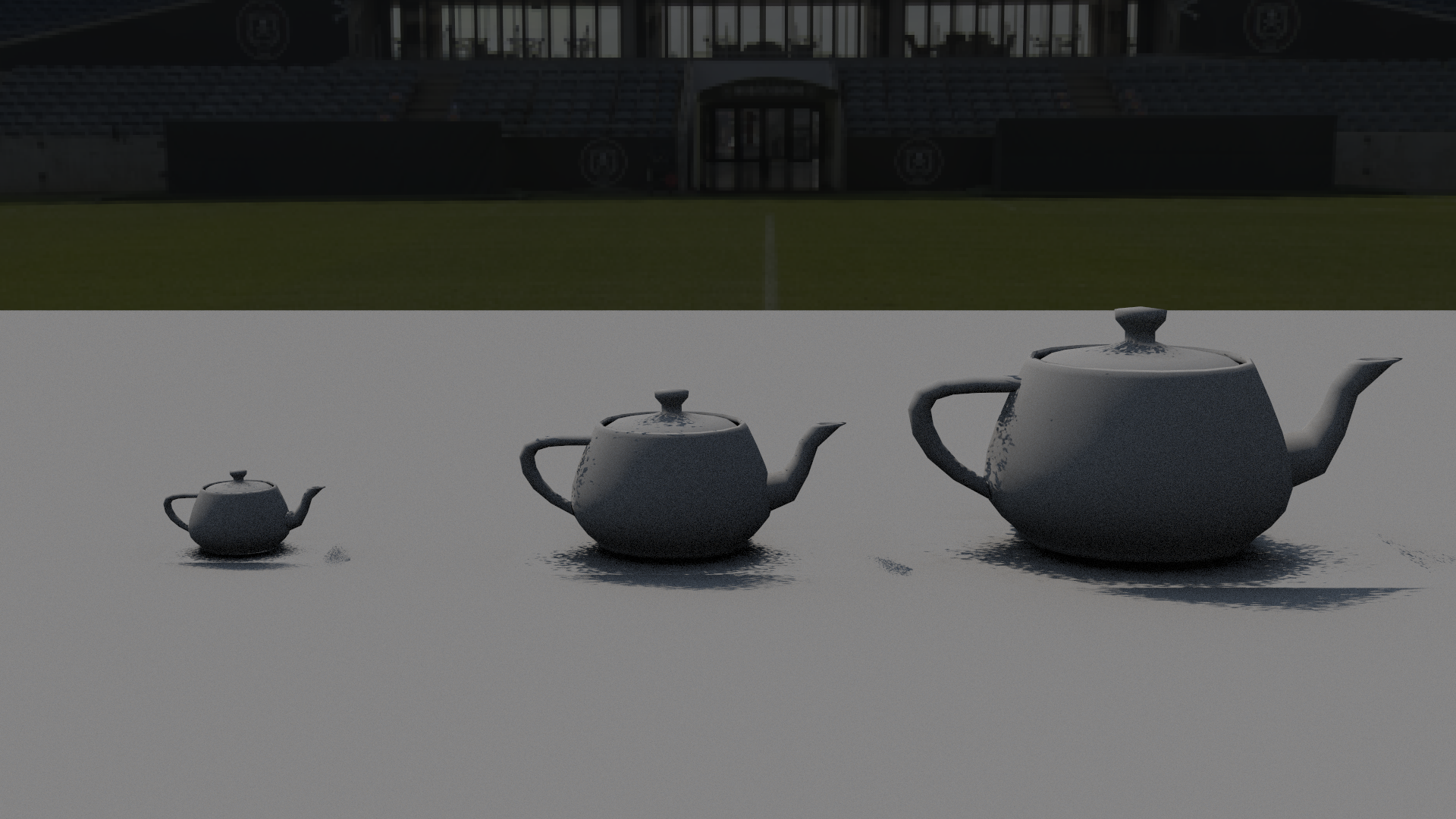} &
\includegraphics[width=0.23\linewidth]{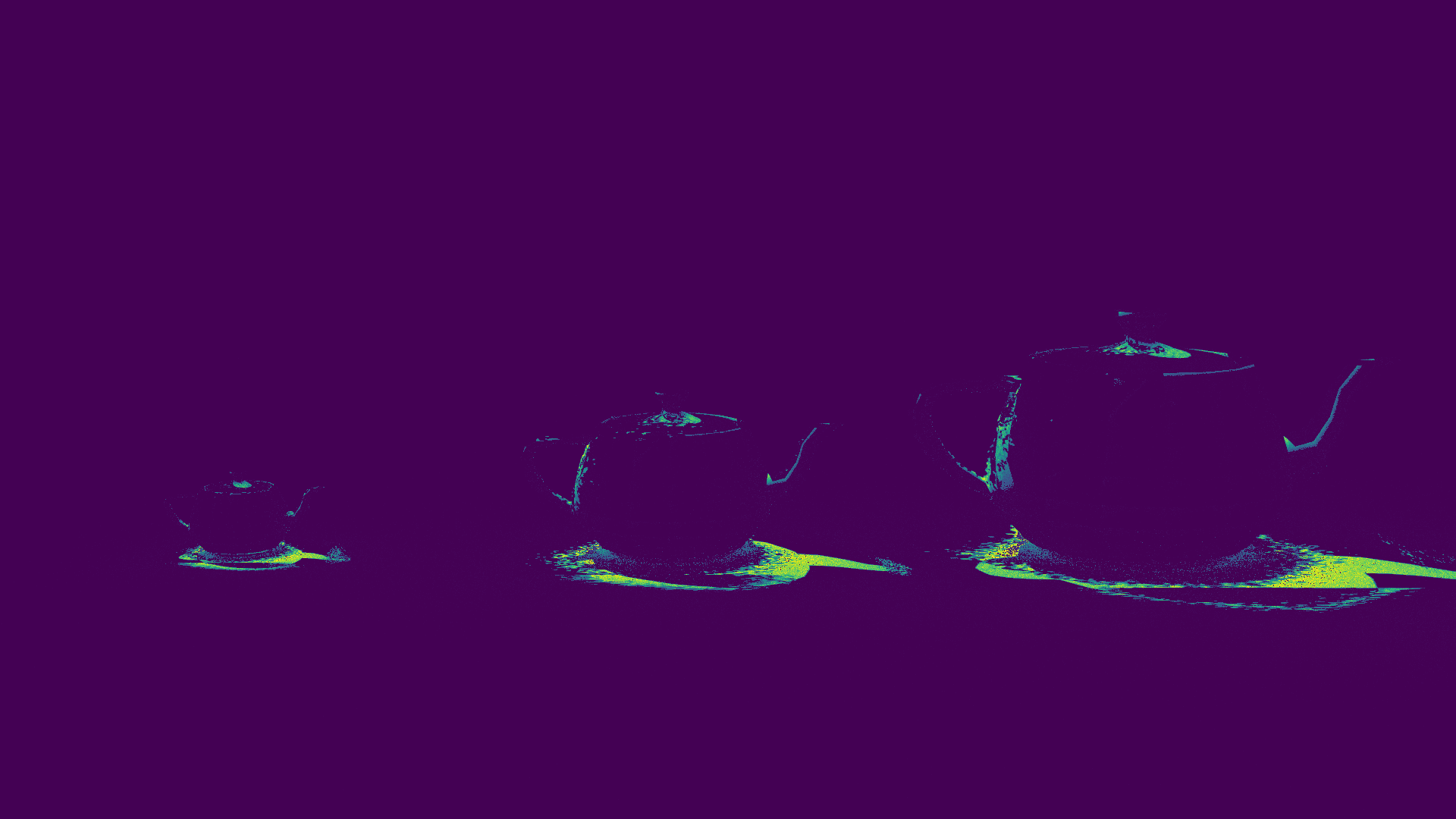} &
\includegraphics[width=0.23\linewidth]{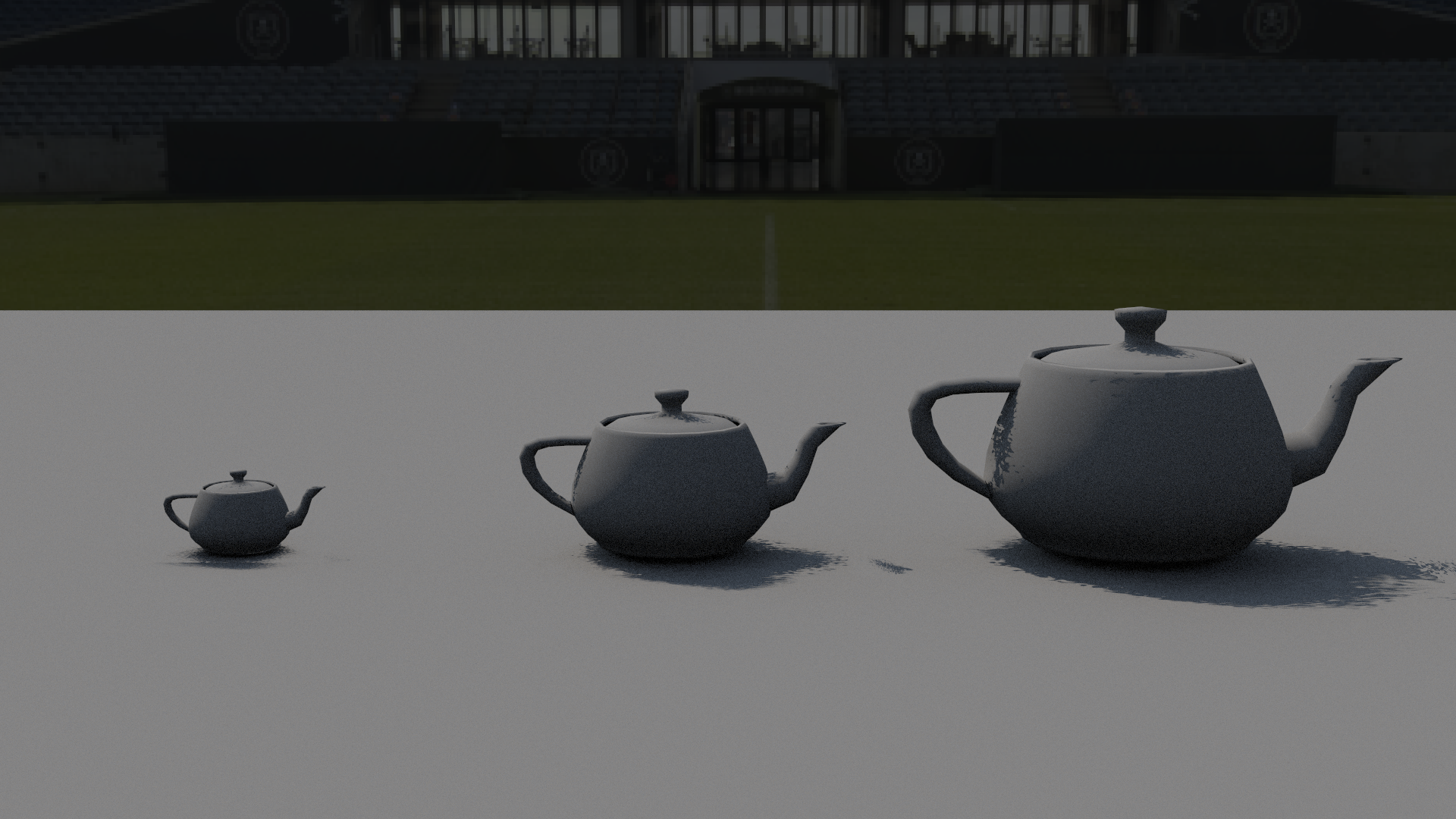} &
\includegraphics[width=0.23\linewidth]{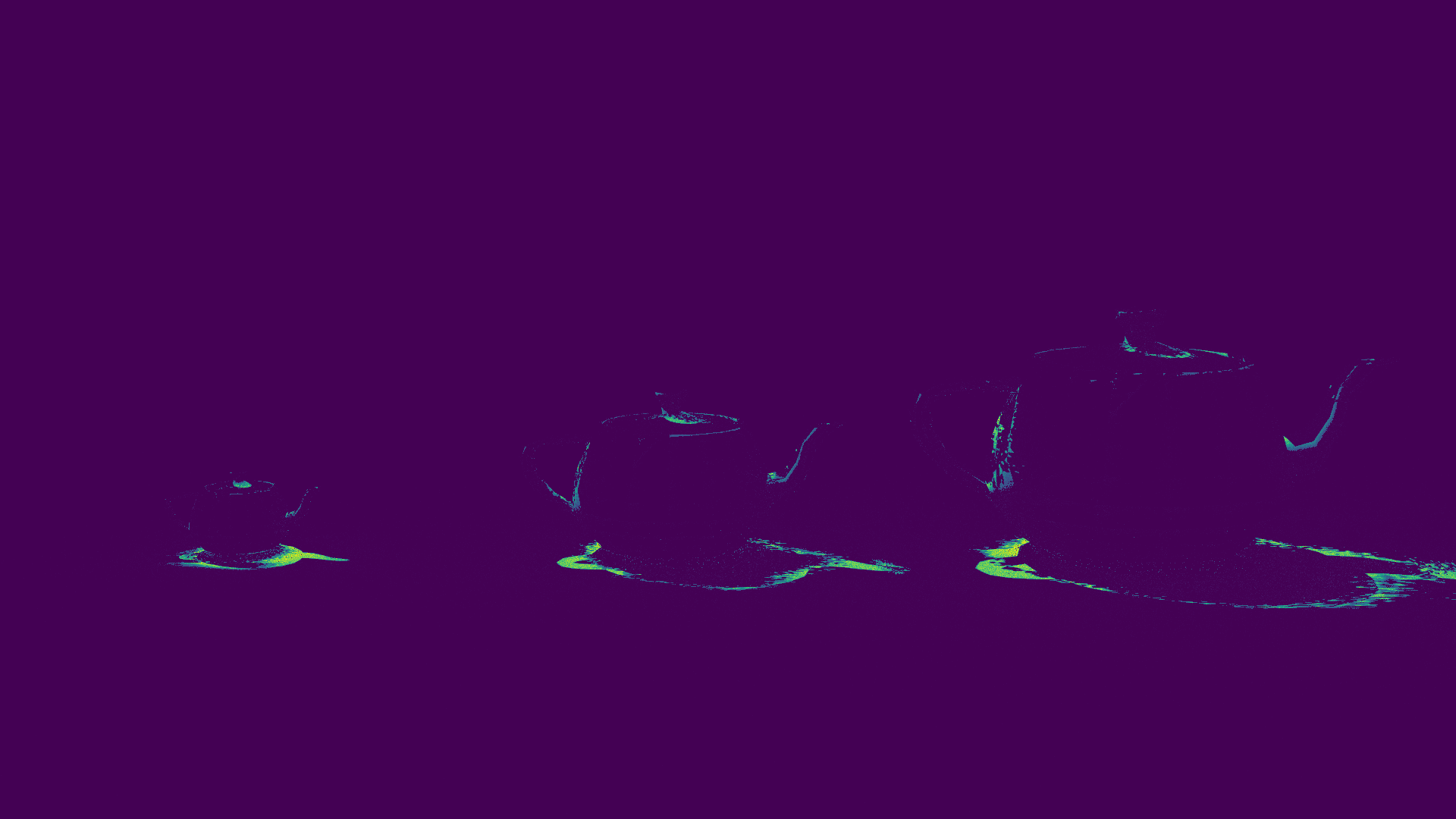}\\

\raisebox{0.06\linewidth}{\rotatebox[origin = c]{-90}{\textsf{\scriptsize{Importance}}}}&
\includegraphics[width=0.23\linewidth]{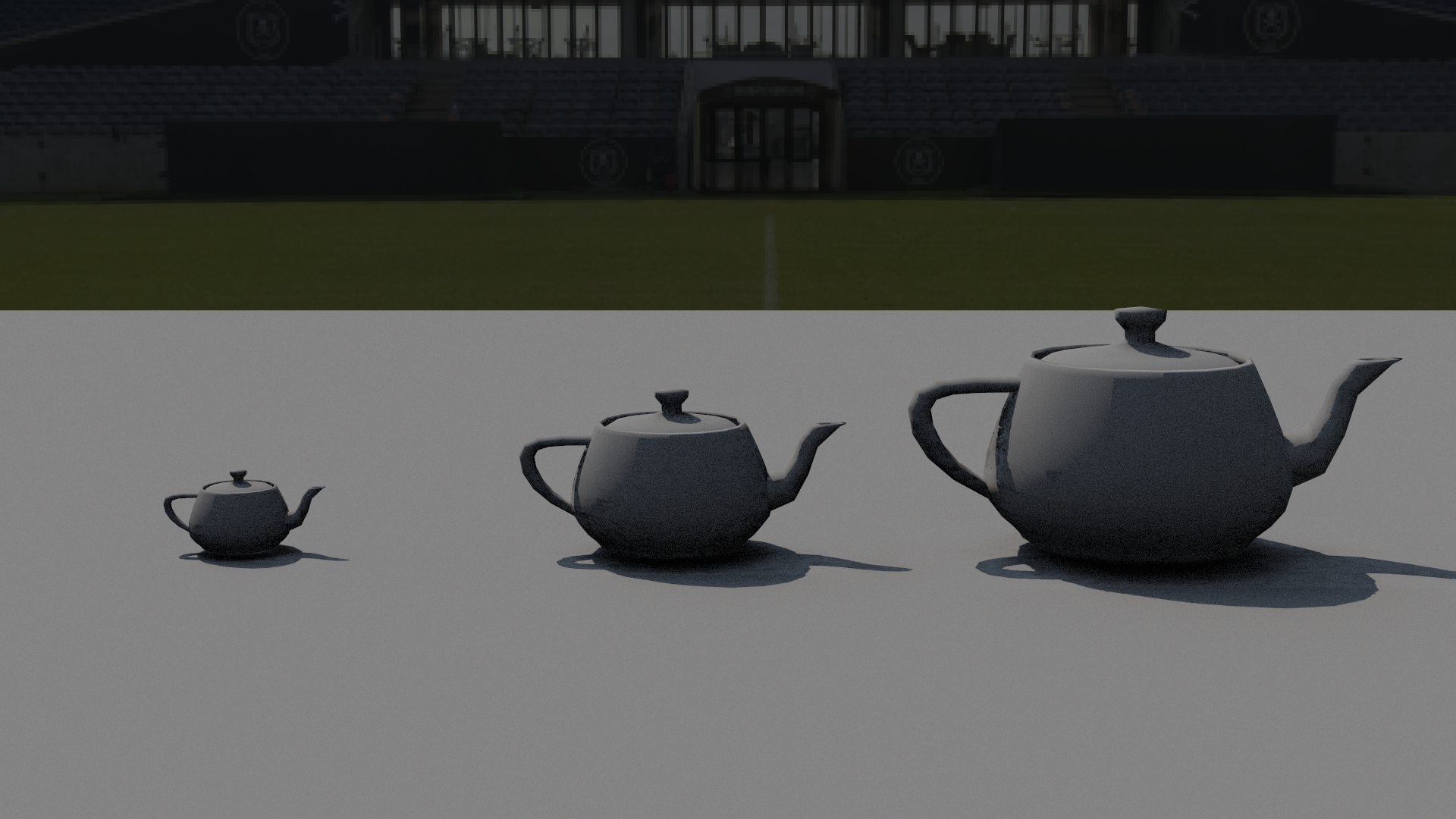} &
\includegraphics[width=0.23\linewidth]{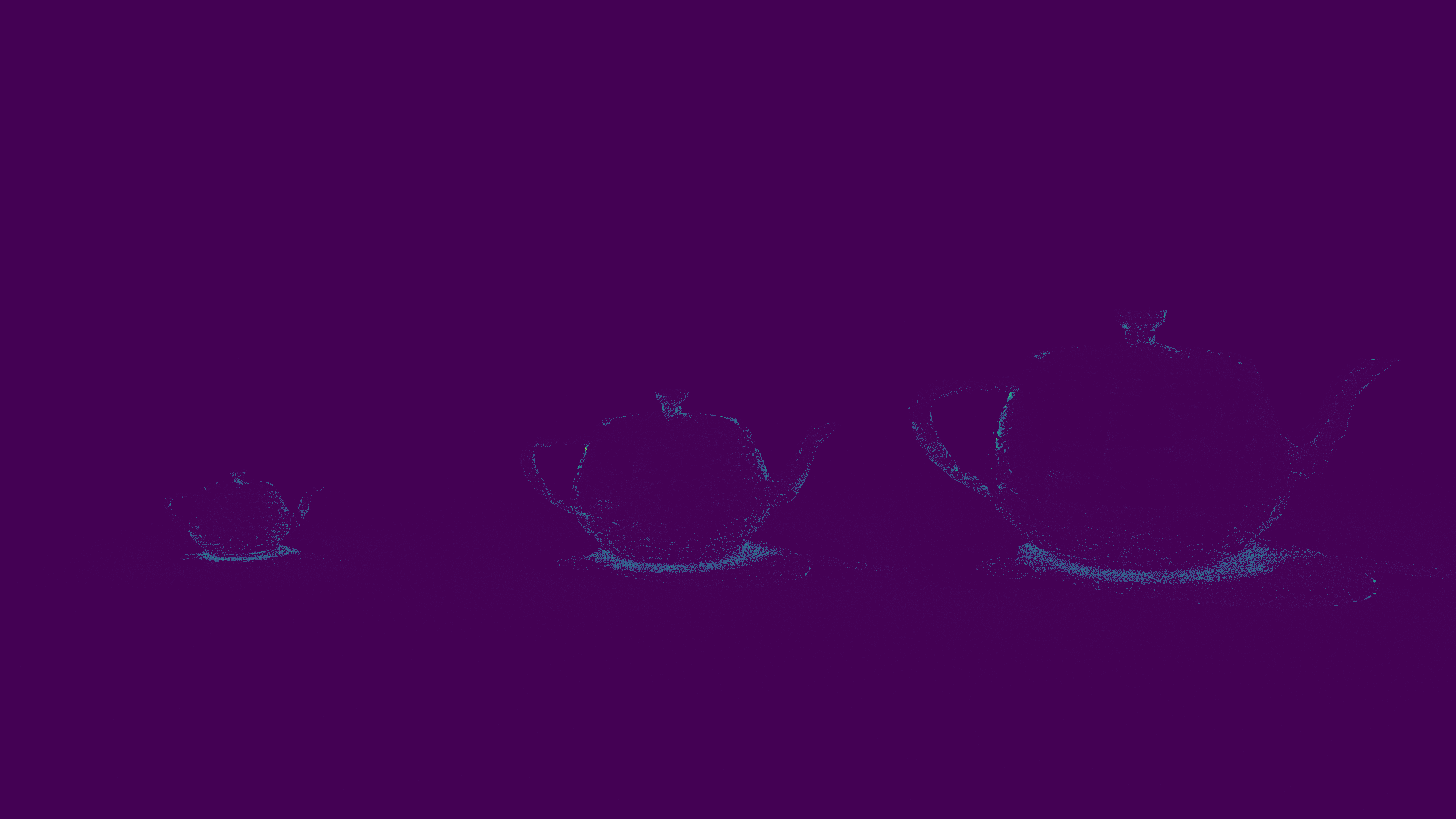} &
\includegraphics[width=0.23\linewidth]{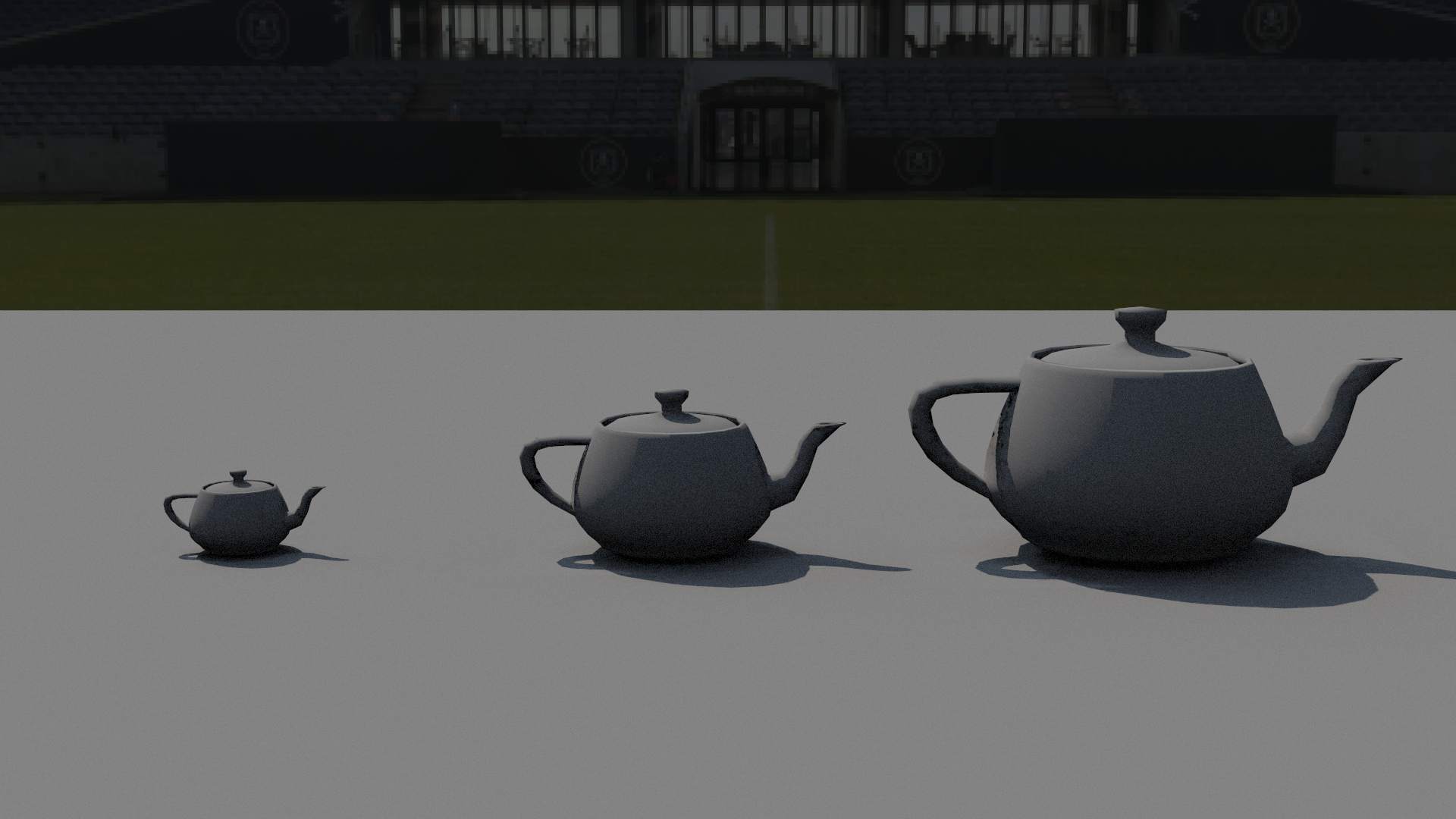} &
\includegraphics[width=0.23\linewidth]{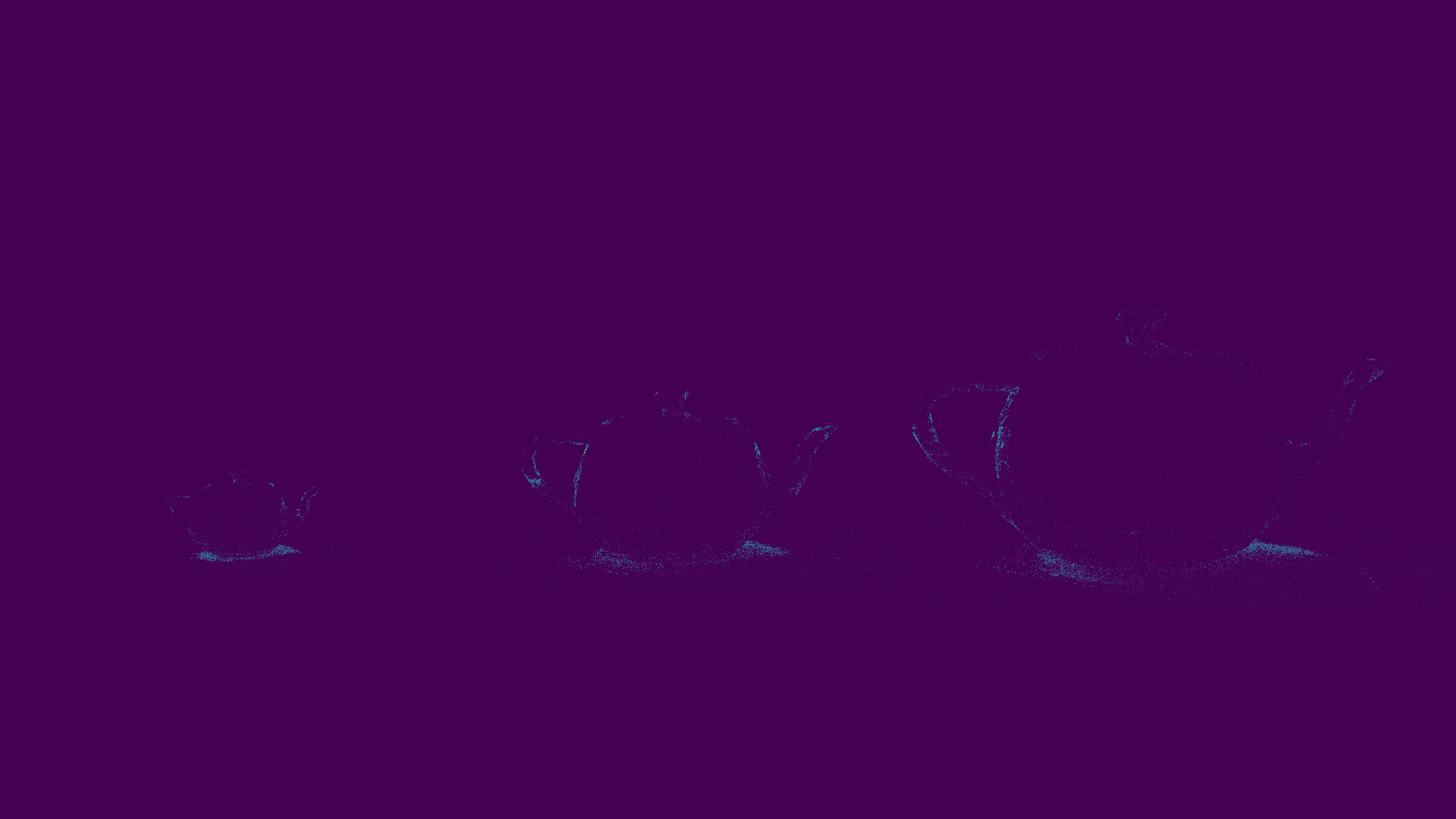}\\

&\scriptsize{(a) 32 training spp}&\scriptsize{(b) Error of (a) $\times 3$}&\scriptsize{(c) 128 training spp}&\scriptsize{(d) Error of (c) $\times 3$}
\end{tabular}
    \caption{Comparison of sampling method used to generate training samples. The first and second rows are the results using uniform and light importance sampling methods, respectively. }
    \label{fig:samplingCmp}
\end{figure}

        

\begin{figure*}
    \centering
    \setlength{\tabcolsep}{0.002\linewidth}
    \begin{tabular}{cccc}
        \raisebox{0.07\linewidth}{\rotatebox[origin = c]{-90}{\textsf{\small{Rendered}}}}&
        \includegraphics[width=0.32\linewidth]{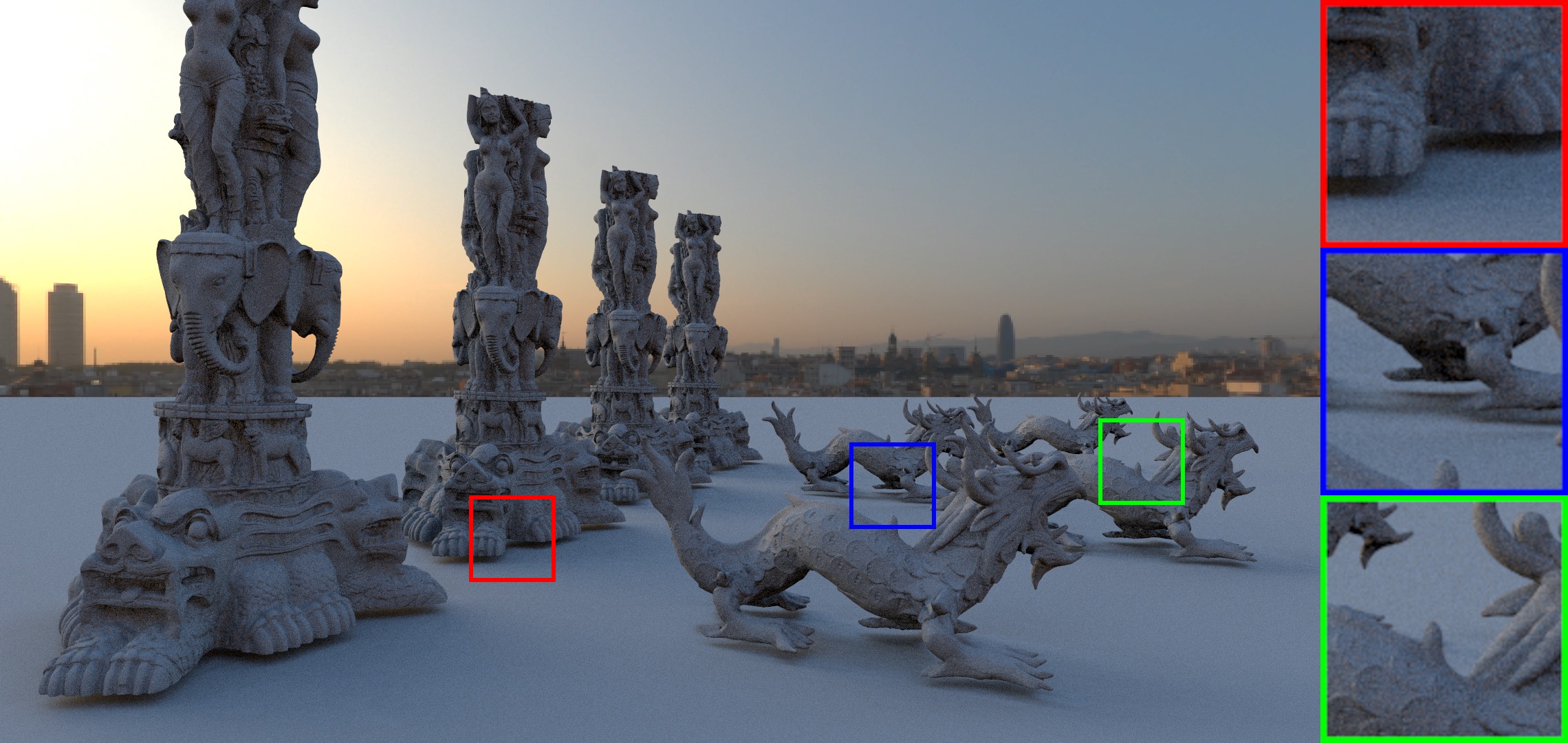} &
        \includegraphics[width=0.32\linewidth]{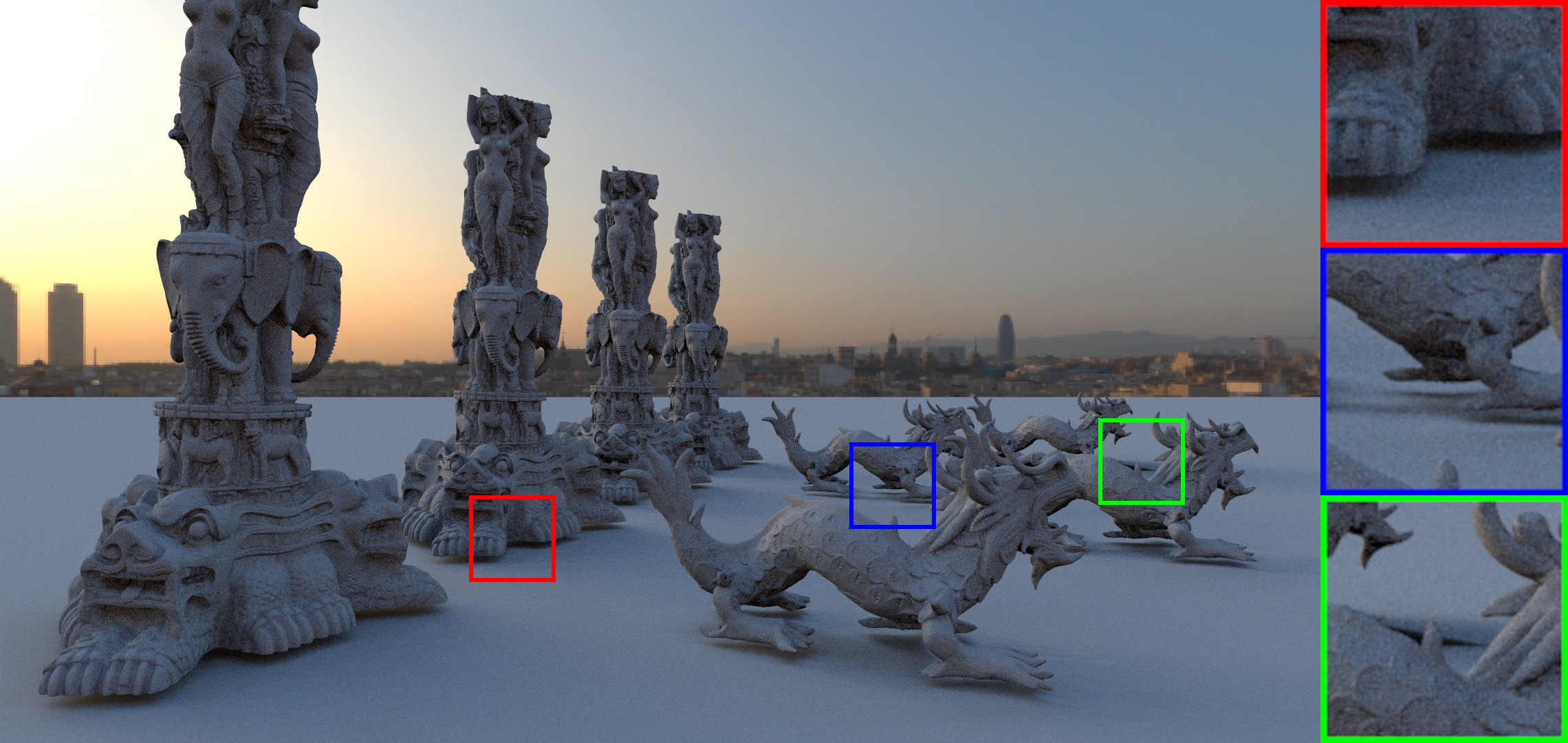} &
        \includegraphics[width=0.32\linewidth]{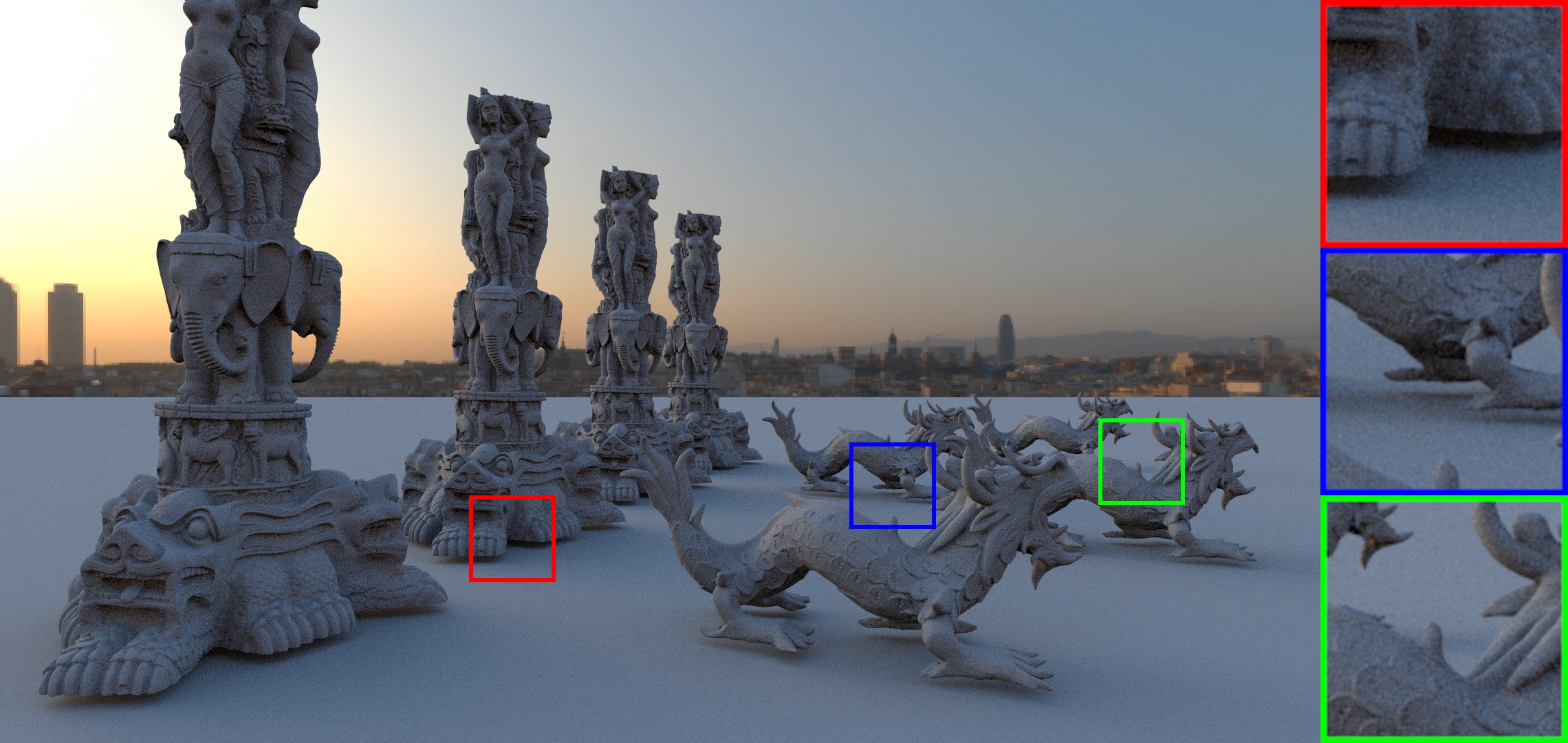} \\
        
        \raisebox{0.07\linewidth}{\rotatebox[origin = c]{-90}{\textsf{\small{Error $\times 3$}}}}&
        \includegraphics[width=0.32\linewidth]{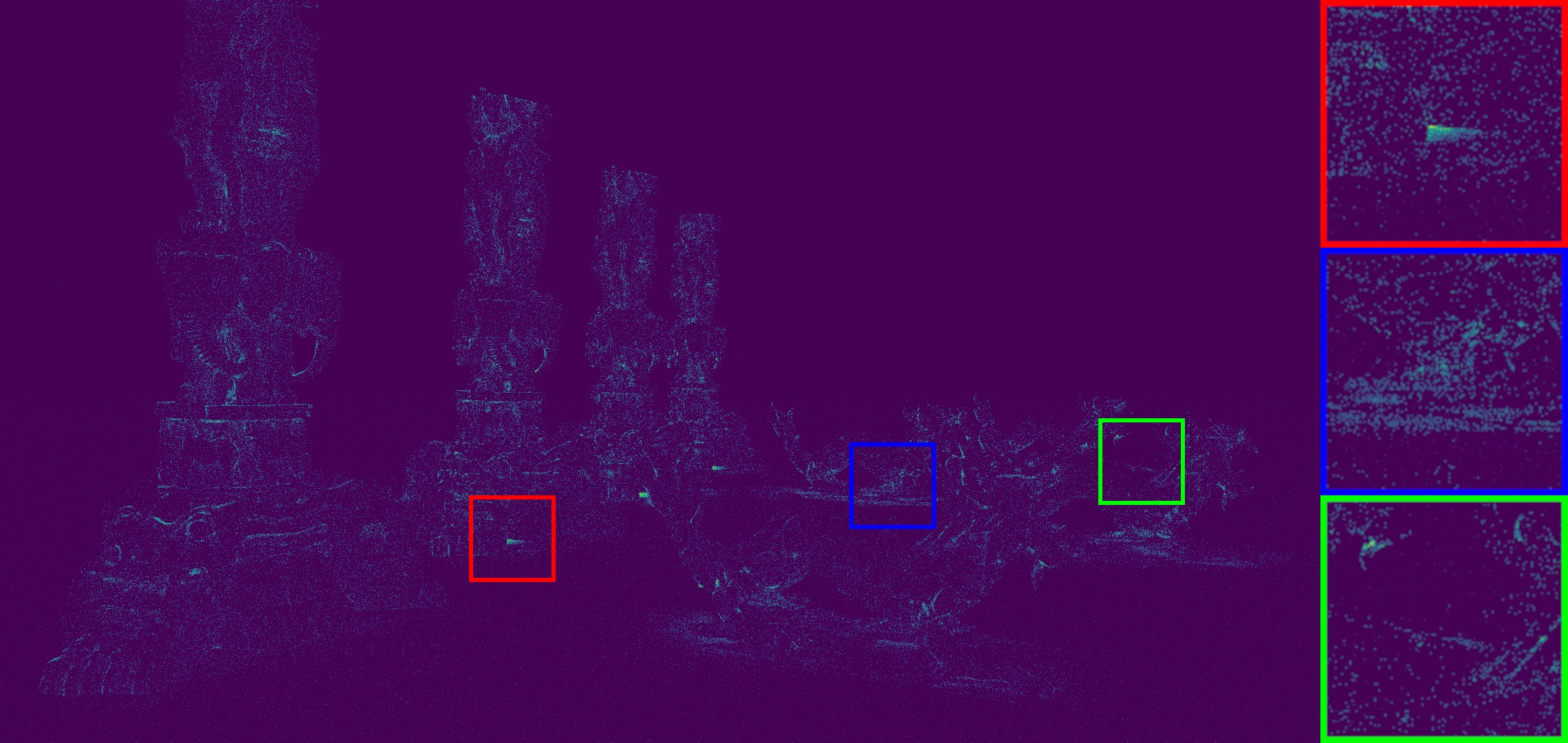} &
        \includegraphics[width=0.32\linewidth]{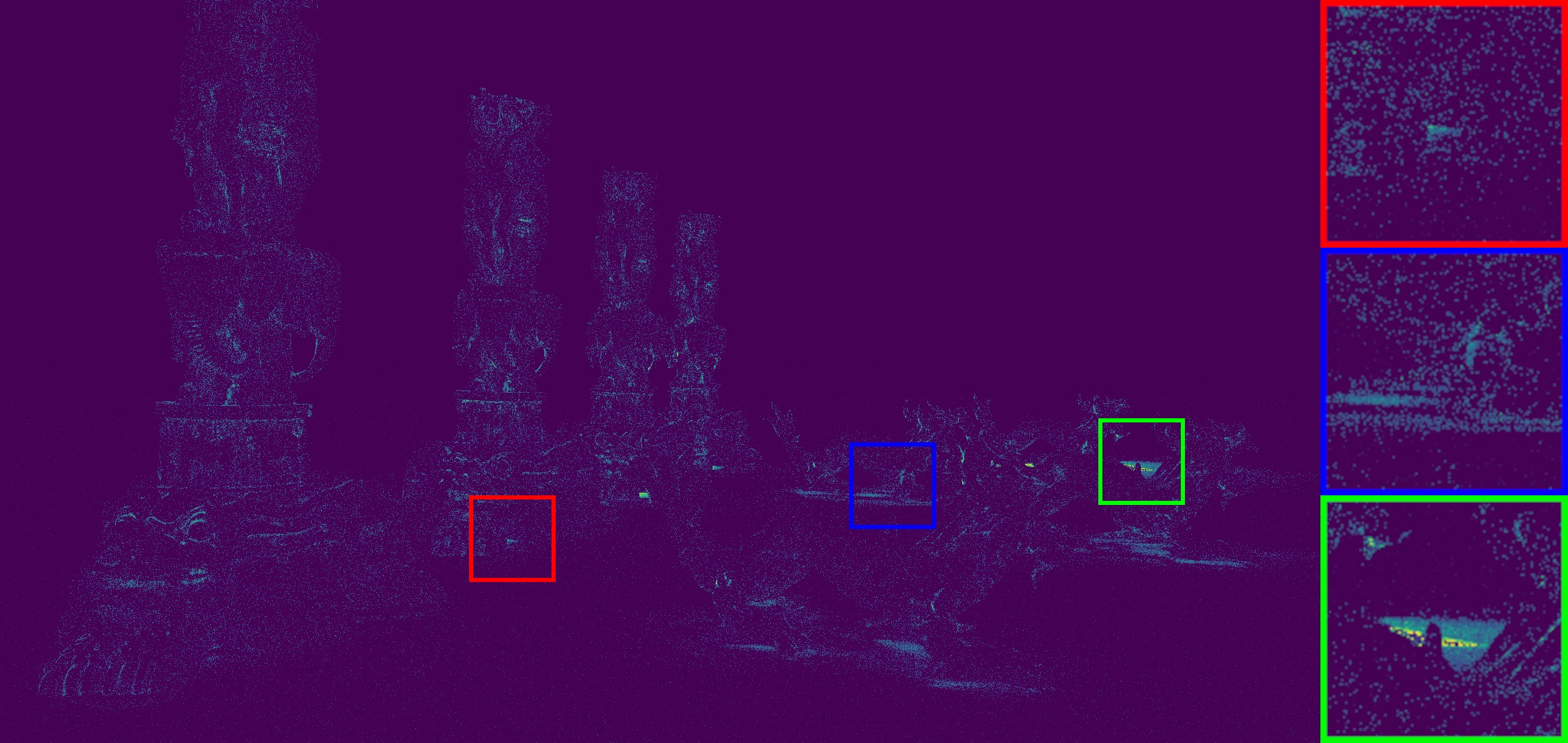} & \\

        & \small{PSNR: $31.36$ dB} & \small{PSNR: $31.32$ dB} & \\
        
        & \small{(a) Single Shared NN} & \small{(b) Single NN per object} & \small{(c) Reference}
    \end{tabular}
    \caption{Comparison between one shared network and one network per object in NIF. All images are rendered after 64 spp training using (a) a single shared NN and (b) a single NN per object. (c) is a reference image.}
    \label{fig:sharedCmp}
\end{figure*}

In NIF for the inner network, both 1D and 2D grids are employed, which results in a lot of combinations for $R$ and $N$.
However, changing the resolutions $R$ of these grids does not drastically affect the quality as those in NIF for the outer network. Though higher resolutions take more inference time and show lower PSNR values without enough training samples, a similar PSNR of about $41$ dB can be achieved both for one and two-dimensional grids with the \textsc{Statuette} scene.
Therefore, for all grids in NIF for the inner network, we use $R = 128$ which shows a better balance between performance and quality.
Also, for the dimension of latent vectors $N$, we examine it with multiple combinations for one and two-dimensional grids by varying $N$ in a one-dimensional grid from $2$ to $5$ and $N$ in a two-dimensional grid from $3$ to $6$.
In this analysis, $N = 3$ in a one-dimensional grid and $N = 5$ in a two-dimensional grid show the well-balanced result between performance and quality.
Thus, we use the configuration described in Table~\ref{tab:arch} in all our experiments because they show nearly the best-balanced results between performance and quality.

\subsection{Comparison of Sampling Method} \label{sampling}
We used ray directions with light importance sampling to generate training samples for the networks in our implementation.
To show the effectiveness of importance sampling, we compare it with uniform sampling in Fig.~\ref{fig:samplingCmp}.
As shown in Fig.~\ref{fig:samplingCmp}, importance-sampled rays significantly reduce errors, especially with low training spp.
Even with $128$ training spp, we can still see significant errors with uniform sampling, especially in the shadow of a small object on the left.
This is because the number of training samples we generate is proportional to the size of the object.

\begin{figure}
    \centering
    \setlength{\tabcolsep}{0.002\linewidth}
    \begin{tabular}{ccc}
        \includegraphics[width=0.33\linewidth]{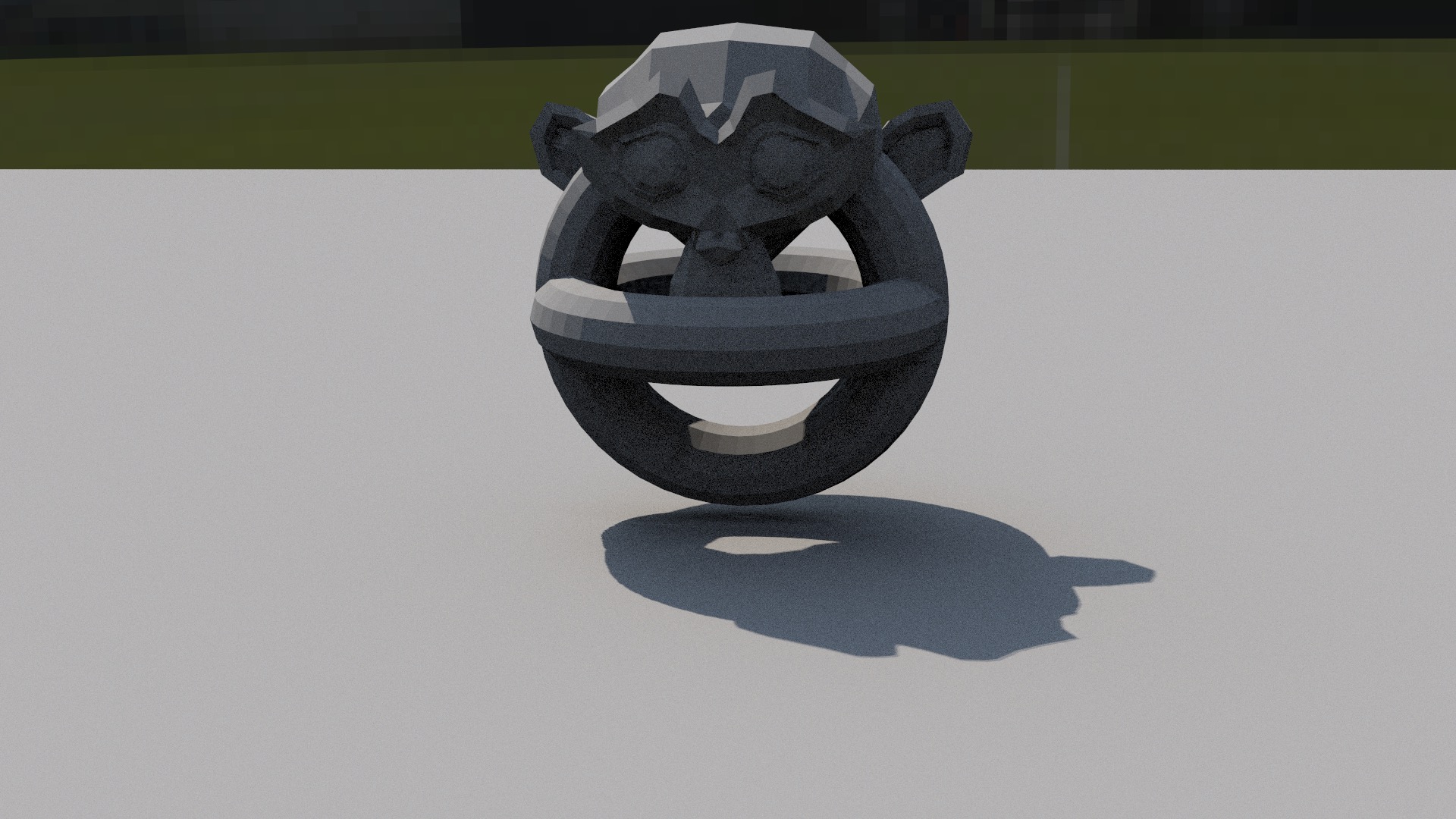} &
        \includegraphics[width=0.33\linewidth]{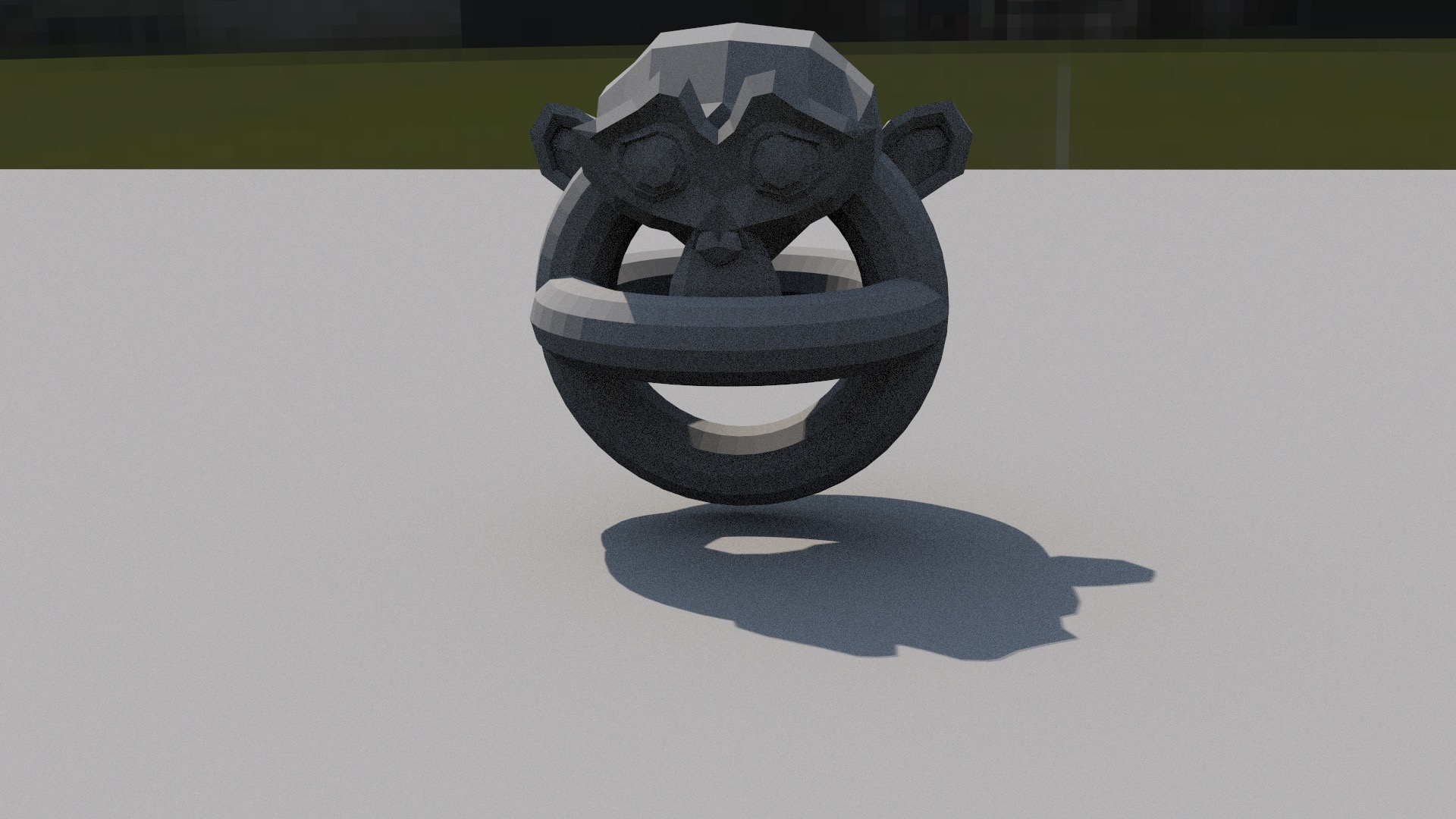} &
        \includegraphics[width=0.33\linewidth]{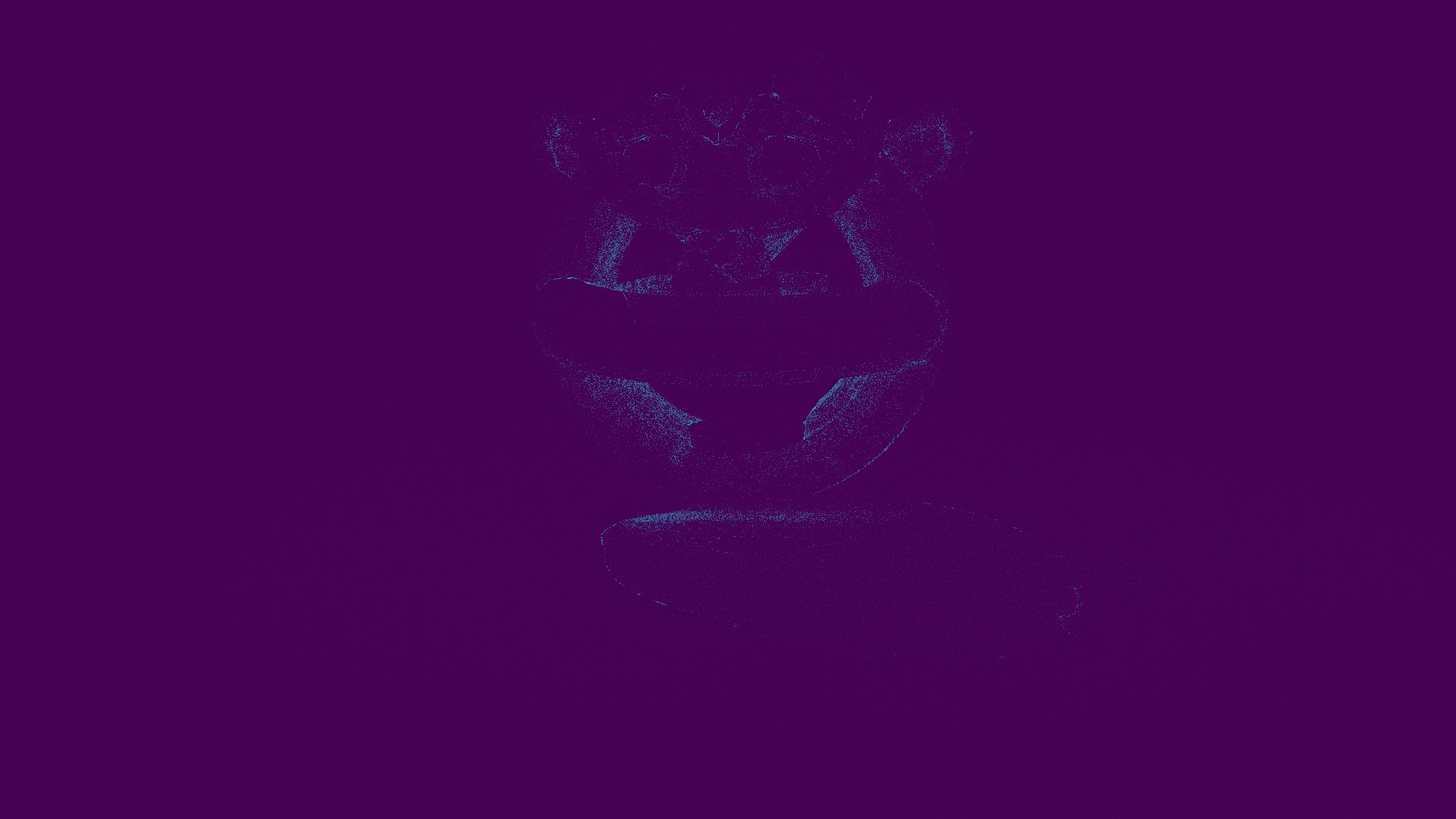} \\

        \small{(a) NIF} &
        \small{(b) Reference } &
        \small{(c) Error of (a) $\times 3$} \\
    \end{tabular}

    \caption{The scene with overlapping objects, rendered after $64$ training spp. PSNR is $40.47$ dB.}

        \label{fig:overlap}
\end{figure}

\subsection{Comparison with Single NN per Object} \label{sharedNet}
As we described in Sec.~\ref{NIFPipeline}, we only need to train a single NN per NIF because the same one can be used for querying the occlusion result of all objects in a scene. We call this type of implementation "the shared NN." Since how a shared NN implementation affects the accuracy of the output needs to be clarified, we compared our implementation to a case where an NN is allocated for each object. Fig.~\ref{fig:sharedCmp} shows a comparison of these two approaches. 
Both methods achieve a similar PSNR of about $31.3$ dB, so our implementation of shared NN is effective. Furthermore, it has another advantage since a shared NN simplifies the complexity of implementation and resource usage on GPUs.
The error images show that the shared NN can also achieve better results, especially in distant areas, while it shows higher errors close to the camera.
The reason is that the number of training samples is proportional to the object's size on the screen, and training a single NN per object requires more training samples compared to the shared NN. This is another advantage of using the single shared NN.

\begin{figure}
    \centering
    \setlength{\tabcolsep}{0.002\linewidth}
    \begin{tabular}{ccc}
        \includegraphics[width=0.33\linewidth]{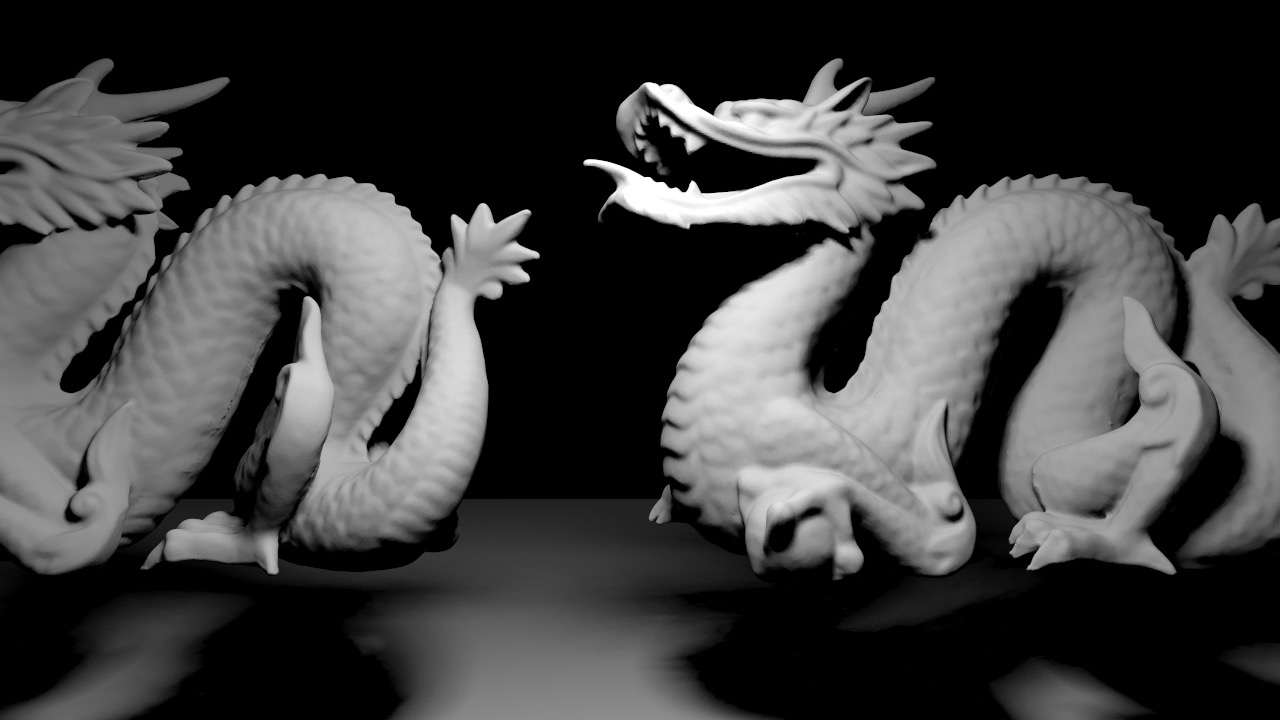} &
        \includegraphics[width=0.33\linewidth]{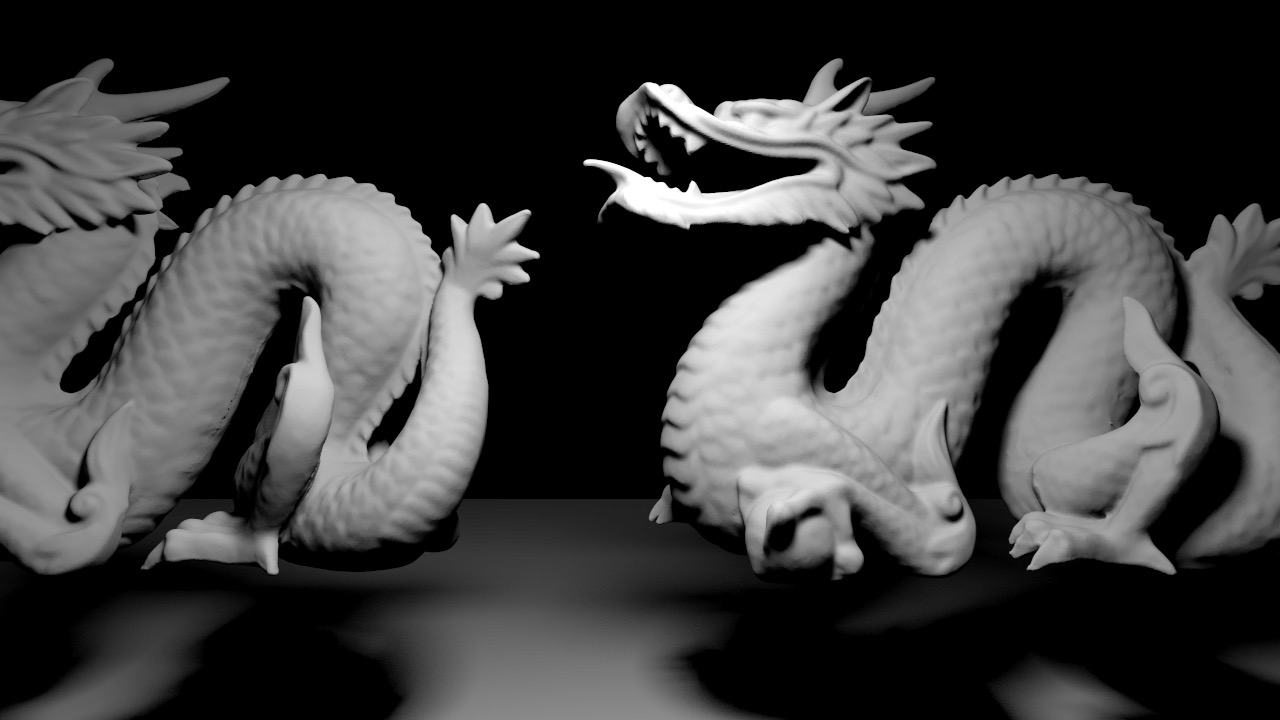} &
        \includegraphics[width=0.33\linewidth]{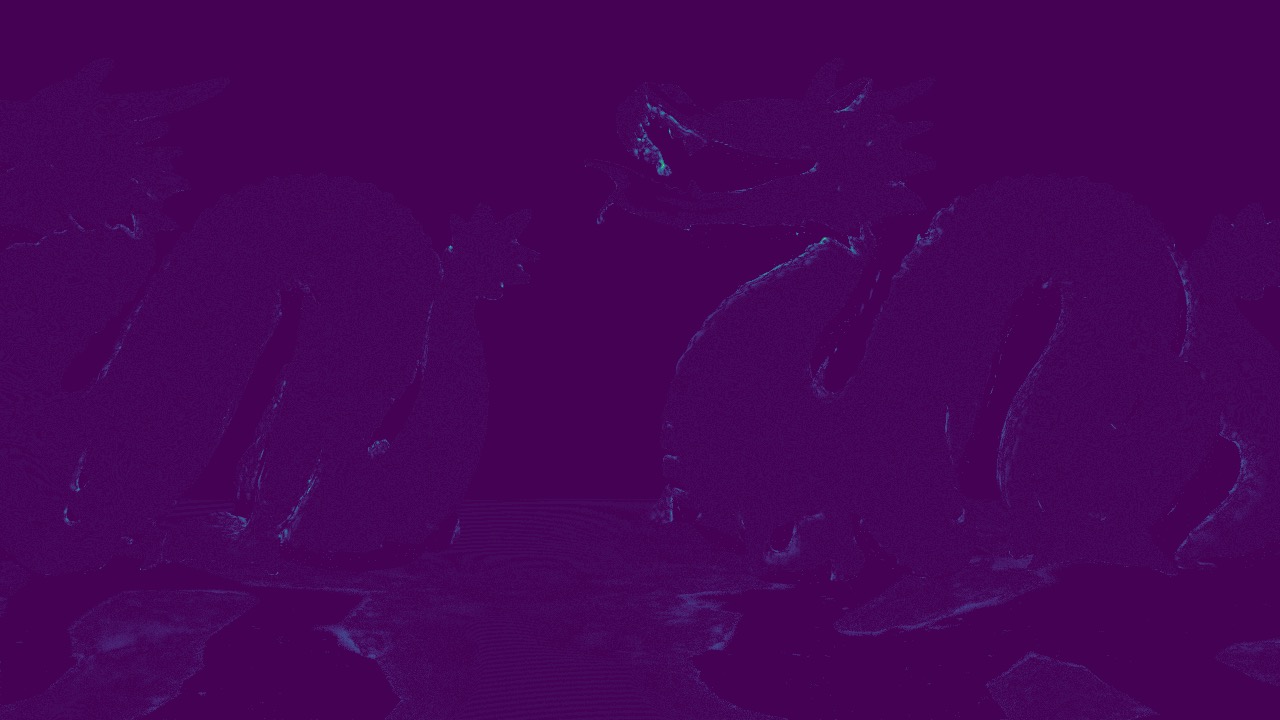} \\

        \small{(a) NIF} &
        \small{(b) Reference } &
        \small{(c) Error of (a) $\times 3$} \\
    \end{tabular}

    \caption{The rendered images of a scene illuminated by area lights, rendered after 128 training spp. PSNR is $43.58$ dB.}

        \label{fig:area_light}
\end{figure}

\subsection{Handling Scenes with Overlapping Objects}
NIF can handle scenes where multiple objects intersect geometrically with each other and have their AABBs overlapped. This is because, during the ray casting stage in training, a single ray can result in multiple data points if it intersects with or is within distinct AABBs. For example, two data points will be generated and will both contribute to the training of the inner network if a shadow ray originated from an intersection region of two AABBs. Fig.~\ref{fig:overlap} illustrates the results of such an example where three objects (two Torus-like shapes and a \textsc{Suzanne} model) intersect with each other.

\subsection{Extension for Various Types of Light Sources}
The NIF framework is designed to support various types of light sources. Not only can it render with IBL, but it is also capable of generating images with other types of lights such as an area light and a point light. It could be further extended to support multiple light sources, provided that a sufficient number of samples are used for training. Fig.~\ref{fig:area_light} shows the rendered result of a scene illuminated by area lights. In this example, we first compute the positions where the primary rays intersect with the objects. Then, we perform importance sampling from those positions directly to the light source during the training and the secondary ray casting phase.

\subsection{Performance Evaluation} \label{sec:perf}
\definecolor{colorbrewer1}{HTML}{E4007E}
\definecolor{colorbrewer2}{HTML}{00AFEC}
\definecolor{colorbrewer3}{HTML}{40DDA8}
\definecolor{colorbrewer4}{HTML}{FFDA24}
\definecolor{colorbrewer5}{HTML}{800080}

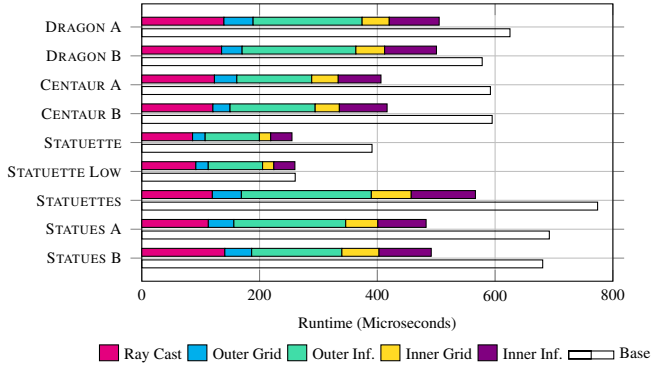
\begin{figure}
\centering
\begin{tikzpicture}
\begin{axis}[
    width=0.93\columnwidth, height=150pt,
    bar width=3pt,
    xmin=0,
    xmax=800,
    grid=major,
    xlabel={Runtime (Microseconds)},
    x tick label style={font=\tiny},
    label style={font=\tiny},
    y dir=reverse,
    ytick=data,
    ytick align=outside,
        yticklabels={\textsc{Dragon A}, \textsc{Dragon B}, \textsc{Centaur A}, \textsc{Centaur B}, \textsc{Statuette}, \textsc{Statuette Low}, \textsc{Statuettes}, \textsc{Statues A}, \textsc{Statues B}},
	y tick label style={align=center, font=\tiny},
	legend style={font=\tiny, legend columns=6, at={(0.5,-0.20)},anchor=north, draw = none},
]
\addplot[xbar stacked,bar shift = -0.2, draw=black,fill=colorbrewer1] plot coordinates {(139.49, 0) (135.53, 1) (123.39, 2)   (120.77, 3) (86.21, 4) (91.8,5) (119.86,6) (113.05,7) (141.01,8)};
\addplot[xbar stacked,bar shift = -0.2, draw=black,fill=colorbrewer2] plot coordinates {(49.55, 0) (34.85, 1) (37.99, 2) (29.09, 3) (21.43, 4) (21.09,5) (49.30,6) (43.34,7) (45.55,8) };
\addplot[xbar stacked,bar shift = -0.2, draw=black,fill=colorbrewer3] plot coordinates {(185.12, 0) (193.19, 1) (127.17, 2)  (144.57, 3)  (92.15, 4) (92.23,5) (220.67,6) (189.91,7) (153.24,8) };
\addplot[xbar stacked,bar shift = -0.2, draw=black,fill=colorbrewer4] plot coordinates {(46.1, 0) (48.78, 1) (45.09, 2) (41.12, 3)  (19.08, 4) (19.08,5) (67.76,6) (54.59,7) (63.07,8) };
\addplot[xbar stacked,bar shift = -0.2, draw=black,fill=colorbrewer5] plot coordinates {(84.75, 0) (88.08, 1) (72.53, 2)  (81.03, 3)  (36.18, 4) (35.81,5) (108.98,6) (81.93,7) (88.67,8) };

\addplot[xbar, bar shift = 0.2, draw=black,fill=none] plot coordinates {(625.26, 0) (578.07, 1) (592.02, 2)   (595.00, 3) (391.00, 4) (260.56,5) (774.05,6) (692.09,7) (680.82,8) };

\legend{Ray Cast, Outer Grid, Outer Inf., Inner Grid, Inner Inf., Base}

\end{axis}

\end{tikzpicture}
\caption{Runtime performance comparison. }
\label{fig:perf}
\end{figure}

We compared the performance of NIF against its BVH-based implementation as a counterpart and measured the runtime differences to demonstrate the speedup of our method. For the baseline BVH-based implementation, we utilized the state-of-the-art SAH BVH builder with the support of the hardware ray tracing cores on the RDNA3 GPU~\cite{navi3}. Also, we used wave matrix multiply accumulate (WMMA) instruction for NN execution. 
Table~\ref{tab:perf} shows the performance breakdown of each stage in NIF. Fig.~\ref{fig:perf} illustrates the data in a stacked bar chart to provide a visualized comparison. From the data, the notation \textit{Grid} indicates the transformation process of preparing the inputs for the NN by featuring the grid as described in Equation~\ref{eq:nif_outer} and \ref{eq:nif_inner}, whereas \textit{Inference} denotes the process of predicting whether the inputs would result in occlusions by performing inference with the NN. As shown in the results, when NIF is adopted in the ray tracing pipeline and thus prevents the GPU from performing the BVH traversal on the most divergent part of the tree (i.e. the bottom-level BVH), the ray casting time is reduced by up to $35 \%$, or the performance improvement of $1.53 \times$ is achieved for the \textsc{Statuette} scene, which has $10$M triangles.  
The performance improvement varies depending on the scenes. Some factors bring performance improvement in NIF. First, NIF does not traverse the bottom-level BVH whose structure and size could be complex and large, especially for an object with a large number of triangles. It causes divergent execution and memory access. At the same time, it is known that some rays can traverse more than others which stalls the entire GPU execution, making such execution inefficient. On the other hand, NIF only traverses the top-level BVH, which is relatively small; thus, the execution is less divergent than the traversal of the entire BVH. NIF runs NN inference whose memory access is coherent and execution is uniform, which is ideal for GPUs. This results in better utilization of GPUs. We can also see that ray casting in NIF is relatively expensive, as shown in Fig.~\ref{fig:perf}. This is mainly due to the amount of data we need to write to global memory and the number of atomic operations required. 

\begin{figure}
     \centering

    \begin{tabular}{c}
 \begin{tikzpicture} \begin{axis} [
            	width=\columnwidth, height=100pt,
            	ymin=0,
                xlabel={Number of Polygons in Millions (M)},
                ylabel={Runtime (Microseconds)},
                xlabel near ticks,
                ylabel near ticks,
                label style={font=\tiny},
                tick label style={font=\tiny},
                grid=major,
                legend style={cells={anchor=west}}
            ]
            
            \addplot[mark=x, only marks, colorbrewer5] coordinates{
            (7.2, 505.01) (7.2, 500.43) (2.5, 406.17) (2.5, 416.58) (10, 255.05) (0.5, 260.01) (50, 566.57) (17.5, 482.82) (52.8, 491.54)};

            \legend{}
            \end{axis}
            \end{tikzpicture} \\

             \small{(a)} \\

     \begin{tikzpicture} \begin{axis} [
            	width=\columnwidth, height=100pt,
            	ymin=0,
                xlabel={Number of Rays processed in Thousands (K)},
                ylabel={Runtime (Microseconds)},
                xlabel near ticks,
                ylabel near ticks,
                label style={font=\tiny},
                tick label style={font=\tiny},
                grid=major,
                legend style={cells={anchor=west}}
            ]
            
            \addplot[mark=*,only marks, colorbrewer1] coordinates{
            (171+44, 255.05) (172+44, 260.01) (243+110, 406.17) (285+126, 416.58) (298+137, 491.54) (366+130,505.01) (376+124, 482.82) (387+141,500.43) (445+176, 566.57)};

            \addplot[dashed,thick,colorbrewer3] coordinates{
            (171+44, 255.05) (445+176, 566.57)
            };

            \legend{}
            \end{axis}

            \end{tikzpicture} \\

             \small{(b)} \\
    
    \end{tabular}
    
        \caption{(a) Scene complexity and execution time. (b) Number of rays processed by NIF and execution time. }
        \label{fig:graph:num_poly_ray_perf}
\end{figure}
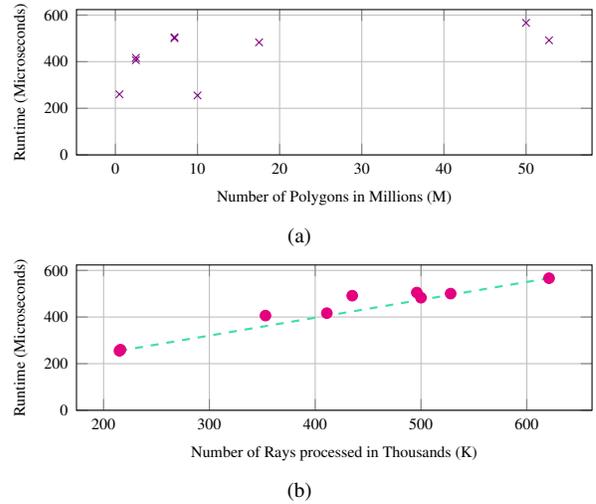

\begin{figure}
    \centering
\setlength{\tabcolsep}{0.002\linewidth}
\begin{tabular}{cccc}
\raisebox{0.08\linewidth}{\rotatebox[origin = c]{-90}{\textsf{\scriptsize{\textsc{Dragon A}}}}}&
\includegraphics[width=0.31\linewidth]{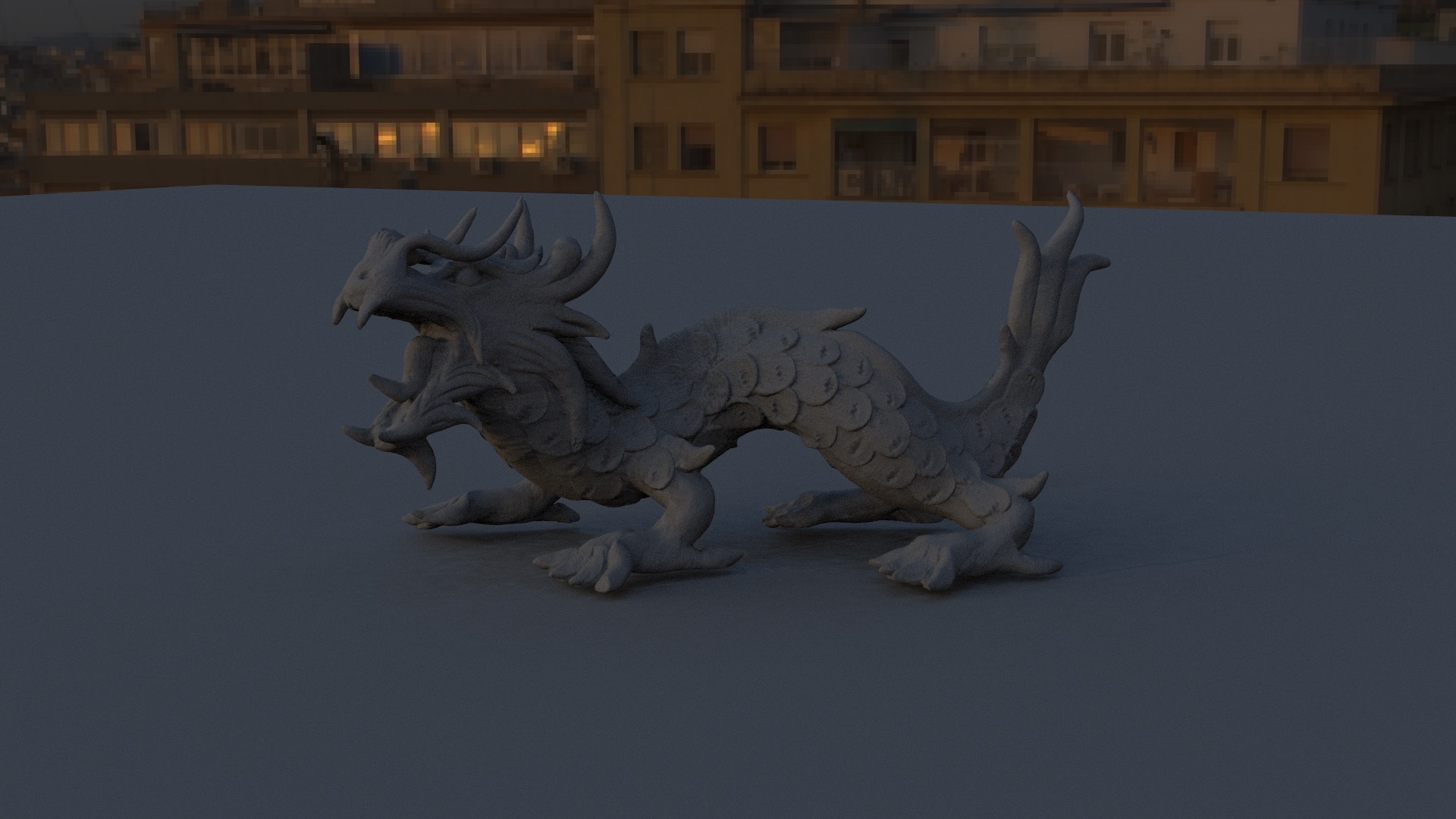} &
\includegraphics[width=0.31\linewidth]{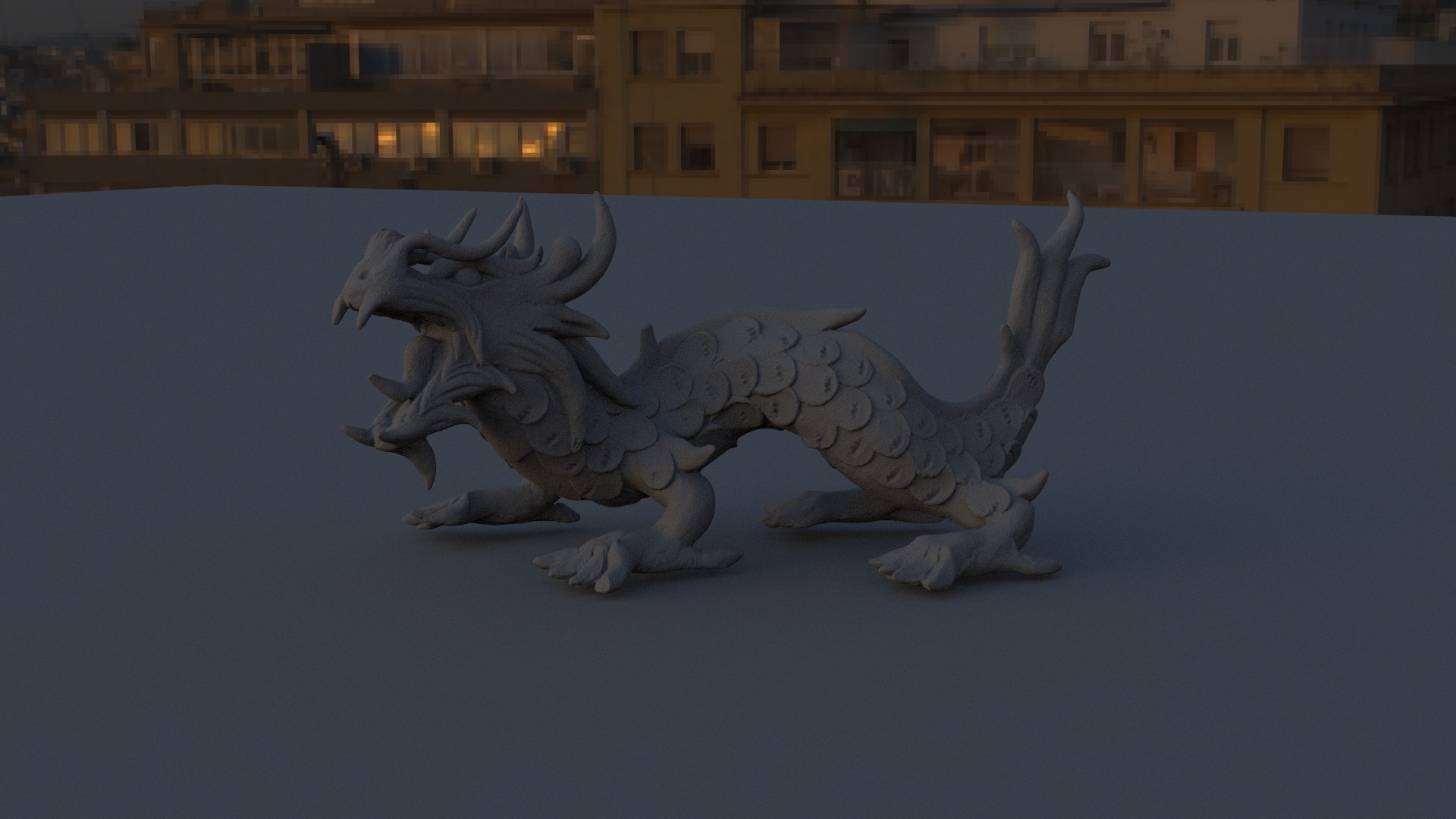} &
\includegraphics[width=0.31\linewidth]{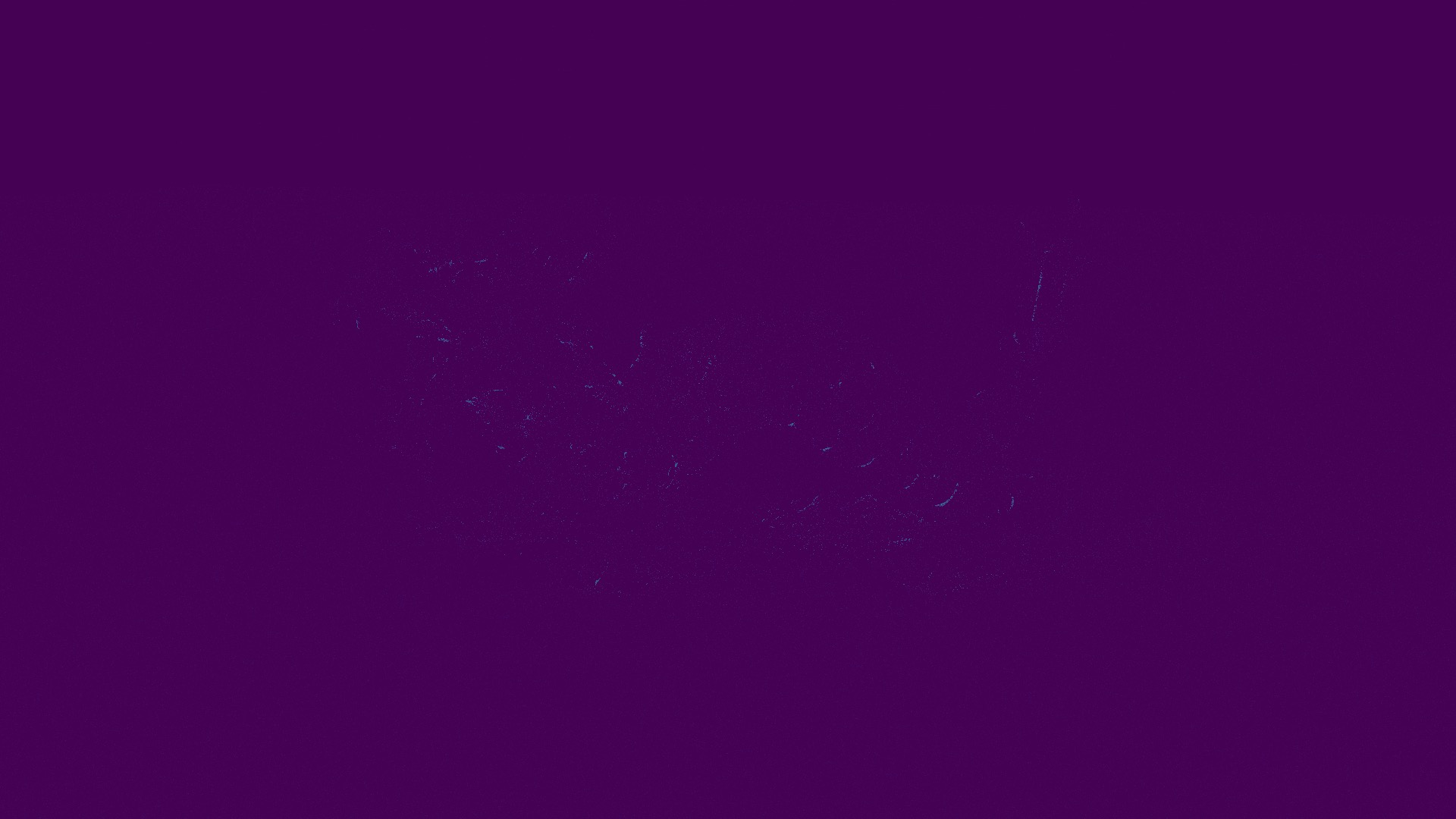} \\

\raisebox{0.08\linewidth}{\rotatebox[origin = c]{-90}{\textsf{\scriptsize{\textsc{Dragon B}}}}}&
\includegraphics[width=0.31\linewidth]{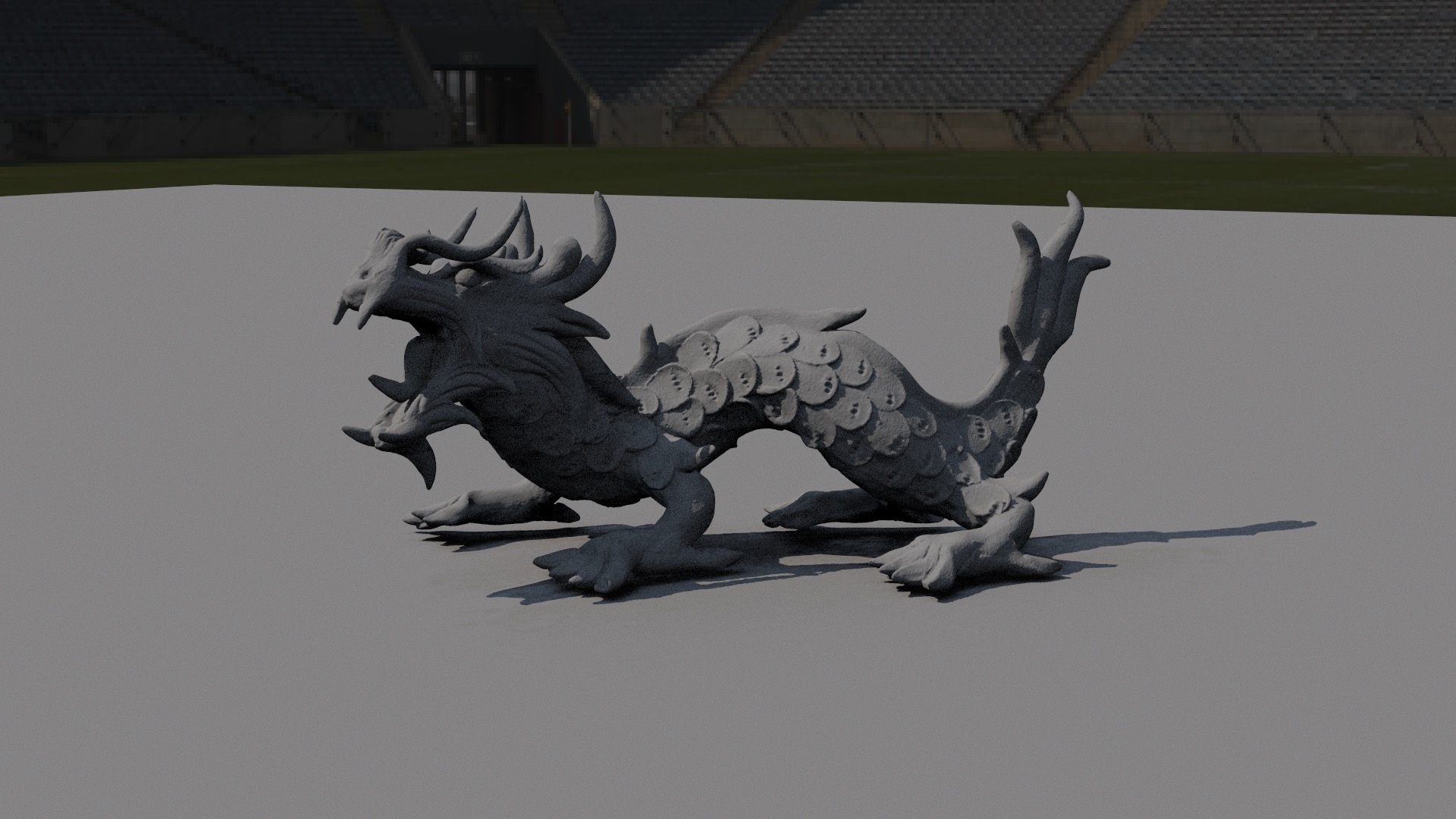} &
\includegraphics[width=0.31\linewidth]{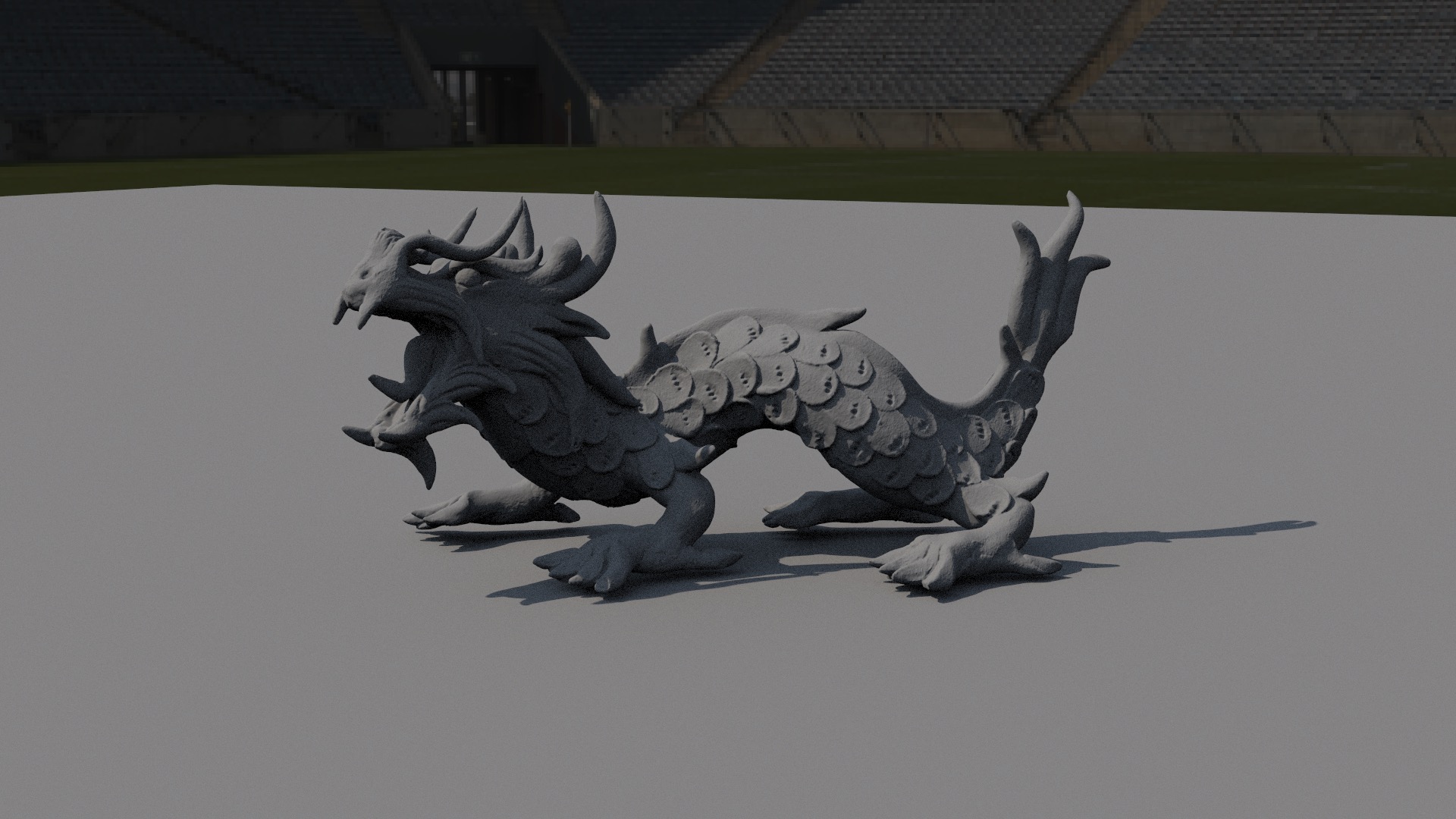} &
\includegraphics[width=0.31\linewidth]{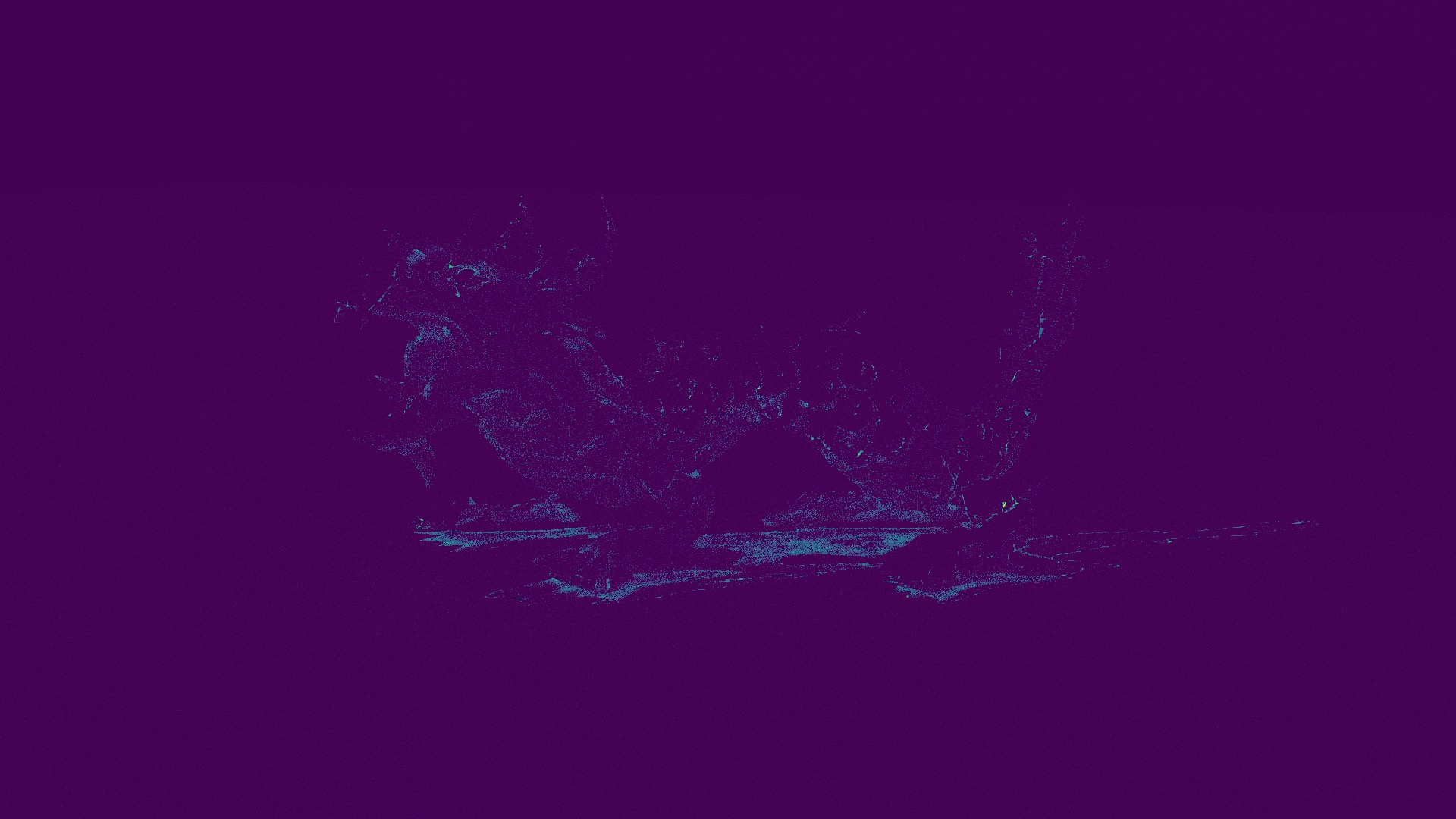} \\

\raisebox{0.08\linewidth}{\rotatebox[origin = c]{-90}{\textsf{\scriptsize{\textsc{Centaur A}}}}}&
\includegraphics[width=0.31\linewidth]{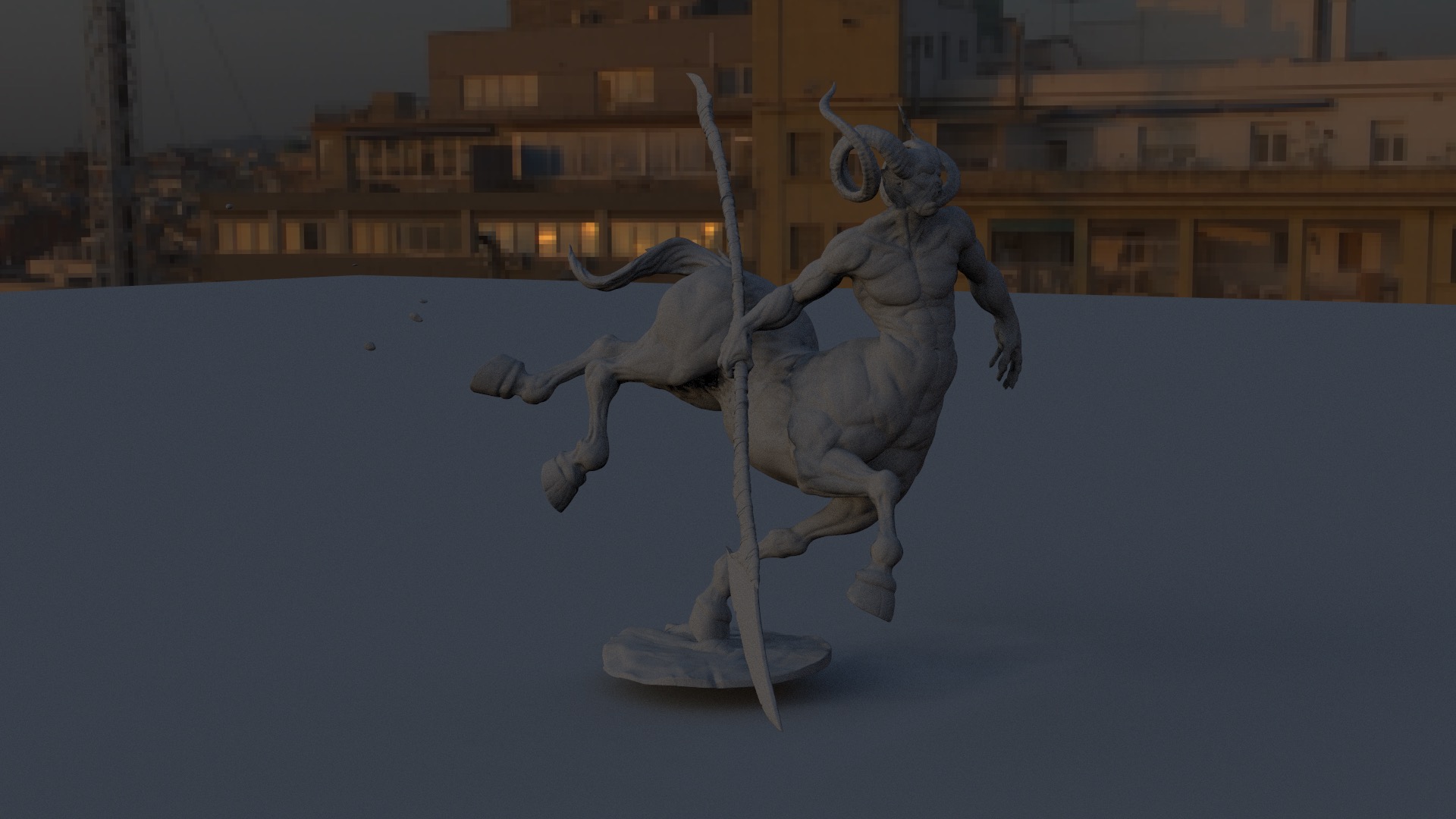} &
\includegraphics[width=0.31\linewidth]{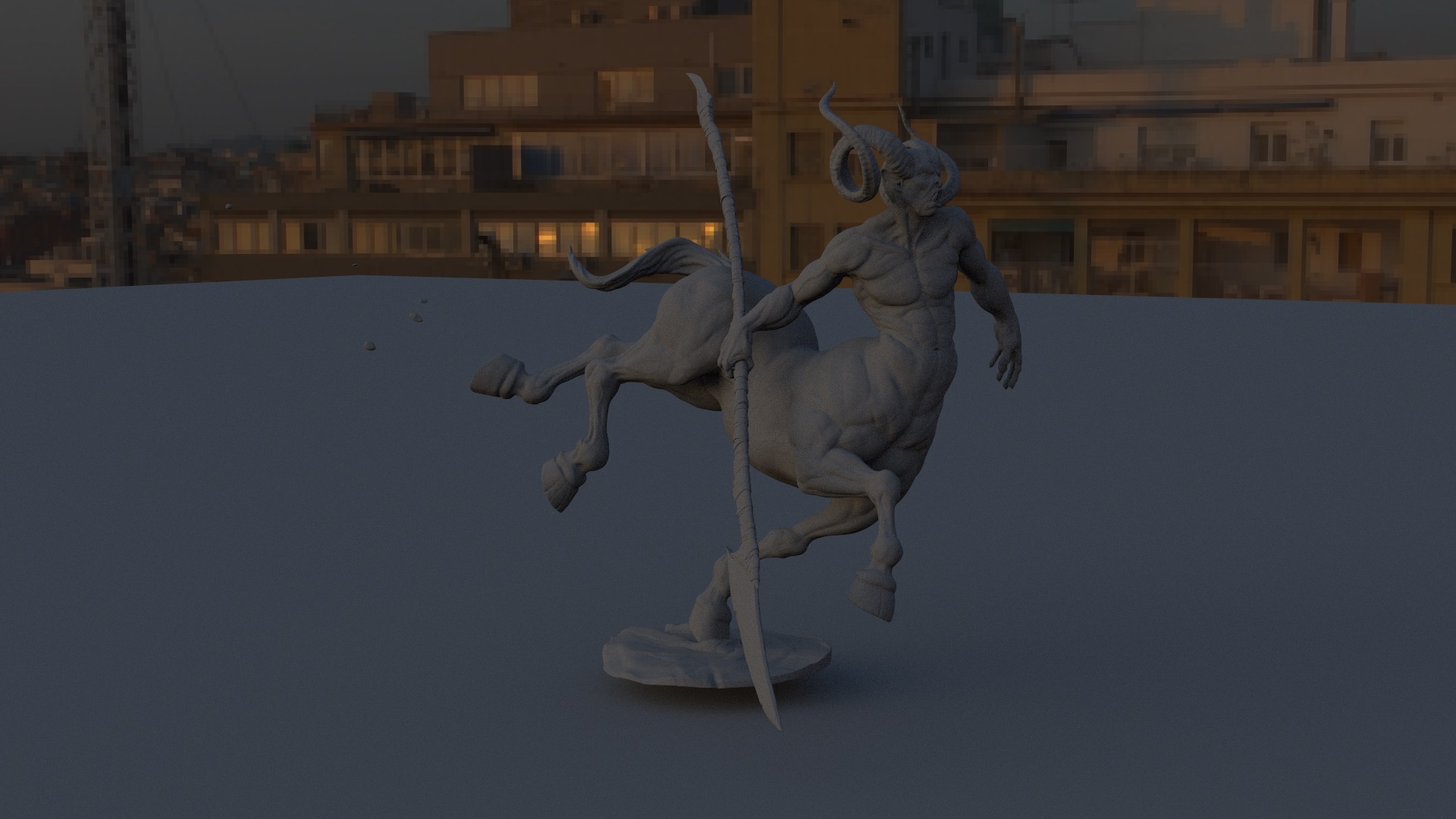} &
\includegraphics[width=0.31\linewidth]{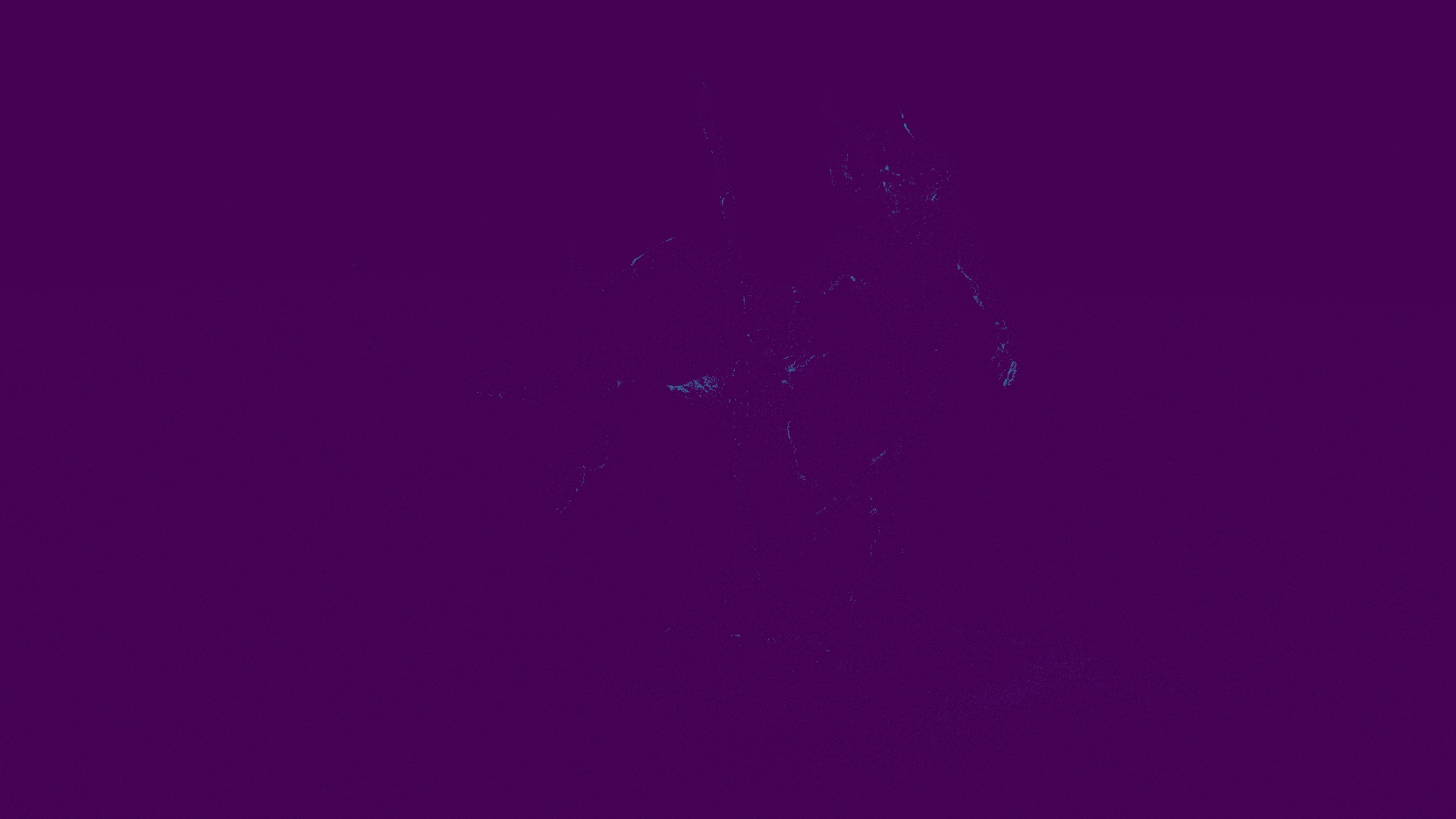} \\

\raisebox{0.08\linewidth}{\rotatebox[origin = c]{-90}{\textsf{\scriptsize{\textsc{Centaur B}}}}}&
\includegraphics[width=0.31\linewidth]{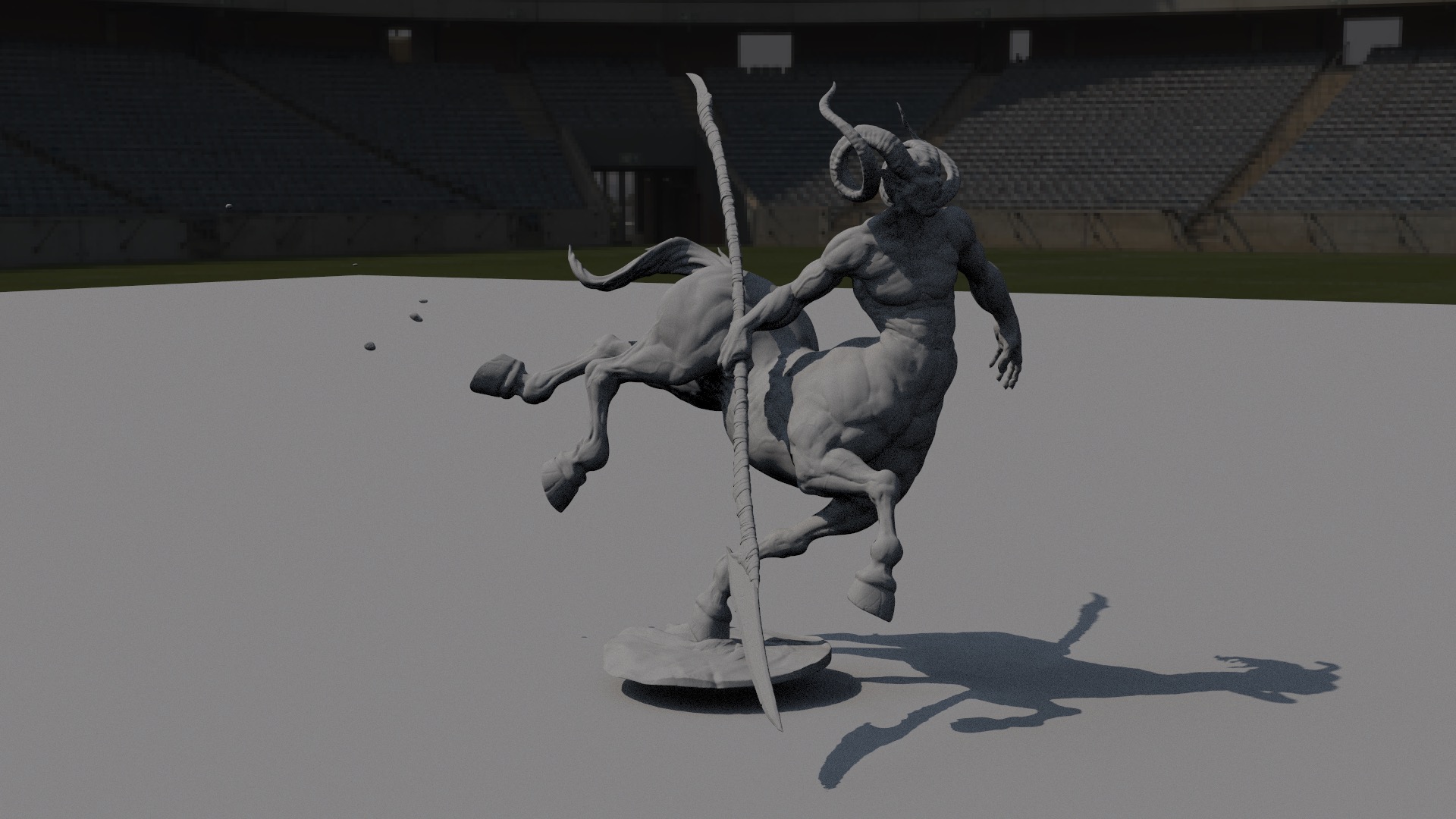} &
\includegraphics[width=0.31\linewidth]{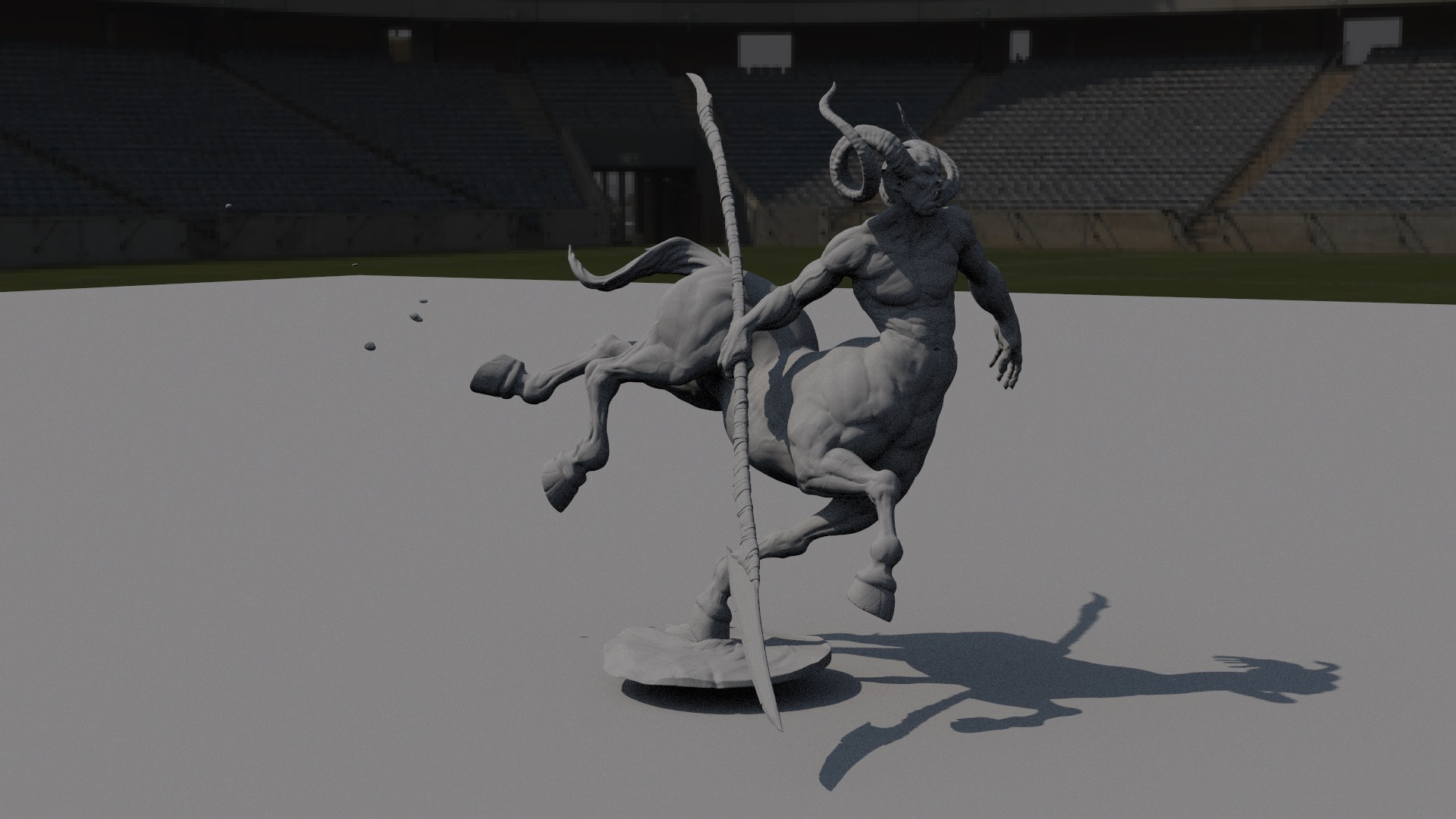} &
\includegraphics[width=0.31\linewidth]{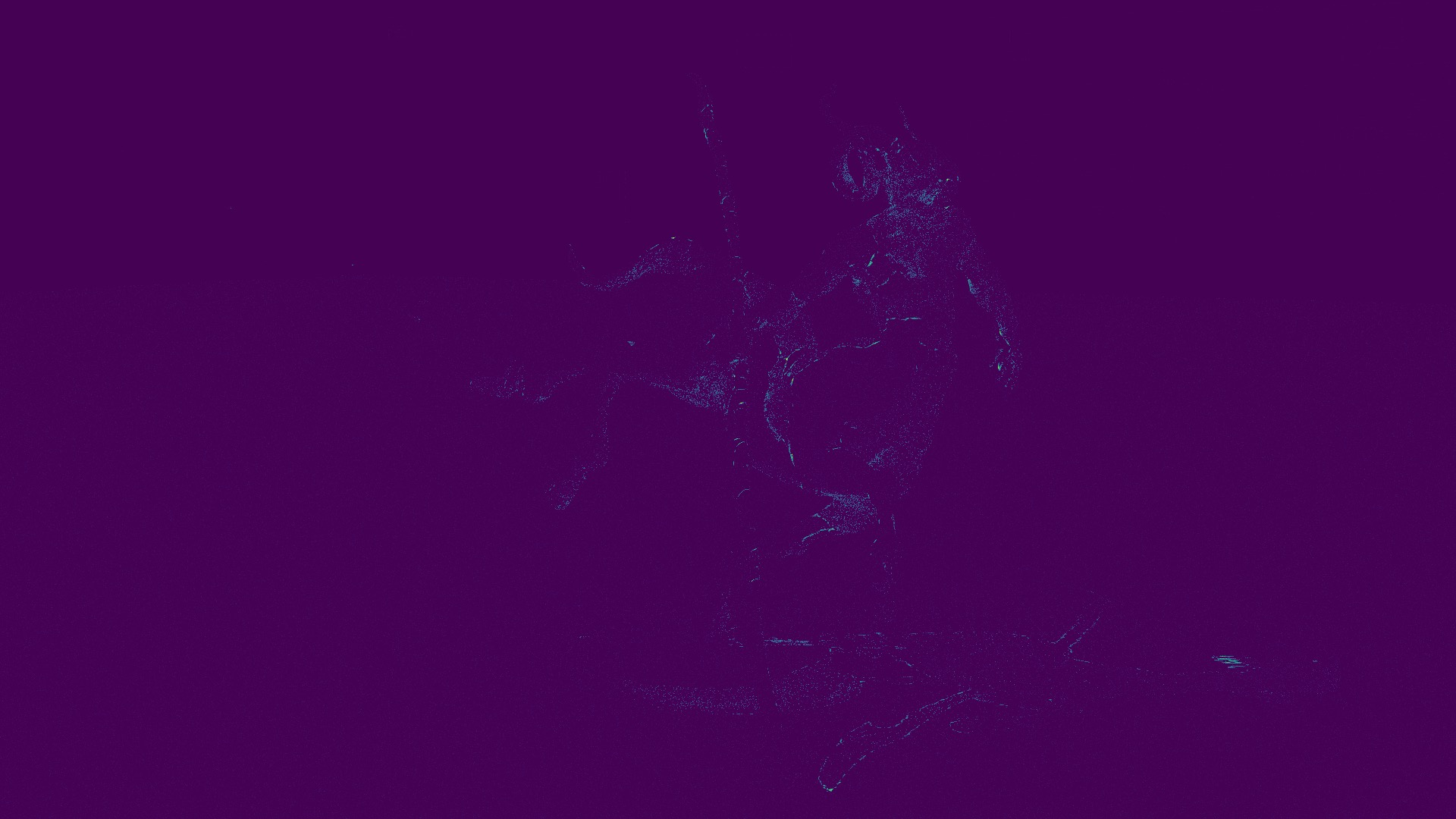} \\

\raisebox{0.08\linewidth}{\rotatebox[origin = c]{-90}{\textsf{\scriptsize{\textsc{Statuette}}}}}&
\includegraphics[width=0.31\linewidth]{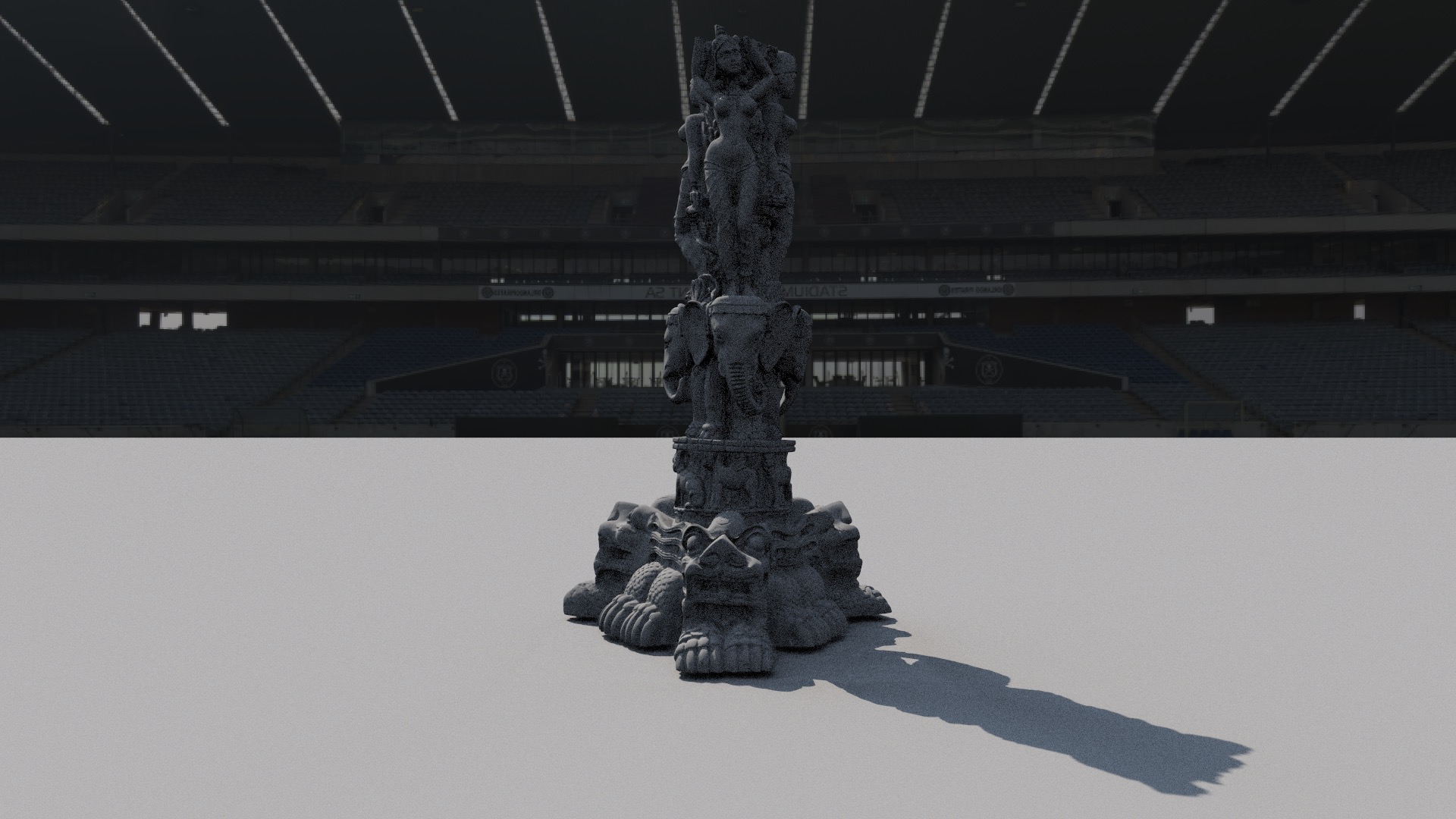} &
\includegraphics[width=0.31\linewidth]{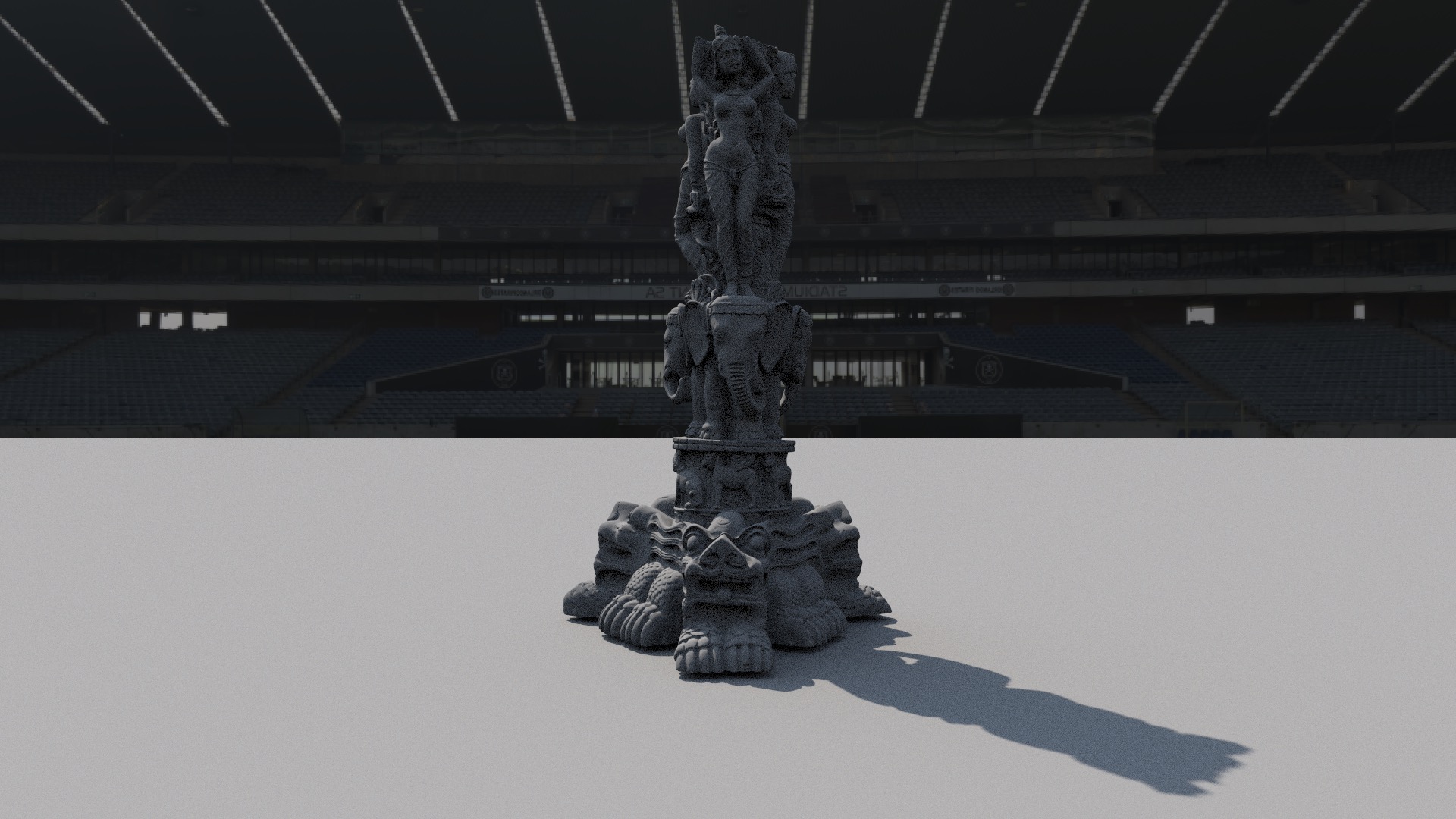} &
\includegraphics[width=0.31\linewidth]{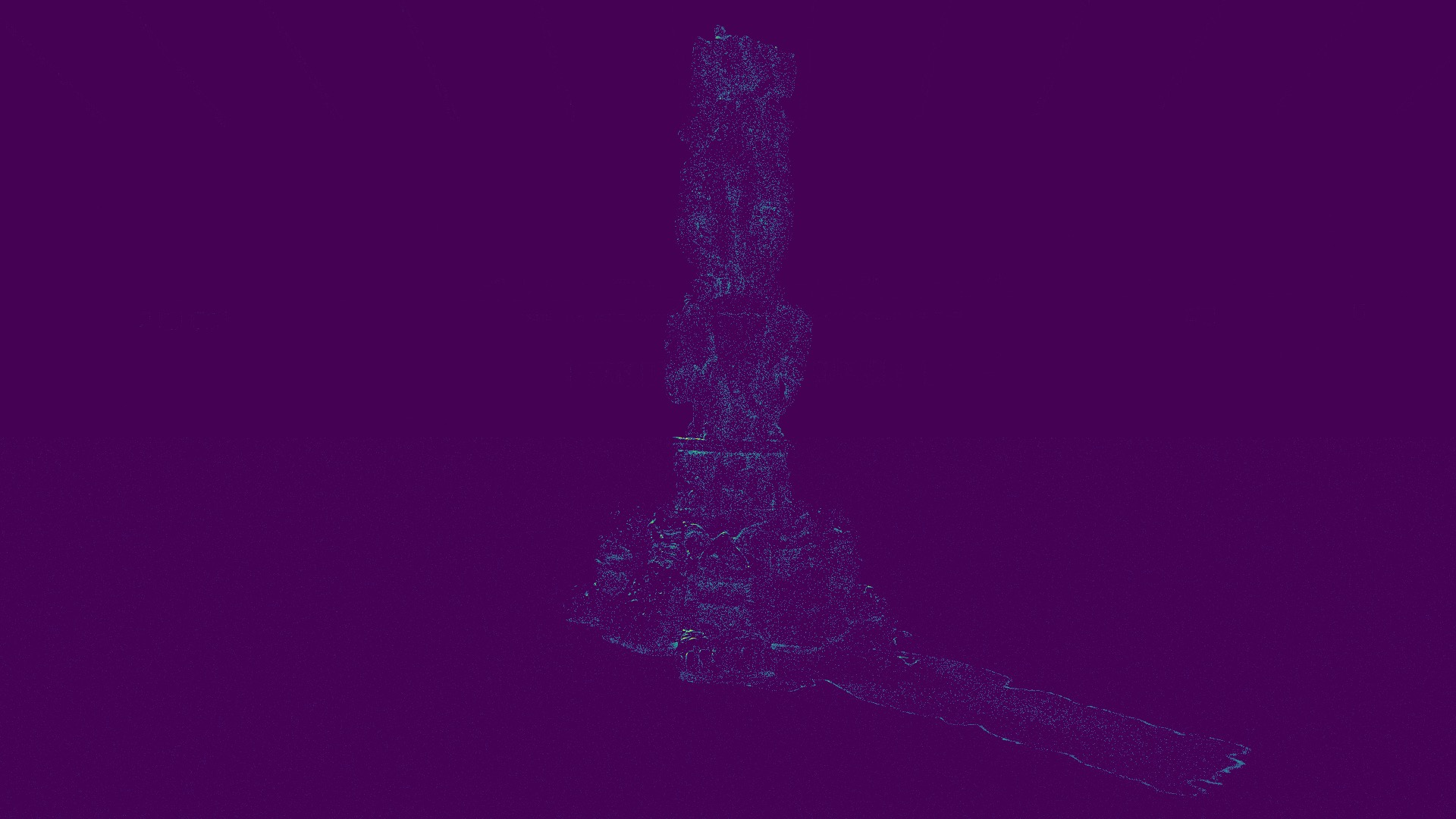} \\

\raisebox{0.08\linewidth}{\rotatebox[origin = c]{-90}{\textsf{\scriptsize{\textsc{Statuette Low}}}}}&
\includegraphics[width=0.31\linewidth]{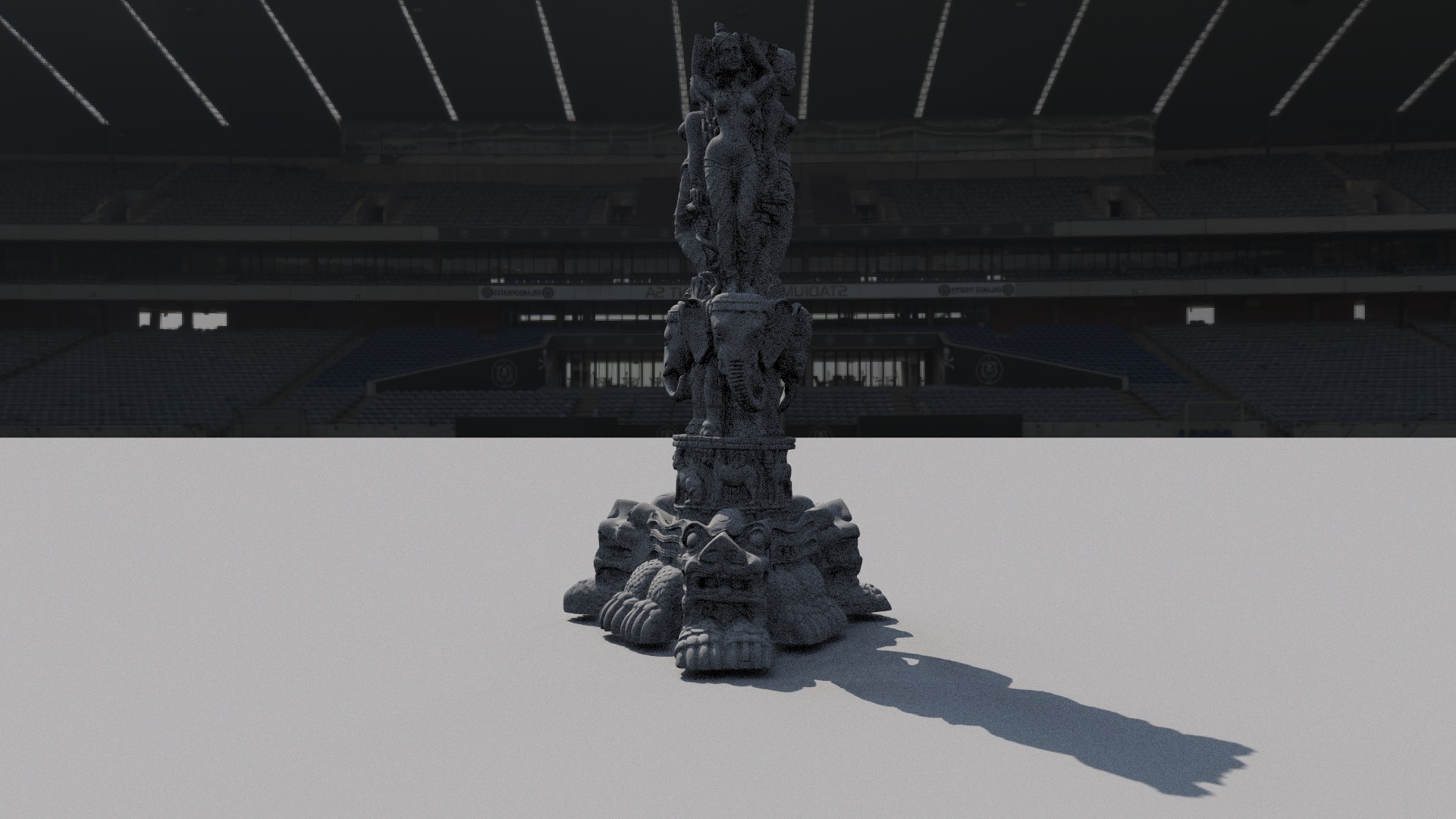} &
\includegraphics[width=0.31\linewidth]{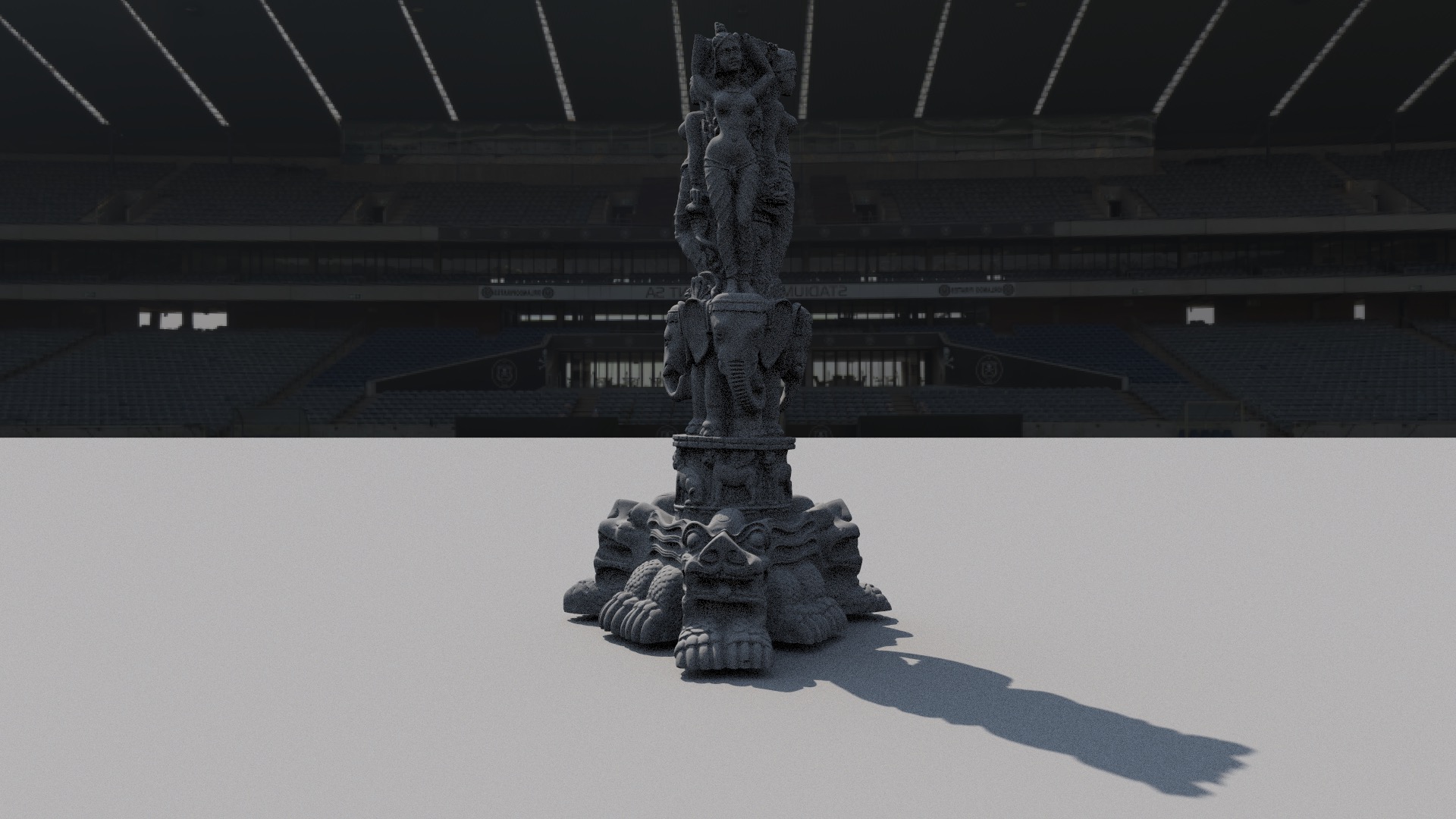} &
\includegraphics[width=0.31\linewidth]{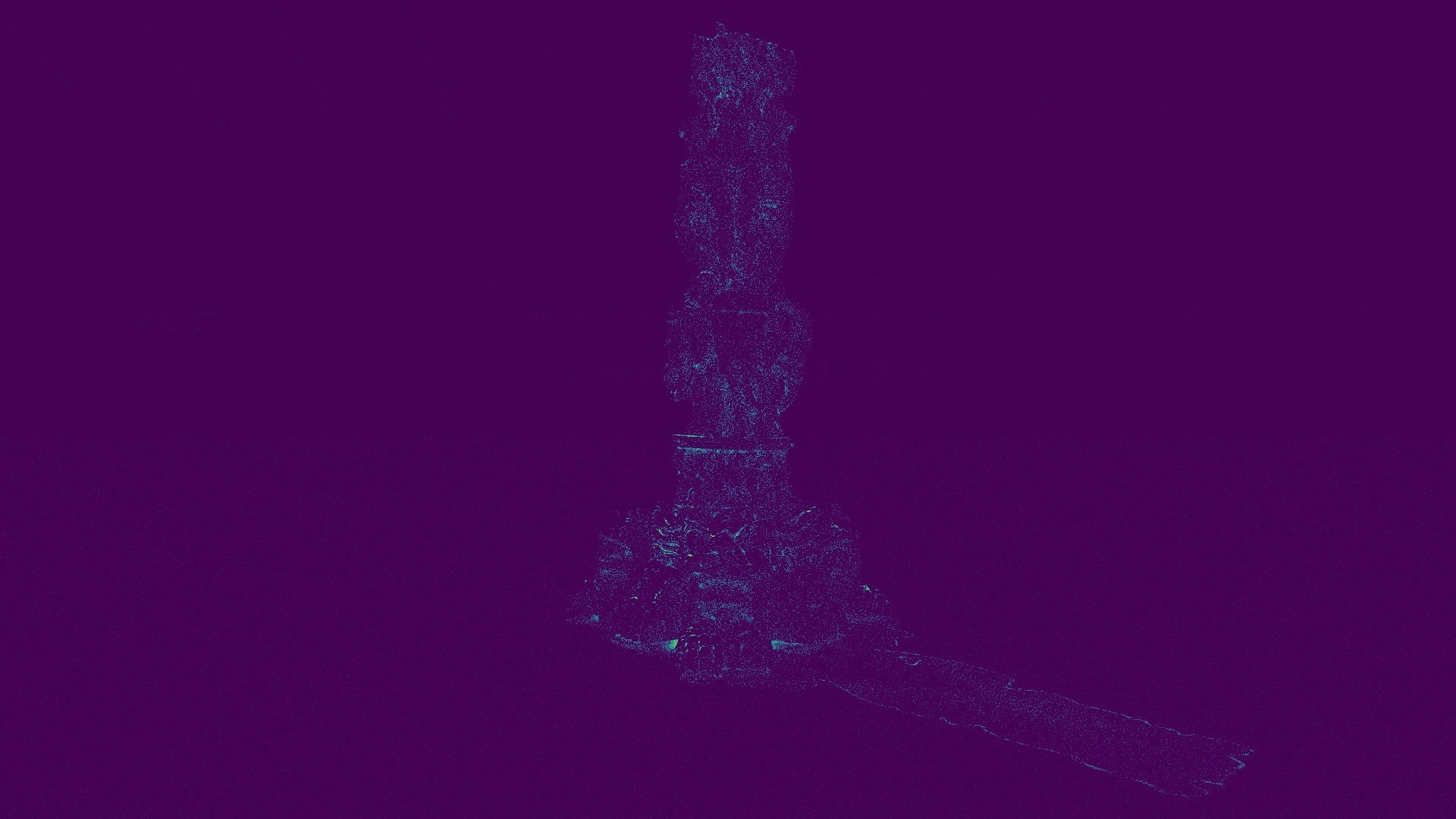} \\

\raisebox{0.08\linewidth}{\rotatebox[origin = c]{-90}{\textsf{\scriptsize{\textsc{Statuettes}}}}}&
\includegraphics[width=0.31\linewidth]{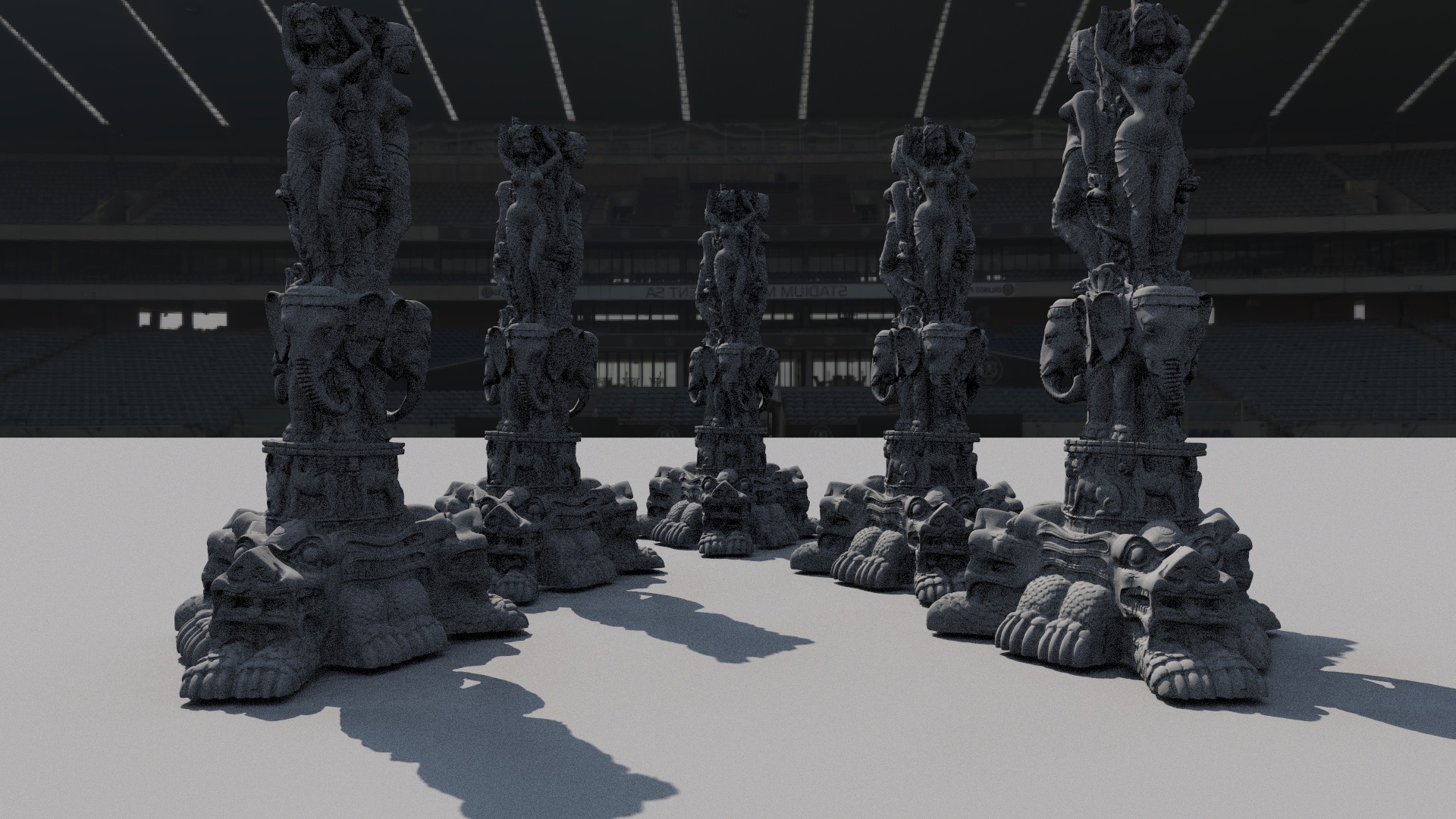} &
\includegraphics[width=0.31\linewidth]{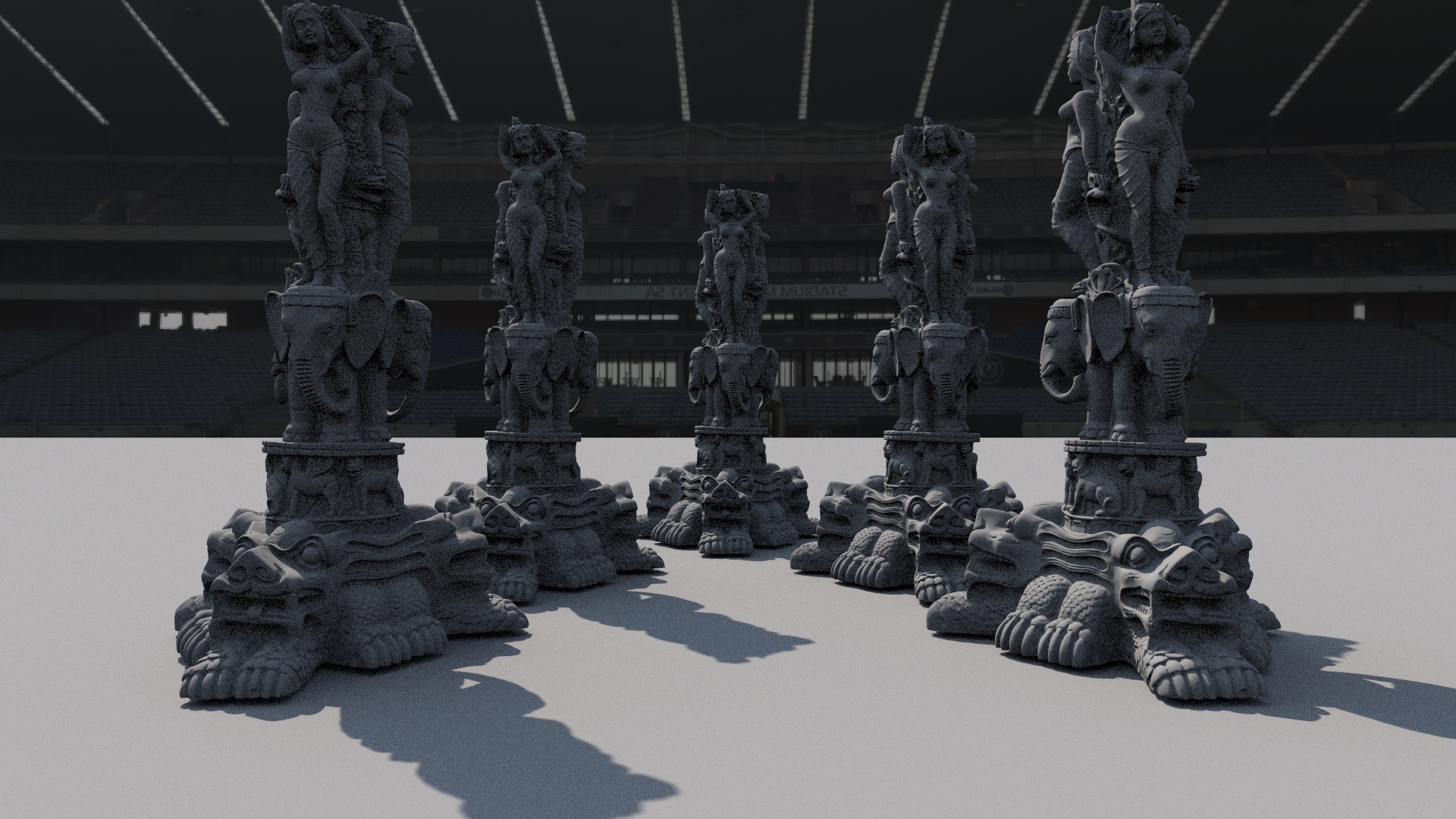} &
\includegraphics[width=0.31\linewidth]{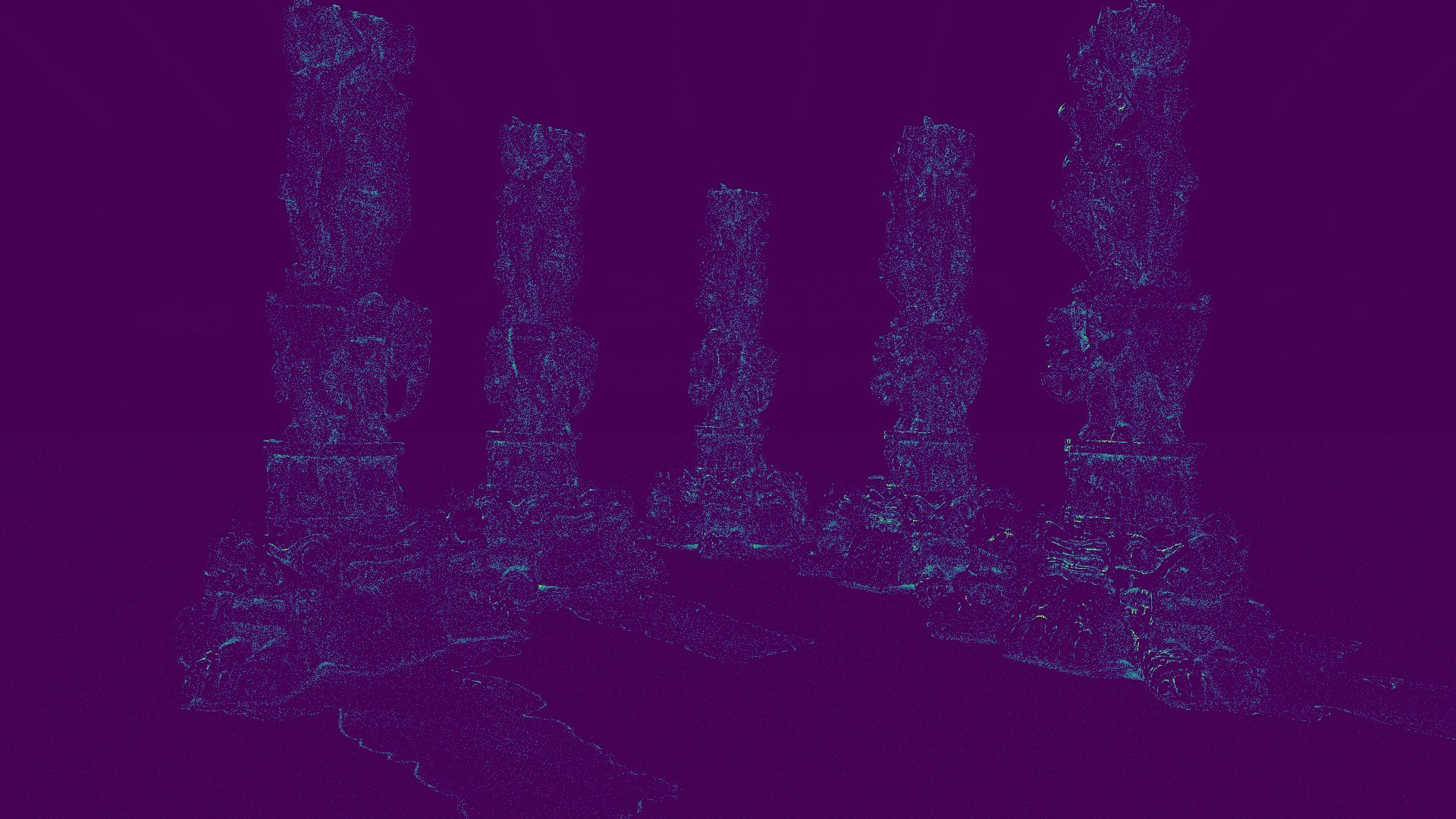} \\

\raisebox{0.08\linewidth}{\rotatebox[origin = c]{-90}{\textsf{\scriptsize{\textsc{Statues A}}}}}&
\includegraphics[width=0.31\linewidth]{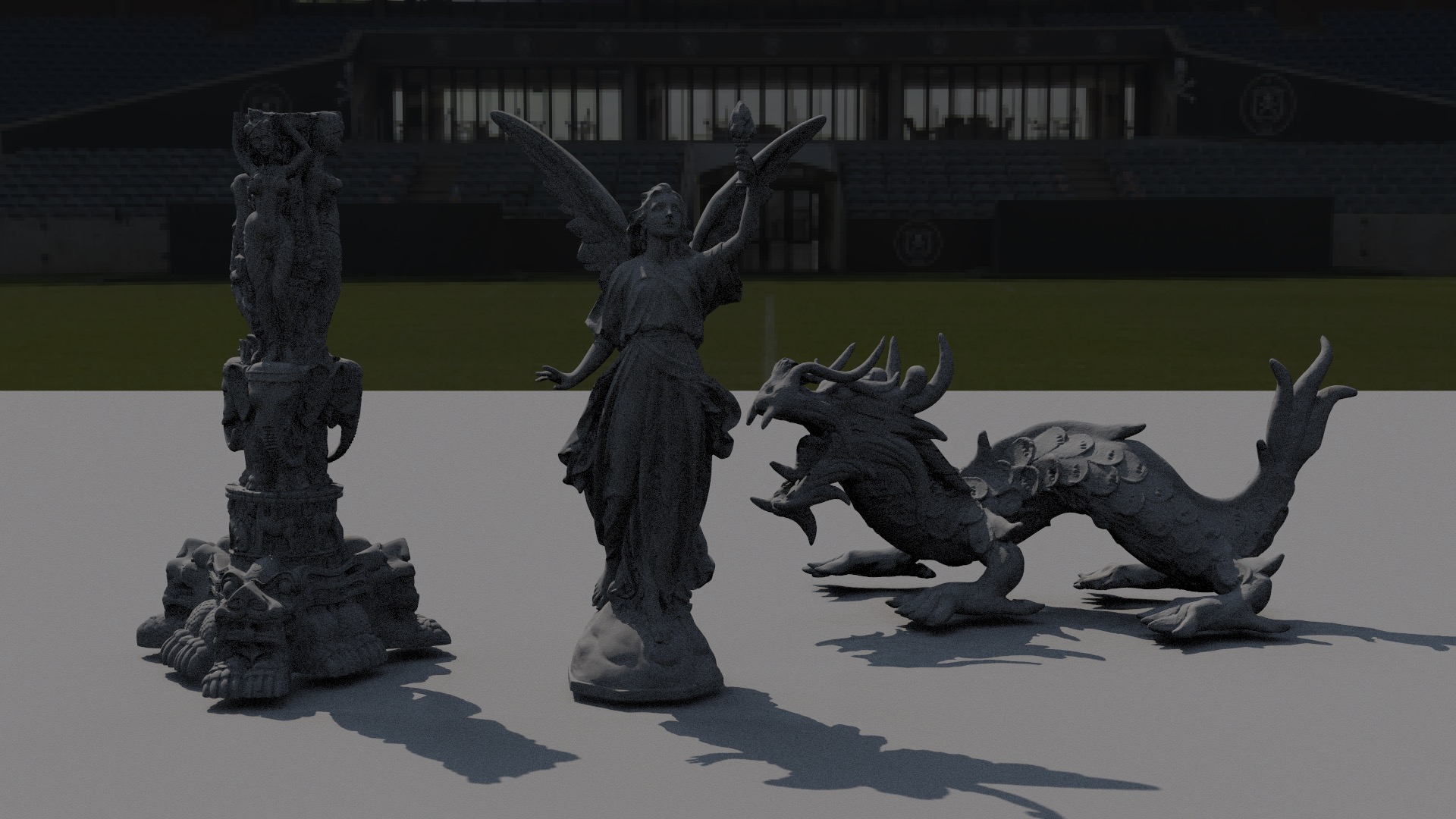} &
\includegraphics[width=0.31\linewidth]{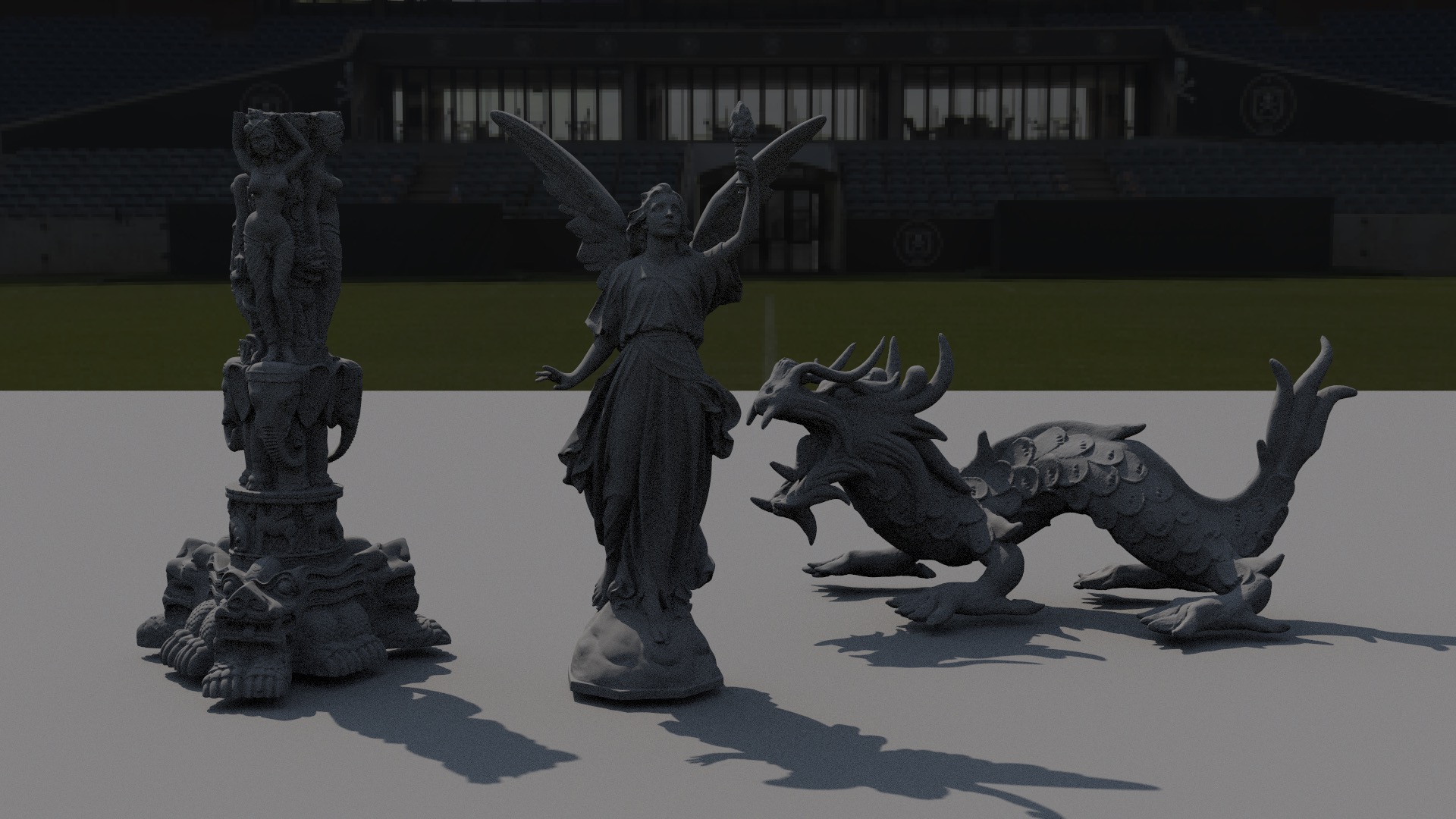} &
\includegraphics[width=0.31\linewidth]{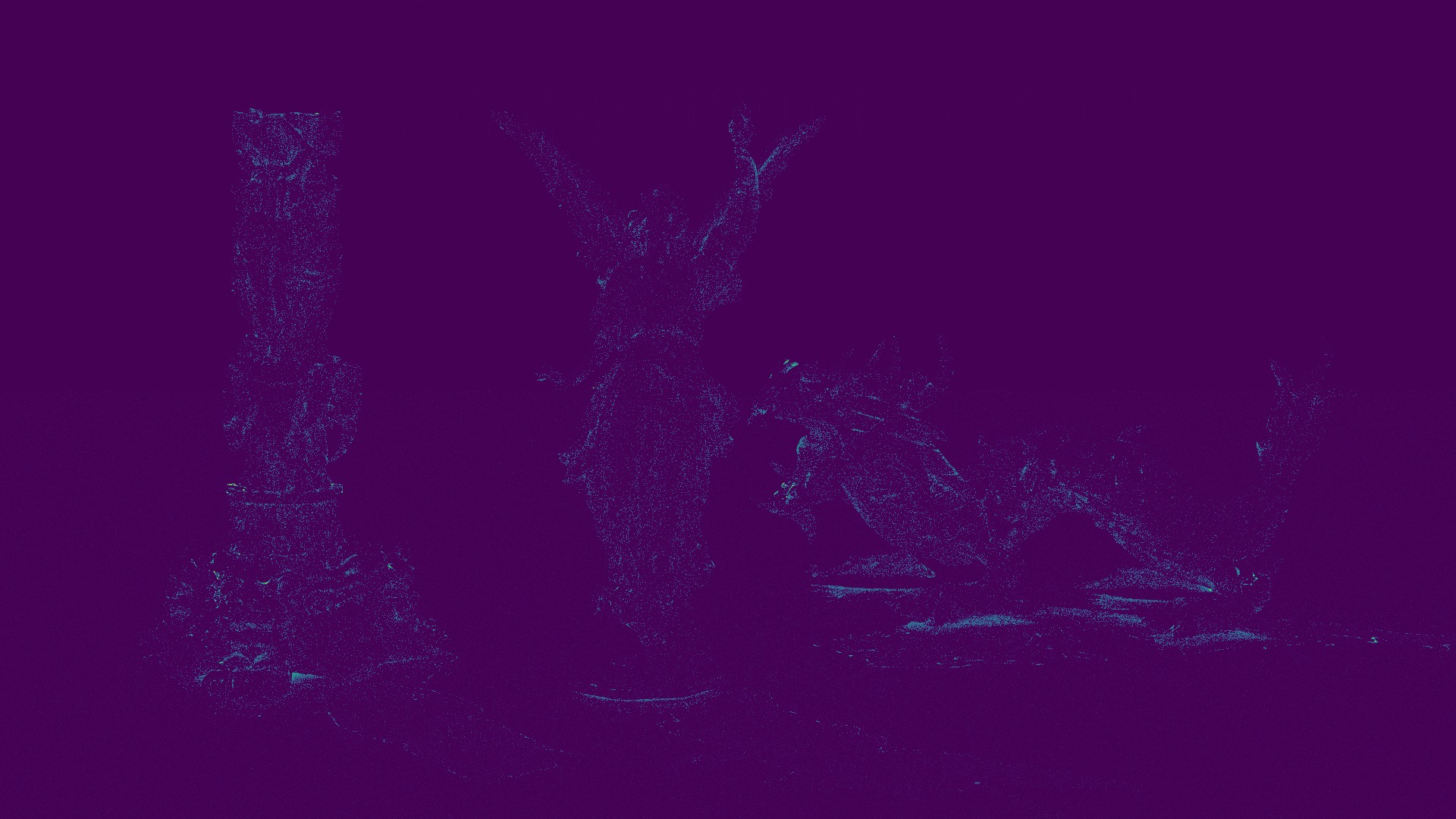} \\

\raisebox{0.08\linewidth}{\rotatebox[origin = c]{-90}{\textsf{\scriptsize{\textsc{Statues B}}}}}&
\includegraphics[width=0.31\linewidth]{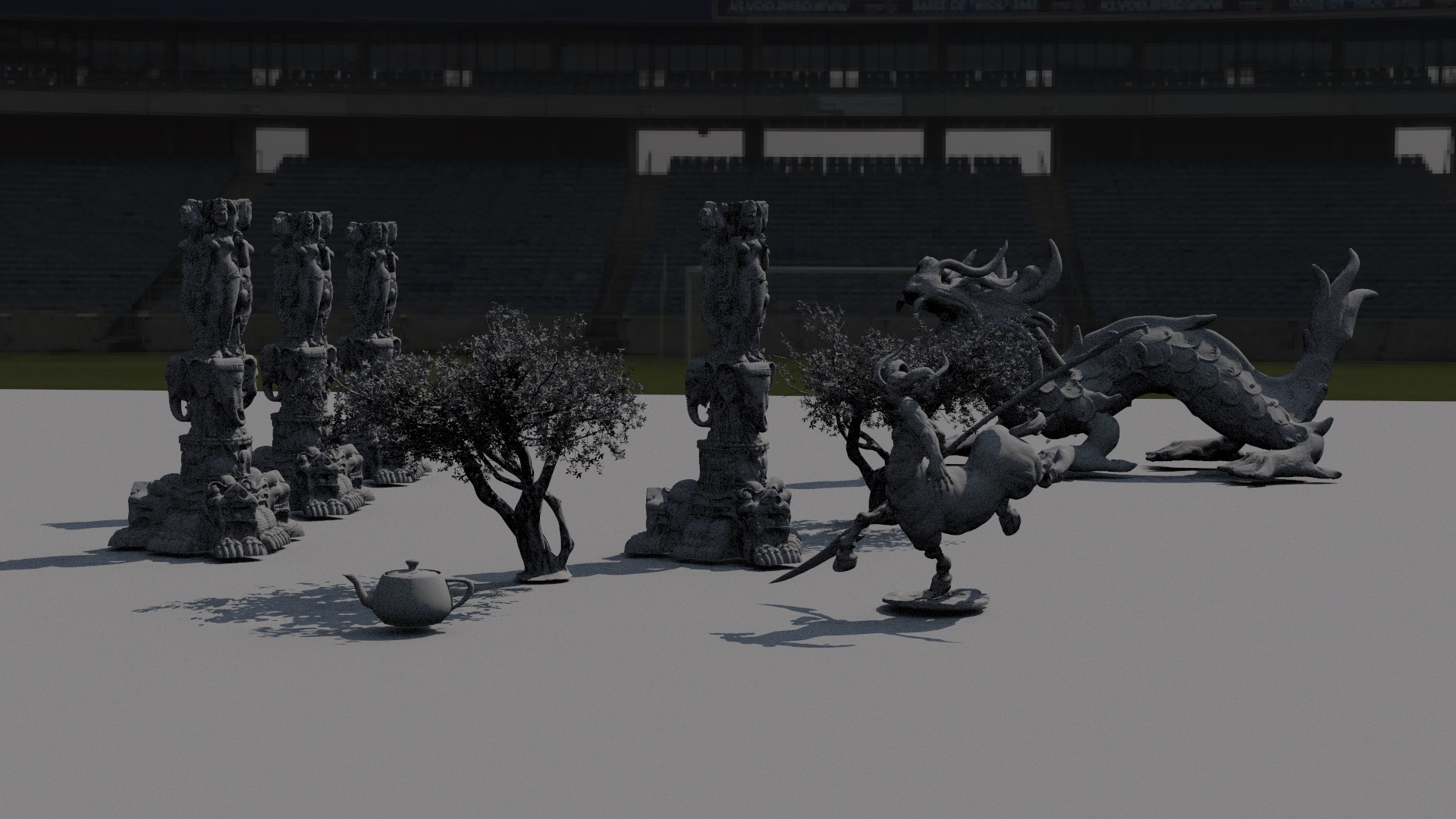} &
\includegraphics[width=0.31\linewidth]{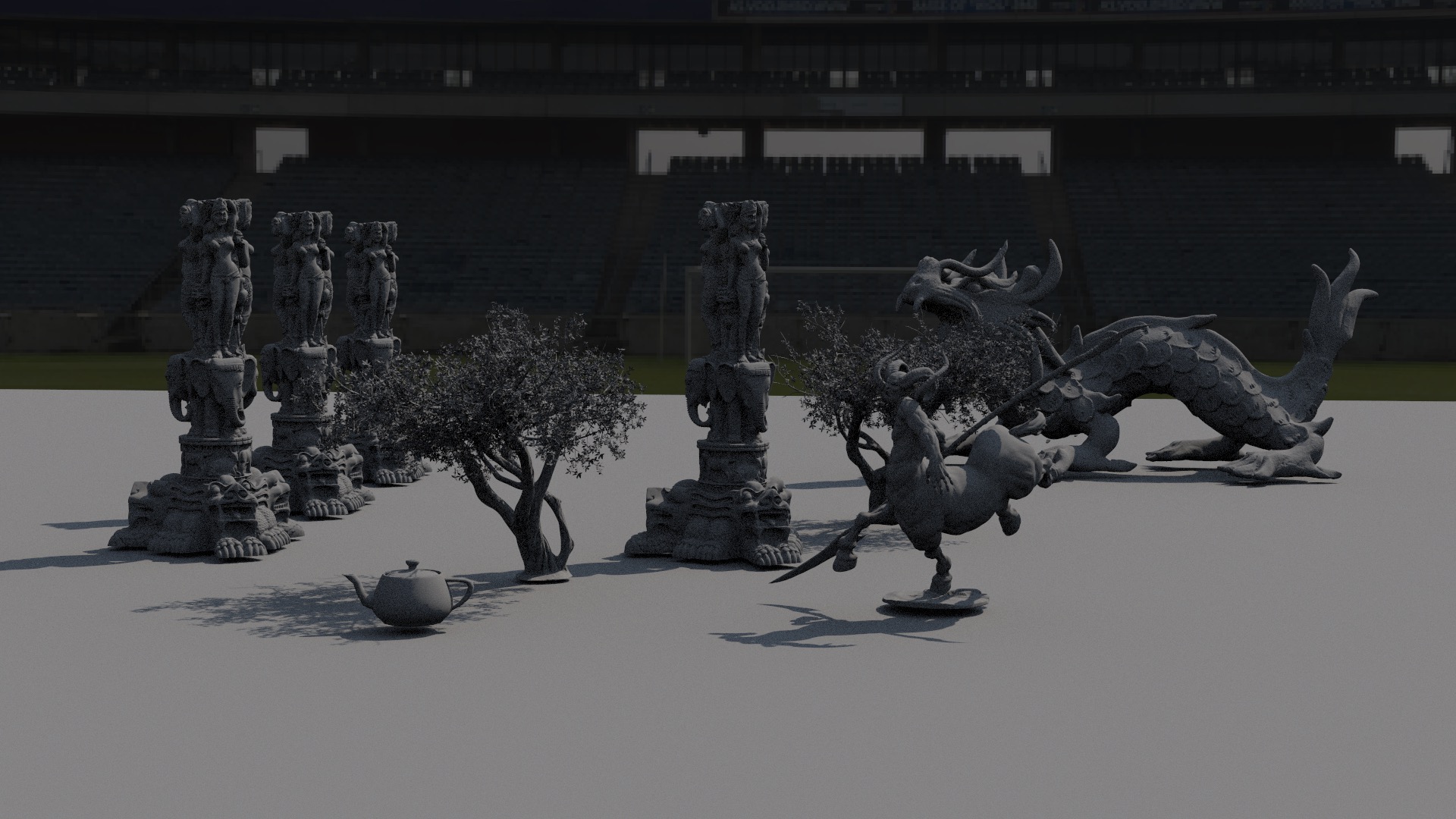} &
\includegraphics[width=0.31\linewidth]{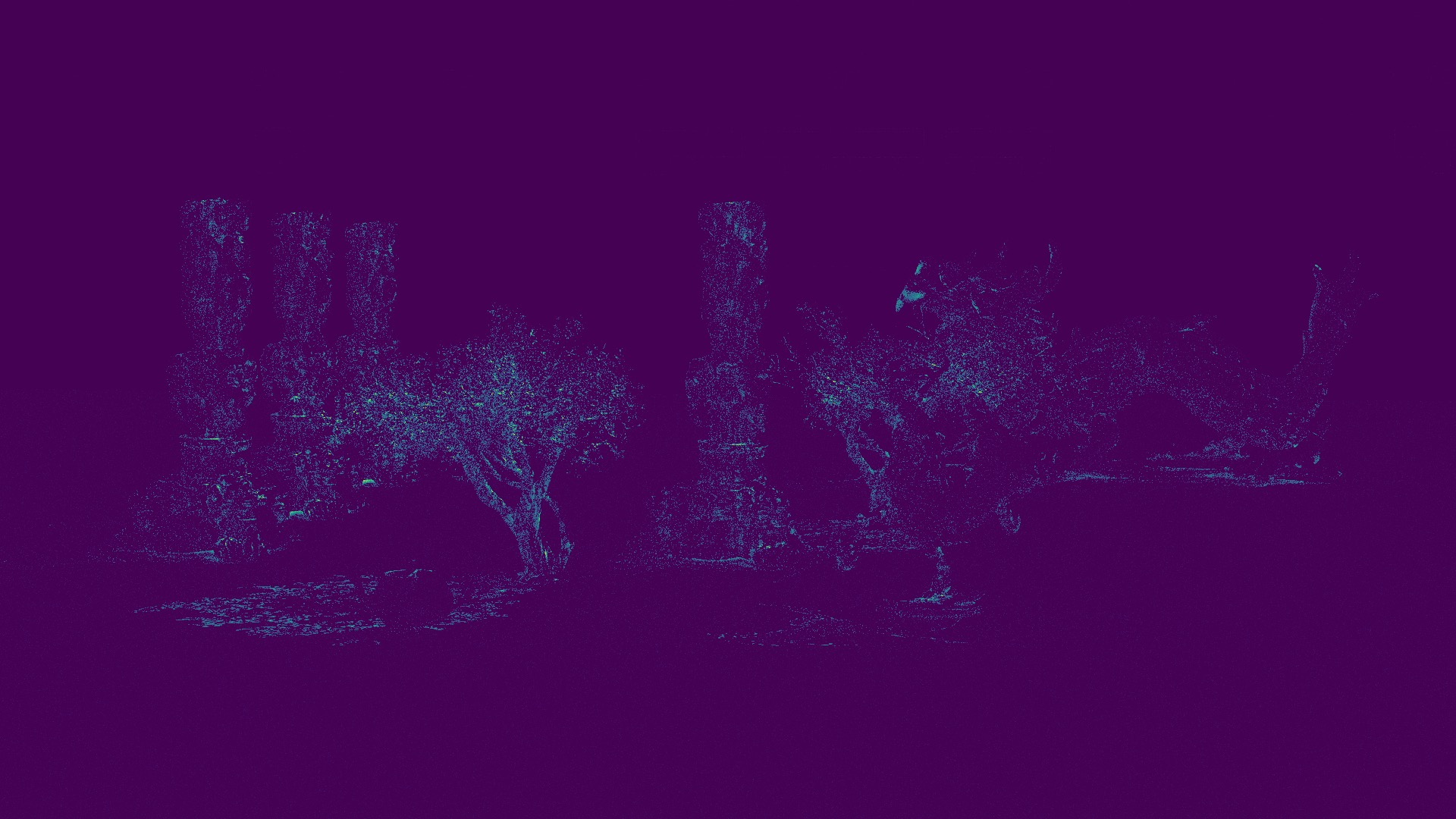} \\

&\small{(a) NIF}&\small{(b) Reference}&\small{(c) Error of (a) $\times 3$}
\end{tabular}
    \caption{The scenes for performance evaluation rendered after 32 training spp with 128 inference spp.}
    \label{fig:examples0}
\end{figure}

Based on the results, we can see that the execution time is not correlated to the scene complexity or the number of polygons in the scene, as depicted in Fig.~\ref{fig:graph:num_poly_ray_perf}a. Instead, as shown in Fig.~\ref{fig:graph:num_poly_ray_perf}b, the execution time is linearly proportional to the number of rays processed by NIF which is the summation of (b) and (c) in Table~\ref{tab:perf}. These results show that NIF is effective when applied to a complex geometry. 
Since the computational cost is not proportional to the model complexity in NIF, applying NIF to a model with low complexity could not gain much benefit from it. This is shown in the \textsc{Statuette Low} scene where there are only $0.5$K triangles, which results in no performance improvement. 
Comparing the \textsc{Statuette} and the \textsc{Statuette Low} scenes also shows this nature.
Although both scenes look almost the same, the BVH ray cast time has a considerable difference between these two scenes due to the bottom-level BVH traversal. On the other hand, NIF only takes almost the same time for these two scenes. 
Note that the training time is also linearly proportional to the amount of input data. In this experiment, it varies from $0.1$ seconds (\textsc{Centaur A}) to $1$ second (\textsc{Statuettes}) for a single spp.
The test scenes and their corresponding differences are depicted in Fig.~\ref{fig:examples0}. Note that the differences are amplified to make them more visible.

\begin{table*}
    \caption{Runtime performance breakdown in microseconds, and statistics. (a) is the number of shadow rays cast. (b) and (c) are the number of rays processed by the outer network and inner network, respectively.}
    \label{tab:perf}
    \centering
    \scriptsize
\begin{tabular}{p{15mm}||c c c c c c c c c c}
\toprule 
Scene           & \textsc{Dragon A}  & \textsc{Dragon B}  & \textsc{Centaur A} & \textsc{Centaur B} & \textsc{Statuette}   & \textsc{Statuette Low}  & \textsc{Statuettes} & \textsc{Statues A} & \textsc{Statues B}   \\ 
\# of Triangles & 7.2M     & 7.2M     & 2.5M     & 2.5M     & 10M     & 0.5M     & 50M     & 17.5M      & 52.8M                   \\ 
(a)      & 1.2M      & 1.3M      & 1.1M      & 1.2M      & 855K      & 855K      & 803K      & 855K      & 942K                    \\
(b)   & 366K      & 387K      & 243K      & 285K      & 171K      & 172K       & 445K      & 376K       & 298K                     \\
(c)   & 130K       & 141K       & 110K       & 126K       & 44K      & 44K       & 176K       & 124K       & 137K                      \\
\midrule
BVH Ray Cast & 625.26 & 578.07 & 592.02 & 595.00 & 391.00 & 260.56 & 774.05 & 692.09 & 680.82 \\
\midrule
NIF: \\
Ray Cast        & 139.49     & 135.53    & 123.39    & 120.77    & 86.21 & 91.8  & 119.86  & 113.05  & 141.01    \\
Outer Grid   & 49.55     & 34.85     & 37.99     & 29.09      & 21.43 & 21.09   & 49.3  & 43.34  & 45.55     \\ 
Outer Inference & 185.12     & 193.19   & 127.17    & 144.57     & 92.15 & 92.23 & 220.67   & 189.91 & 153.24   \\
Inner Grid   & 46.1     & 48.78     & 45.09      & 41.12     & 19.08 & 19.08  & 67.76   & 54.59  & 63.07     \\
Inner Inference	&  84.75    & 88.08    & 72.53    & 81.03     & 36.18 & 35.81  & 108.98  & 81.93  & 88.67    \\
\midrule
NIF Total       & 505.01    & 500.43  & 406.17    & 416.58    & 255.05 & 260.01 & 566.57  & 482.82 & 491.54     \\
\midrule
Speedup & 1.24 & 1.16	& 1.46 & 1.43 & \underline{1.53} & 1.00	& 1.37 & 1.43 & 1.39  \\
\bottomrule
\end{tabular}

\end{table*}

    




Additionally, we demonstrated that our method can also accelerate ray casting in a more complex scene and be used in one unified rendering framework while preserving the conventional ray-tracing pipeline with BVHs, which is shown in Fig.~\ref{fig:teaser} where five complex models are placed in the \textsc{Bistro} scene.
As discussed above, applying NIF to a model with low complexity is ineffective, thus in this scene, we execute NIF only for the complex models which have more than $100$K triangles while we traverse BVHs for other simple geometries.
Therefore, we can embed NIF into a regular ray tracing pipeline and convert some parts of highly divergent workloads (i.e. the bottom-level BVH) into more coherent workloads. 
Even for this complex case, our method can reduce the ray casting time by about $15 \%$ while preserving the image quality.

\section{Conclusions and Future Work}
In this paper, we introduced a novel approach called Neural Intersection Function to accelerate ray casting. It acts as an alternative to the bottom-level BVH traversal and can be integrated into the existing ray tracing pipeline. Together with its two neural networks, outer and inner, and the specialized input parameterization featuring grid encoding to eradicate aliased rays, we showed that NIF can represent geometry with sufficient accuracy, so the shadow computed using the proposed method is close enough to the one computed using BVH ray casting. Instead of using NN to reconstruct a single object which is all the previous work has focused on, we showed that it is possible to embed NN into a rendering pipeline which has never been shown before. 
The experiments demonstrate that our method can significantly improve performance while preserving image quality, especially for a scene with a large number of triangles. Depending on the scene and how complex the object model is, a maximum of $1.53 \times$ speedup is observed. This is the performance as of today. However, the relative performance in the future hardware depends on the architecture. A lot of effort has been invested to accelerate NN and matrix multiplication which also benefits NIF, making it likely that the relative performance of NIF to increase in the future. 

Although NIF provides several advantages, it also bears some limitations in our current implementation.
Currently, we only use rays based on the current viewpoint to train NIF.
The accuracy of NIF would be lower if used for queries from arbitrary rays that deviate too much; therefore, NIF is required to be trained again in order to achieve better accuracy for other viewpoints and light sources.
This is a drawback of the simplification and a design trade-off we did for our implementation.
Extension to dynamic scenes efficiently with our training methodology remains our future work.
Additionally, the proposed method can be further optimized to improve the efficiency of ray querying by reducing redundancies. For example, theoretically, we do not need to query the same ray from NIF multiple times if any of the intersection points to an object has been found already. Therefore, despite being considered marginal, there is an opportunity for optimization, although it comes with the complexity in the execution logic and the GPU kernel design which is likely not worth the effort as of today.
Last but not least, to evaluate the performance in our paper, we used NIF to compute the visibility of secondary rays.
However, as shown in Fig.~\ref{fig:gridResCmp}, NIF can be also used for primary ray casting, encoding other information such as shading normal and depth than visibility.
NIF should be extended to encode more geometric information, but we still need further analysis for more practical use to investigate if NIF is applicable to other types of rays such as deeper paths of shadow rays.

\section*{Acknowledgments}
We are grateful to our team members for their help with the neural network implementation and optimizations, and proofreading.
We would also like to thank the Amazon Lumberyard team for the \textsc{Bistro} scene~\cite{ORCAAmazonBistro}, the Stanford Computer Graphics Laboratory for \textsc{Stanford Bunny}, \textsc{Dragon}, \textsc{Lucy}, \textsc{Asian Dragon}, and \textsc{Thai Statue} models, Michael Milano for \textsc{Centaur} model.

\bibliographystyle{eg-alpha-doi} 
\bibliography{main}        


\end{document}